\documentclass[preprint]{ptephy_v1}

\preprintnumber{LHCb-PROC-2023-003, Nikhef-2023-005} 
\newboolean{articletitles}
\setboolean{articletitles}{true} 

\newboolean{uprightparticles}
\setboolean{uprightparticles}{false} 

\let\eg\undefined

\usepackage{mciteplus}
\usepackage{maybemath}
\usepackage{colortbl}
\usepackage{xcolor}
\usepackage{multirow}
\usepackage{hyperref}
\usepackage{wrapfig}

\usepackage{xspace} 
\usepackage{upgreek}

\def\MagUp {\mbox{\em Mag\kern -0.05em Up}\xspace}

\ifthenelse{\boolean{uprightparticles}}%
{
 
 \def\Pgamma      {\ensuremath{\upgamma}\xspace}

 \def\Pmu         {\ensuremath{\upmu}\xspace}                 
 \def\Pnu         {\ensuremath{\upnu}\xspace}                 
                  
 \def\Ppi         {\ensuremath{\uppi}\xspace}                 
                  
 \def\Prho        {\ensuremath{\uprho}\xspace}                 
                  
 \def\Ptau        {\ensuremath{\uptau}\xspace}                 
                  
 \def\Pphi        {\ensuremath{\upphi}\xspace}                 
                  
 \def\Pchi        {\ensuremath{\upchi}\xspace}                 
 \def\Ppsi        {\ensuremath{\uppsi}\xspace}                 
 \def\Pomega      {\ensuremath{\upomega}\xspace}                 

 \def\PDelta      {\ensuremath{\Delta}\xspace}                 
 \def\PXi         {\ensuremath{\Xi}\xspace}                 
 \def\PLambda     {\ensuremath{\Lambda}\xspace}                 
 \def\PSigma      {\ensuremath{\Sigma}\xspace}                 
 \def\POmega      {\ensuremath{\Omega}\xspace}                 
 \def\PUpsilon    {\ensuremath{\Upsilon}\xspace}
 \let\oldPi\Pi
 \def\PPi         {\ensuremath{\oldPi}\xspace}

 \def\PB      {\ensuremath{\mathrm{B}}\xspace}                 
                  
 \def\PD      {\ensuremath{\mathrm{D}}\xspace}

 \def\PJ      {\ensuremath{\mathrm{J}}\xspace}                 
 \def\PK      {\ensuremath{\mathrm{K}}\xspace}

 \def\PW      {\ensuremath{\mathrm{W}}\xspace}

 \def\PZ      {\ensuremath{\mathrm{Z}}\xspace}                 
                  
 \def\Pb      {\ensuremath{\mathrm{b}}\xspace}                 
 \def\Pc      {\ensuremath{\mathrm{c}}\xspace}                 
 \def\Pd      {\ensuremath{\mathrm{d}}\xspace}                 
 \def\Pe      {\ensuremath{\mathrm{e}}\xspace}

 \def\Pi      {\ensuremath{\mathrm{i}}\xspace}

 \def\Pp      {\ensuremath{\mathrm{p}}\xspace}                 
 \def\Pq      {\ensuremath{\mathrm{q}}\xspace}                 
                  
 \def\Ps      {\ensuremath{\mathrm{s}}\xspace}                 
 \def\Pt      {\ensuremath{\mathrm{t}}\xspace}                 
 \def\Pu      {\ensuremath{\mathrm{u}}\xspace}

 \def\thebaroffset{0.0em}
}
{
 
 \def\Pgamma      {\ensuremath{\gamma}\xspace}

 \def\Pmu         {\ensuremath{\mu}\xspace}                 
 \def\Pnu         {\ensuremath{\nu}\xspace}                 
                  
 \def\Ppi         {\ensuremath{\pi}\xspace}                 
                  
 \def\Prho        {\ensuremath{\rho}\xspace}                 
                  
 \def\Ptau        {\ensuremath{\tau}\xspace}                 
                  
 \def\Pphi        {\ensuremath{\phi}\xspace}                 
                  
 \def\Pchi        {\ensuremath{\chi}\xspace}                 
 \def\Ppsi        {\ensuremath{\psi}\xspace}                 
 \def\Pomega      {\ensuremath{\omega}\xspace}                 
 \mathchardef\PDelta="7101
 \mathchardef\PXi="7104
 \mathchardef\PLambda="7103
 \mathchardef\PSigma="7106
 \mathchardef\POmega="710A
 \mathchardef\PUpsilon="7107
 \mathchardef\PPi="7105
                  
 \def\PB      {\ensuremath{B}\xspace}                 
                  
 \def\PD      {\ensuremath{D}\xspace}

 \def\PJ      {\ensuremath{J}\xspace}                 
 \def\PK      {\ensuremath{K}\xspace}

 \def\PW      {\ensuremath{W}\xspace}

 \def\PZ      {\ensuremath{Z}\xspace}                 
                  
 \def\Pb      {\ensuremath{b}\xspace}                 
 \def\Pc      {\ensuremath{c}\xspace}                 
 \def\Pd      {\ensuremath{d}\xspace}                 
 \def\Pe      {\ensuremath{e}\xspace}

 \def\Pi      {\ensuremath{i}\xspace}

 \def\Pp      {\ensuremath{p}\xspace}                 
 \def\Pq      {\ensuremath{q}\xspace}                 
                  
 \def\Ps      {\ensuremath{s}\xspace}                 
 \def\Pt      {\ensuremath{t}\xspace}                 
 \def\Pu      {\ensuremath{u}\xspace}

 \def\thebaroffset{0.18em}
}
\newcommand{\offsetoverline}[2][\thebaroffset]{\kern #1\overline{\kern -#1 #2}}%

\makeatletter
\ifcase \@ptsize \relax
  \newcommand{\miniscule}{\@setfontsize\miniscule{4}{5}}
\or
  \newcommand{\miniscule}{\@setfontsize\miniscule{5}{6}}
\or
  \newcommand{\miniscule}{\@setfontsize\miniscule{5}{6}}
\fi
\makeatother

\DeclareRobustCommand{\optbar}[1]{\shortstack{{\miniscule (\rule[.5ex]{1.25em}{.18mm})}
  \\ [-.7ex] $#1$}}

\def\epem       {{\ensuremath{\Pe^+\Pe^-}}\xspace}

\def\muon       {{\ensuremath{\Pmu}}\xspace}
\def\mup        {{\ensuremath{\Pmu^+}}\xspace}
\def\mun        {{\ensuremath{\Pmu^-}}\xspace} 

\def\mumu       {{\ensuremath{\Pmu^+\Pmu^-}}\xspace}

\def\taup       {{\ensuremath{\Ptau^+}}\xspace}

\def\ellm       {{\ensuremath{\ell^-}}\xspace}
\def\ellp       {{\ensuremath{\ell^+}}\xspace}

\def\ellell     {\ensuremath{\ell^+ \ell^-}\xspace}

\def\neu        {{\ensuremath{\Pnu}}\xspace}
\def\neub       {{\ensuremath{\overline{\Pnu}}}\xspace}

\def\g      {{\ensuremath{\Pgamma}}\xspace}

\def\W      {{\ensuremath{\PW}}\xspace}

\def\Z      {{\ensuremath{\PZ}}\xspace}

\def\quark     {{\ensuremath{\Pq}}\xspace}
\def\quarkbar  {{\ensuremath{\overline \quark}}\xspace}

\def\uquark    {{\ensuremath{\Pu}}\xspace}
\def\uquarkbar {{\ensuremath{\overline \uquark}}\xspace}

\def\dquark    {{\ensuremath{\Pd}}\xspace}
\def\dquarkbar {{\ensuremath{\overline \dquark}}\xspace}

\def\squark    {{\ensuremath{\Ps}}\xspace}
\def\squarkbar {{\ensuremath{\overline \squark}}\xspace}

\def\cquark    {{\ensuremath{\Pc}}\xspace}
\def\cquarkbar {{\ensuremath{\overline \cquark}}\xspace}

\def\bquark    {{\ensuremath{\Pb}}\xspace}
\def\bquarkbar {{\ensuremath{\overline \bquark}}\xspace}

\def\tquark    {{\ensuremath{\Pt}}\xspace}

\def\pion   {{\ensuremath{\Ppi}}\xspace}
\def\piz    {{\ensuremath{\pion^0}}\xspace}
\def\pip    {{\ensuremath{\pion^+}}\xspace}
\def\pim    {{\ensuremath{\pion^-}}\xspace}
\def\pipm   {{\ensuremath{\pion^\pm}}\xspace}
\def\pimp   {{\ensuremath{\pion^\mp}}\xspace}

\def\rhomeson {{\ensuremath{\Prho}}\xspace}
\def\rhoz     {{\ensuremath{\rhomeson^0}}\xspace}

\def\kaon    {{\ensuremath{\PK}}\xspace}

\def\KorKbar {\kern \thebaroffset\optbar{\kern -\thebaroffset \PK}{}\xspace}
\def\Kz      {{\ensuremath{\kaon^0}}\xspace}

\def\Kp      {{\ensuremath{\kaon^+}}\xspace}
\def\Km      {{\ensuremath{\kaon^-}}\xspace}
\def\Kpm     {{\ensuremath{\kaon^\pm}}\xspace}

\def\KS      {{\ensuremath{\kaon^0_{\mathrm{S}}}}\xspace}

\def\Kstarz  {{\ensuremath{\kaon^{*0}}}\xspace}

\def\Kstar   {{\ensuremath{\kaon^*}}\xspace}

\def\Kstarp  {{\ensuremath{\kaon^{*+}}}\xspace}

\def\Dbar    {{\ensuremath{\offsetoverline{\PD}}}\xspace}
\def\D       {{\ensuremath{\PD}}\xspace}

\def\DorDbar {\kern \thebaroffset\optbar{\kern -\thebaroffset \PD}\xspace}
\def\Dz      {{\ensuremath{\D^0}}\xspace}
\def\Dzb     {{\ensuremath{\Dbar{}^0}}\xspace}
\def\Dp      {{\ensuremath{\D^+}}\xspace}
\def\Dm      {{\ensuremath{\D^-}}\xspace}
\def\Dpm     {{\ensuremath{\D^\pm}}\xspace}
\def\Dmp     {{\ensuremath{\D^\mp}}\xspace}
\def\DpDm    {\ensuremath{\Dp {\kern -0.16em \Dm}}\xspace}
\def\Dstar   {{\ensuremath{\D^*}}\xspace}

\def\Dstarz  {{\ensuremath{\D^{*0}}}\xspace}

\def\Ds      {{\ensuremath{\D^+_\squark}}\xspace}

\def\Dspm    {{\ensuremath{\D^{\pm}_\squark}}\xspace}
\def\Dsmp    {{\ensuremath{\D^{\mp}_\squark}}\xspace}

\def\B       {{\ensuremath{\PB}}\xspace}
\def\Bbar    {{\ensuremath{\offsetoverline{\PB}}}\xspace}
\def\Bb      {{\ensuremath{\Bbar}}\xspace}
\def\BorBbar {\kern \thebaroffset\optbar{\kern -\thebaroffset \PB}\xspace}
\def\Bz      {{\ensuremath{\B^0}}\xspace}

\def\Bd      {{\ensuremath{\B^0}}\xspace}
\def\Bdb     {{\ensuremath{\Bbar{}^0}}\xspace}
\def\BdorBdbar {\kern \thebaroffset\optbar{\kern -\thebaroffset \Bd}\xspace}
\def\Bu      {{\ensuremath{\B^+}}\xspace}
\def\Bub     {{\ensuremath{\B^-}}\xspace}
\def\Bp      {{\ensuremath{\Bu}}\xspace}
\def\Bm      {{\ensuremath{\Bub}}\xspace}

\def\Bs      {{\ensuremath{\B^0_\squark}}\xspace}

\def\BsorBsbar {\kern \thebaroffset\optbar{\kern -\thebaroffset \Bs}\xspace}

\def\jpsi     {{\ensuremath{{\PJ\mskip -3mu/\mskip -2mu\Ppsi}}}\xspace}
\def\psitwos  {{\ensuremath{\Ppsi{(2S)}}}\xspace}

\def\Y#1S{\ensuremath{\PUpsilon{(#1S)}}\xspace}

\def\FourS {{\Y4S}\xspace}

\def\theX     {{\ensuremath{\Pchi_{c1}(3872)}}\xspace}

\def\proton      {{\ensuremath{\Pp}}\xspace}

\def\Lz          {{\ensuremath{\PLambda}}\xspace}
\def\Lbar        {{\ensuremath{\offsetoverline{\PLambda}}}\xspace}
\def\LorLbar     {\kern \thebaroffset\optbar{\kern -\thebaroffset \PLambda}\xspace}

\def\Xires       {{\ensuremath{\PXi}}\xspace}

\def\Lb           {{\ensuremath{\Lz^0_\bquark}}\xspace}

\def\BF         {{\ensuremath{\mathcal{B}}}\xspace}

\newcommand{\decay}[2]{\ensuremath{#1\!\to #2}\xspace} 

\def\to                 {\ensuremath{\rightarrow}\xspace}

\def\order   {{\ensuremath{\mathcal{O}}}\xspace}

\def\qsq       {{\ensuremath{q^2}}\xspace}

\def\CP                {{\ensuremath{C\!P}}\xspace}

\def\Vtd  {{\ensuremath{V_{\tquark\dquark}^{\phantom{\ast}}}}\xspace}

\def\Vts  {{\ensuremath{V_{\tquark\squark}^{\phantom{\ast}}}}\xspace}
\def\Vub  {{\ensuremath{V_{\uquark\bquark}^{\phantom{\ast}}}}\xspace}
\def\Vcb  {{\ensuremath{V_{\cquark\bquark}^{\phantom{\ast}}}}\xspace}

\def\Vtbs  {{\ensuremath{V_{\tquark\bquark}^\ast}}\xspace}

\newcommand{\phis}{{\ensuremath{\phi_{\squark}}}\xspace}

\def\bsll     {\decay{\bquark}{\squark \ell^+ \ell^-}}

\def\AT#1     {\ensuremath{A_{\mathrm{T}}^{#1}}\xspace}           

\def\Bsmm     {\decay{\Bs}{\mup\mun}}
\def\Bdmm     {\decay{\Bd}{\mup\mun}}

\def\C#1      {\ensuremath{\mathcal{C}_{#1}}\xspace}                       
\def\Cp#1     {\ensuremath{\mathcal{C}_{#1}^{'}}\xspace}                    
\def\Ceff#1   {\ensuremath{\mathcal{C}_{#1}^{\mathrm{(eff)}}}\xspace}        
\def\Cpeff#1  {\ensuremath{\mathcal{C}_{#1}^{'\mathrm{(eff)}}}\xspace}       
\def\Ope#1    {\ensuremath{\mathcal{O}_{#1}}\xspace}                       
\def\Opep#1   {\ensuremath{\mathcal{O}_{#1}^{'}}\xspace}                    

\newcommand{\nospaceunit}[1]{\ensuremath{\text{#1}}}       
\newcommand{\aunit}[1]{\ensuremath{\text{\,#1}}}       

\newcommand{\tev}{\aunit{Te\kern -0.1em V}\xspace}
\newcommand{\gev}{\aunit{Ge\kern -0.1em V}\xspace}
\newcommand{\mev}{\aunit{Me\kern -0.1em V}\xspace}
\newcommand{\kev}{\aunit{ke\kern -0.1em V}\xspace}
\newcommand{\ev}{\aunit{e\kern -0.1em V}\xspace}
 
\newcommand{\mevc}{\ensuremath{\aunit{Me\kern -0.1em V\!/}c}\xspace}
\newcommand{\gevc}{\ensuremath{\aunit{Ge\kern -0.1em V\!/}c}\xspace}
\newcommand{\mevcc}{\ensuremath{\aunit{Me\kern -0.1em V\!/}c^2}\xspace}
\newcommand{\gevcc}{\ensuremath{\aunit{Ge\kern -0.1em V\!/}c^2}\xspace}
\newcommand{\gevgevcccc}{\ensuremath{\gev^2\!/c^4}\xspace} 

\def\km   {\aunit{km}\xspace}

\def\mbarn{\aunit{mb}\xspace}
\def\mub{\ensuremath{\,\upmu\nospaceunit{b}}\xspace}

\def\fb   {\ensuremath{\aunit{fb}}\xspace}
\def\invfb   {\ensuremath{\fb^{-1}}\xspace}

\def\ps   {\ensuremath{\aunit{ps}}\xspace}

\def\mhz  {\ensuremath{\aunit{MHz}}\xspace}
\def\khz  {\ensuremath{\aunit{kHz}}\xspace}

\def\invps{\ensuremath{\ps^{-1}}\xspace}

\def\order{{\ensuremath{\mathcal{O}}}\xspace}

\def\gsim{{~\raise.15em\hbox{$>$}\kern-.85em
          \lower.35em\hbox{$\sim$}~}\xspace}
\def\lsim{{~\raise.15em\hbox{$<$}\kern-.85em
          \lower.35em\hbox{$\sim$}~}\xspace}

\def\pt         {\ensuremath{p_{\mathrm{T}}}\xspace}

\def\degrees{\ensuremath{^{\circ}}\xspace}

\def\mrad{\aunit{mrad}\xspace}
\def\rad{\aunit{rad}\xspace}

\def\betastar {\ensuremath{\beta^*}}

\def\evtgen     {\mbox{\textsc{EvtGen}}\xspace}

\def\tell1  {TELL1\xspace}
\def\ukl1   {UKL1\xspace}

\newcommand{\eg}{\mbox{\itshape e.g.}\xspace}

\newcommand{\lhcborcid}[1]{\href{https://orcid.org/#1}{\hspace*{0.1em}\raisebox{-0.45ex}{\includegraphics[width=1em]{figs/orcidIcon.pdf}}}}

\usepackage{columntypes}
\usepackage{commands}
\usepackage{weakstuff}
\usepackage{decays}
\usepackage{LHCb}
\usepackage[export]{adjustbox}
\usepackage{verbatim}

\def\CKMconventionvalue{j}
\newlength\wrapfigL
\newcommand{\mywrapfig}[3][r]{
\setlength{\wrapfigL}{#2}
\begin{wrapfigure}{#1}{\wrapfigL}
  \hskip 0.1\wrapfigL %
  \begin{minipage}{0.9\wrapfigL}
  #3
  \vskip -3ex ~
  \end{minipage}
\end{wrapfigure}
}

\begin{document}

\title{Flavour Physics at LHCb --- 50 years of the \CKM paradigm}


\author{Patrick Koppenburg\footnote{on behalf of the LHCb collaboration}}
\affil{Nikhef, Amsterdam, The Netherlands \email{patrick.koppenburg@nikhef.nl}}


\begin{abstract}%
  The LHCb experiment is in operation since 2009 and has provided measurements 
  of the \CKM matrix with unprecedented precision.
  50 years after the original paper we are in the position of pinning down
  the parameters of the theory, or possibly to show its limitations.
  In these proceedings the status of LHCb is shown in a historical
  perspective, along with some anecdotes.
\end{abstract}

\subjectindex{xxxx, xxx}

\maketitle

\begin{figure}[tb]\centering
  \includegraphics[width=\textwidth]{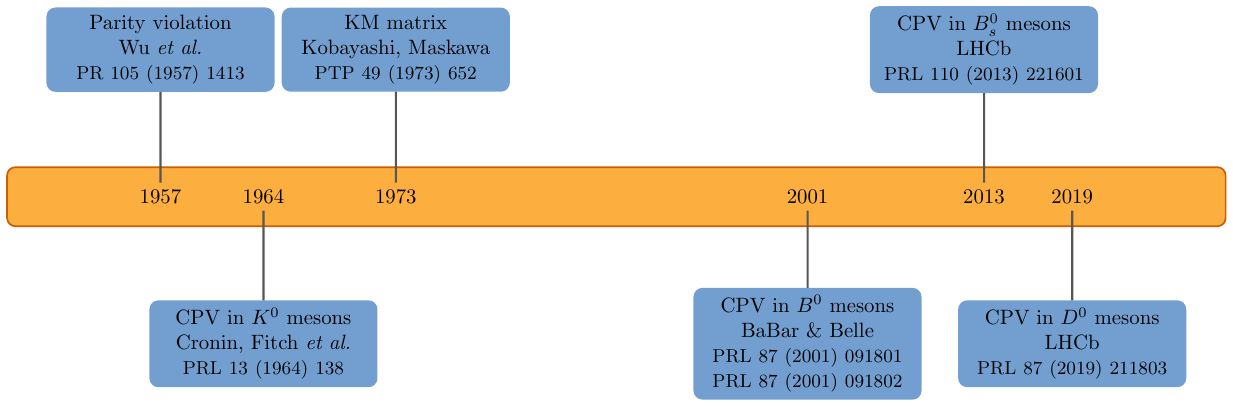}
  \caption{The briefest history of \CP violation \cite{Kobayashi:1973fv,Christenson:1964fg,Wu:1957,Aubert:2001nu,Abe:2001xe,LHCb-PAPER-2013-018,LHCb-PAPER-2019-006}}\label{Fig:CPV}
\end{figure}
\section{Introduction}
It took Kobayashi and Maskawa~\cite{Kobayashi:1973fv} a bit short of a
decade since the observation of \CP violation in neutral
kaons~\cite{Christenson:1964fg} to provide a theoretical description, the \CKM model.
In hindsight, this time-lapse seems relatively short as one needed to wait until the twenty-first century
to get confirmation of the model in the
\bquark and \cquark sectors (Fig.~\ref{Fig:CPV}).
Since then however, \CP violation has been observed in interference of mixing
and decay (also known as ``indirect'') in the \Bd~\cite{Aubert:2001nu,Abe:2001xe} and \Bs~\cite{LHCb-PAPER-2013-018}
systems, and \CP violation in decays (also known as ``direct'')
of \Kz~\cite{KTeV:1999kad,NA48:2002tmj},
\Bd~\cite{BaBar:2004gyj,Chao:2004mn},
\Bs~\cite{LHCb-PAPER-2020-029},
\Bp~\cite{LHCb-PAPER-2012-001} and
\Dz~\cite{LHCb-PAPER-2019-006} mesons.

The original observations of \CP violation in the \Bz system that led
to the Nobel Prize awarded to Kobayashi and Maskawa in 2008
were the work of the BaBar and Belle experiments.
Since then all of the following ``firsts'' were performed by LHCb.
Barring the still missing observation of \CP violation in baryons
(and leptons, but this is another story), we
are now entering the precision regime in which the \CKM
paradigm no longer needs to be established, but is tested to
its ultimate precision.

\begin{figure}[tb]\centering
  \includegraphics[width=\textwidth]{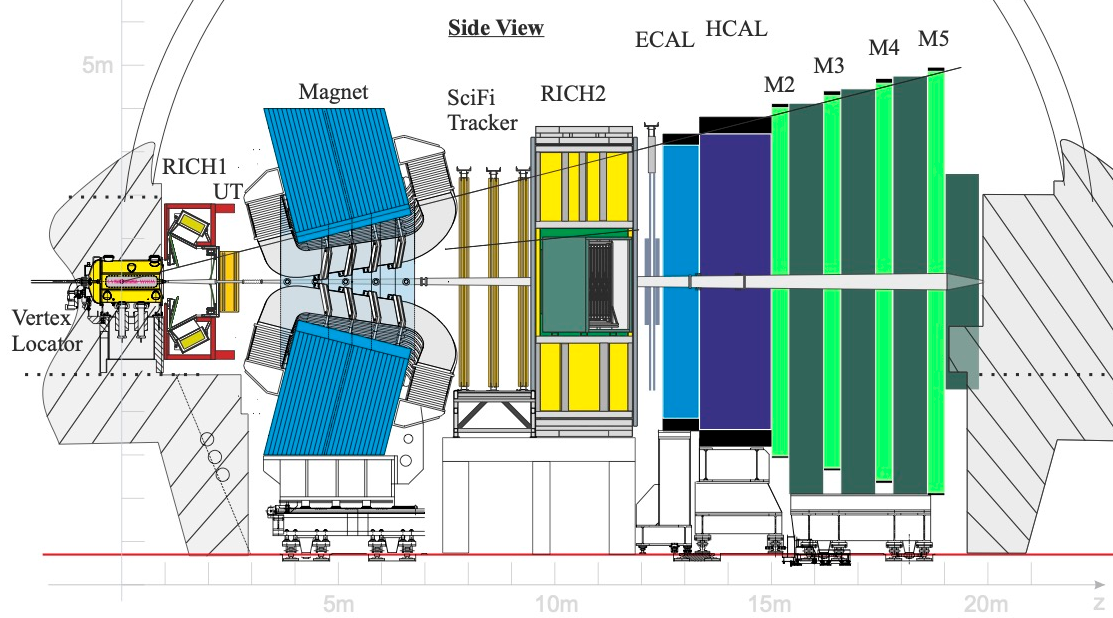}
  \caption{The LHCb detector. Figure adapted from Ref.~\cite{LHCb-DP-2022-002}.}\label{Fig:LHCb}
\end{figure}

\section{LHCb}
The LHCb experiment was designed at a time before the
\B factories Belle and BaBar came into operations,
and when the amount of \CP violation
in the SM was poorly constrained~\cite{Schmidtler:1991tv}.
The benchmark observables were the $\sin2\ckmbeta$ \CP asymmetry amplitude\footnote{See Sec.~\ref{Sec:UT}.
  \CKMconvention{}{As these proceedings are published
    in PTEP, the Belle convention is used in the text: $\phi_1=\beta$, $\phi_2=\alpha$, $\phi_3=\gamma$. Readers who are allergic to this convention can download the source, change the argument of {\tt$\backslash$def$\backslash$CKMconventionvalue\{j\}} to anything but {\tt j} and recompile.}
} obtained
from the decay \decay{\Bd}{\jpsi\KS} --- {\it because it may be that
  the \B factories wouldn't be able to measure \CP violation}~\cite{Anonymous} --- and the frequency of
\Bs oscillations
--- as it was hoped that LHCb would see them before Tevatron.
Incidentally both measurements were performed before LHCb came into operation~\cite{Aubert:2001nu,Abe:2001xe,Abulencia:2006ze},
but these benchmarks set constraints on
the design of the experiment from which the physics programme largely profits today.
In particular the excellent vertex resolution is a legacy of the requirement that
\Bs oscillations should be resolvable for frequencies up to 60\invps.

The LHCb detector~\cite{LHCb-DP-2008-001,LHCb-DP-2014-002}  depicted in Fig.~\ref{Fig:LHCb} is a
single-arm forward spectrometer covering the pseudorapidity range $2 < \eta < 5$, designed for
the study of particles containing \bquark\ or \cquark\ quarks.
It consists of a silicon-strip vertex detector surrounding the $pp$ interaction
region that allows \cquark\ and \bquark\ hadrons to be identified from their characteristically long
flight distance; a tracking system that provides a measurement of the momentum of charged
particles; and two ring-imaging Cherenkov detectors that are able to discriminate between
different species of charged hadrons.

Lessons from past experiments have taught LHCb to have a versatile trigger (Sec.~\ref{Sec:U1}), which
allowed the collaboration to quickly adapt to varying experimental conditions.
These variations are kept minimal by offsetting the LHC beams in order
to keep the luminosity constant throughout a fill, and
throughout a data-taking period. The nominal value was around $2$ to $4\times10^{32}\rm \lumiunit$.

\begin{table}[tb]
  \caption{Operational statistics of LHCb running conditions in Runs 1 and 2.
    The Run 1, Run 2 and last column contain
    sums or weighted averages depending on which is more appropriate
    (or none if neither makes sense).
    $f_\text{LHC}\simeq27\km/c\simeq11\khz$ is the LHC frequency.
  }\label{Tab:operations}
\resizebox{\textwidth}{!}{\begin{tabular}{l@{}C@{}C||C||CC|C||CCCC|C|C}
        \multicolumn{2}{l}{Quantity} & \text{unit} & \text{TDR} & 2011 & 2012 & \text{Run 1} & 2015 & 2016 & 2017 & 2018 & \text{Run 2} & \text{Tot/Avg}\\\hline\hline
Peak Luminosity & {\cal L}_\text{peak} & \mub^{-1}/\rm s &   280  &   461  &   492  &  &   302  &   422  &   453  &   493  &  &  \\ 
Average Luminosity & {\cal L}_\text{avg} & \mub^{-1}/\rm s &   200  &   250  &   330  &   298  &   140  &   240  &   280  &   310  &   268  &   278  \\ 
Seconds of running & t & 10^6\:\rm s &  10.0  &   4.3  &   6.2  &  10.5  &   1.6  &   6.9  &   4.6  &   6.9  &  20.0  &  30.5  \\ 
Integrated luminosity \hskip -4 em~ & \intlumi & \ifb &   2.0  &   1.1  &   2.1  &   3.2  &   0.5  &   1.9  &   1.5  &   2.5  &   6.4  &   9.6  \\ 
Bunches & N_\text{b} &  &  2600  &  1320  &  1320  &  &  1710  &  2036  &  2332  &  2332  &  &  \\ 
Energy & E & \tev &    14  &     7  &     8  &  &    13  &    13  &    13  &    13  &  &  \\ 
Inelastic cross-section \hskip -4 em~ & \sigma_\text{inel} & \mbarn &    80  &    64  &    67  &  &    77  &    77  &    77  &    77  &  &  \\ 
Charged multiplicity & \frac{\rmd N_\text{ch}}{\rmd\eta} &  &     6  &     6  &     6  &     6  &     6  &     6  &     6  &     6  &     6  &     6  \\ 
\bquark{}\bquarkbar cross-section (acc.) \hskip -4 em~ & \sigma_{\bquark{}\bquarkbar} & \mub &   150  &    72  &    83  &  &   144  &   144  &   144  &   144  &  &  \\ 
\hline
\proton{}\proton interactions/BX & \mu = \frac{{\cal L}\sigma_\text{inel}}{f_\text{LHC}N_\text{b}} &  &  0.55  &  1.08  &  1.49  &  1.32  &  0.56  &  0.81  &  0.82  &  0.91  &  0.83  &  0.99  \\ 
Non-empty rate & f_\text{LHC}N_\text{b}(1-e^{-\mu}) & \mhz &  12.3  &   9.8  &  11.5  &  10.8  &   8.3  &  12.7  &  14.7  &  15.7  &  14.0  &  12.9  \\ 
Avg. MB rate  &  {\sigma_\text{inel}{\cal L}_\text{avg}} & \mhz &  16.0  &  16.0  &  22.1  &  19.7  &  10.8  &  18.5  &  21.6  &  23.9  &  20.7  &  20.3  \\ 
MB events & {\sigma_\text{inel}\int{\cal L}_\text{avg}\rmd t} & 10^{12} &   160  &    70  &   141  &   211  &    38  &   146  &   116  &   192  &   493  &   704  \\ 
Peak particle flow & {\frac{\rmd N_\text{ch}}{\rmd\eta}\sigma_\text{inel}{\cal L}_\text{peak}} & 10^6 &   134  &   177  &   198  &  &   140  &   195  &   209  &   228  &  &  \\ 
Irradiation & {\frac{\rmd N_\text{ch}}{\rmd\eta}\sigma_\text{inel}\intlumi} & 10^{15} &   1.0  &   0.4  &   0.8  &   1.3  &   0.2  &   0.9  &   0.7  &   1.2  &   3.0  &   4.2  \\ 
\bquark{}\bquarkbar rate & {\sigma_\text{\bquark{}\bquarkbar}}{\cal L}_\text{avg} & \khz &    30  &    18  &    27  &    24  &    20  &    35  &    40  &    45  &    39  &    34  \\ 
\bquark{}\bquarkbar yield & {\sigma_\text{\bquark{}\bquarkbar}\intlumi} & 10^{9} &   300  &    79  &   174  &   254  &    72  &   274  &   216  &   360  &   922  &  1175  \\ 
\hline
Output rate & \lambda_\text{HLT} & \khz &   2.0  &   2.6  &   4.5  &   3.7  &  10.4  &   6.1  &   7.5  &   5.8  &   6.6  &   5.7  \\ 
Stored events (bkk) \hskip -5 em~ & \lambda_\text{HLT}t & 10^9 &    20  &    11  &    28  &    39  &    17  &    42  &    35  &    40  &   133  &   172  \\ 
Event size & S_\text{ev} & \rm kB &     2  &    53  &    59  &    56  &    48  &    55  &    58  &    58  &    56  &    56  \\ 
HLT B/W & S_\text{ev}\lambda_\text{HLT} & \rm MB/s &     5  &   136  &   263  &   212  &   501  &   333  &   438  &   333  &   371  &   319  \\ 
Total storage & S_\text{ev}\lambda_\text{HLT}t & \rm EB &   0.1  &   0.6  &   1.6  &   2.2  &   0.8  &   2.3  &   2.0  &   2.3  &   7.4  &   9.6  \\ 
\hline
\end{tabular}
}
\end{table}
In the startup year of 2010, LHCb was able to identify the first \B meson at the LHC\footnote{I
  had the honour of presenting the LHC's first \B meson at the LHCC open session in May 2010~\cite{LHCC2010},
  and then killed it by declaring the 1~hour run period during which it was recorded as of too low physics quality.
  It is thus lost for posterity.} 
with low luminosity and then cope with a rapid increase of the collision rate.
Eventually LHCb capped the luminosity and 
collected data corresponding to 1, 2 and 6\invfb at 7, 8 and 13\tev collision energies, respectively.
Together they form the 9\invfb ``legacy'' sample.
The huge cross-section at the LHC provided more than $10^{12}$ \bquark{}\bquarkbar pairs in the LHCb
acceptance, in rough proportion of 4:4:2:1 \Bp:\Bz:\Lb:\Bs hadrons~\cite{LHCb-PAPER-2020-046}.
In total LHCb collected more than $10^{11}$ events between 2010 and 2018,
a number that can be compared to the 770 million \B\Bb pairs collected by Belle.
More numbers can be found in Table~\ref{Tab:operations}.

\begin{figure}[tb]\centering
\CKMconvention{\includegraphics[width=0.8\textwidth]{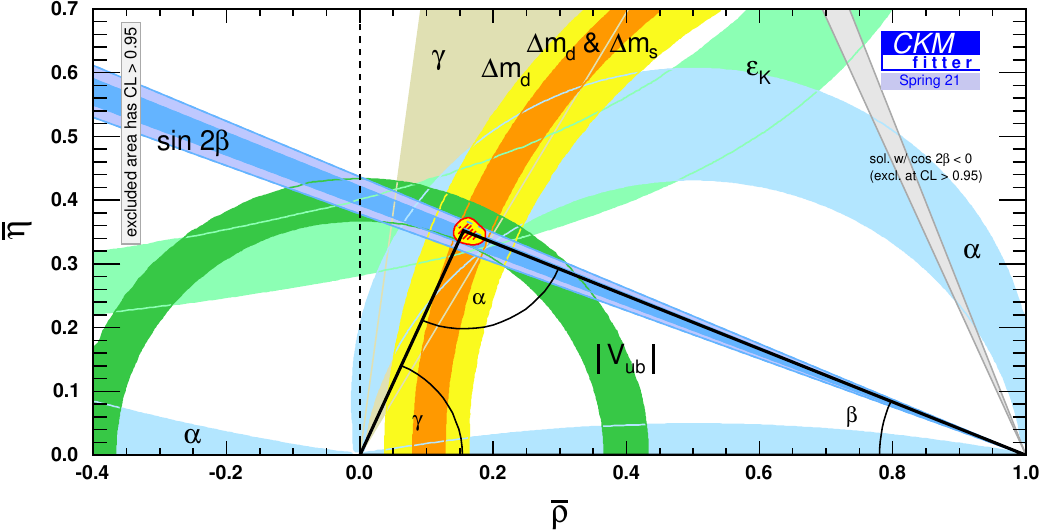}}{\includegraphics[width=\textwidth]{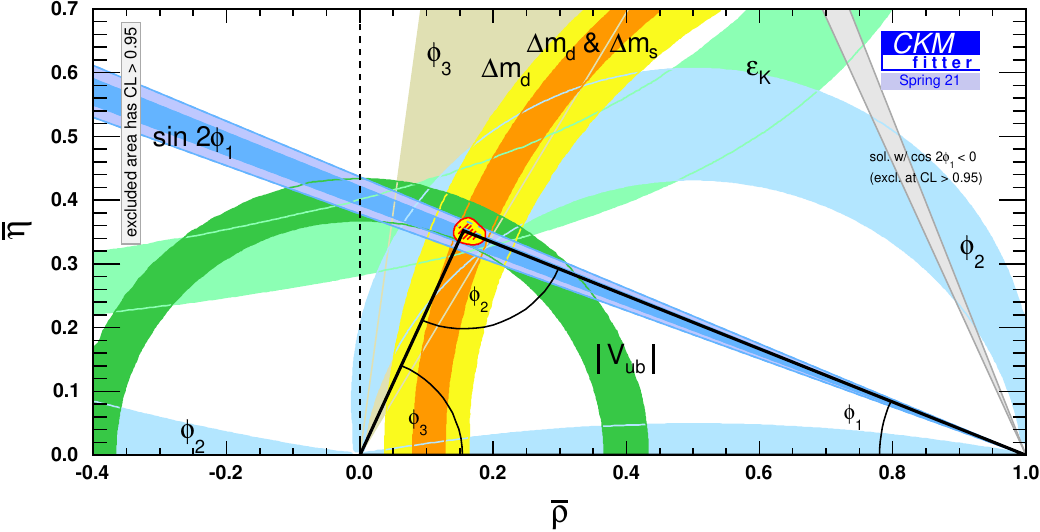}}~%
\caption{Status of the \protect\CKM\ \Bd unitarity triangle as of March 2021~\cite{CKMfitter2005}.}\label{Fig:UT}
\end{figure}
\section{Unitarity triangle}\label{Sec:UT}
Unitarity relations of the \CKM matrix are exploited to draw triangles
in the complex plane, the most renowned being the ``\Bd'' unitarity triangle (``UT'', Fig.~\ref{Fig:UT})
formed from the first and last columns of the \CKM matrix.
The two non-unit sides intersect at an apex, which defines
three angles: \ckmbeta, \ckmalpha, and \ckmgamma. All sides and angles are observables while only
two are independent.\footnote{There is also a constraint from kaon physics, noted $\epsilon_K$, which is beyond the scope of these proceedings.}
This opens the way to over-constraining the
triangle and thus putting the \CKM paradigm to test.

\begin{figure}[tb]\centering
  \includegraphics[height=0.25\textheight,valign=c]{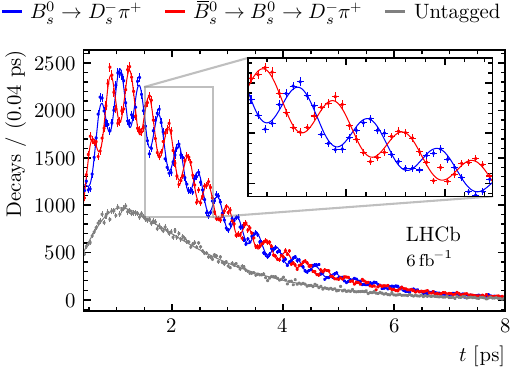}
  \includegraphics[height=0.25\textheight,valign=c]{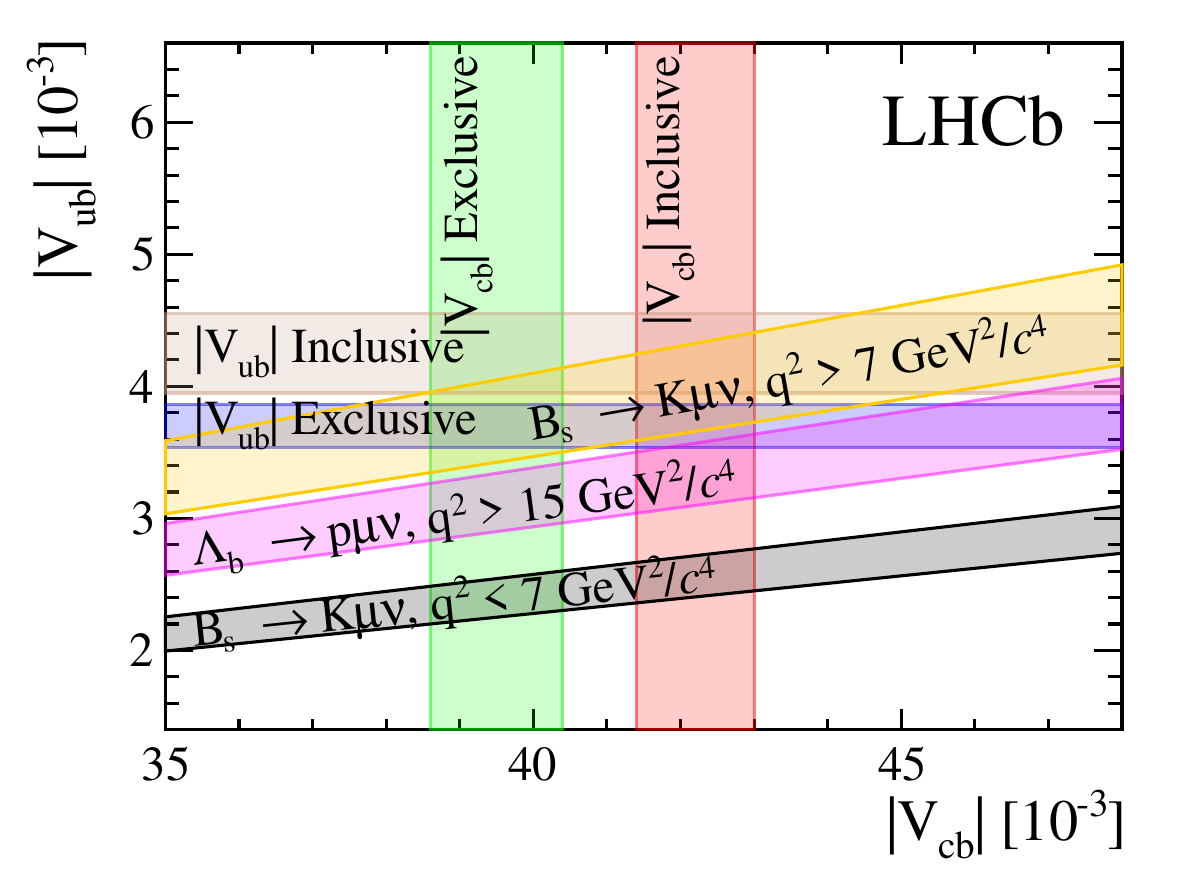}
  \caption{(left) Oscillation pattern of \decay{\Bs}{\Dspm\pimp} decays. Figure from Ref.~\cite{LHCb-PAPER-2021-005}. (right) Constraints on \protect\Vub and \protect\Vcb from Refs.~\cite{LHCb-PAPER-2020-038,LHCb-PAPER-2015-031}.}\label{Fig:Vub}\label{Fig:DMS}
\end{figure}

LHCb has measured the top-right side --- proportional to $|\Vtd\Vtbs|$ --- by
precision measurements of the \Bd and \Bs mixing frequencies.
Such a mixing plot is shown in Fig.~\ref{Fig:DMS}.
LHCb obtained the most precise values of
$\Delta m_s=17.766 \pm 0.006\invps$~\cite{LHCb-PAPER-2021-005}
and $\Delta m_d=505.0 \pm2.3\invps$~\cite{LHCb-PAPER-2015-031}.
Their conversion into constraints
on the \CKM UT is however limited by hadronic uncertainties.

The other side is obtained from semileptonic \bquark decays to
\cquark and \uquark quarks to determine \Vcb and \Vub, respectively.
Such processes are members of a category known colloquially as
``things LHCb cannot do (but still does)''.
LHCb was never designed for decays with missing neutrinos.
However the precise vertexing allows the determination
of the primary production (PV) and secondary decay
vertices (SV) with resolution of a few 10 microns.

\mywrapfig{0.4\textwidth}{
  \includegraphics[width=\textwidth]{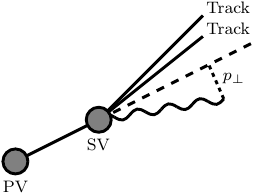}
  \caption{Corrected mass inputs.}\label{Fig:mcorr}
}
The corrected mass $m_\text{corr}=\sqrt{m^2+p_\perp^2}+p_\perp$
is determined from the momentum missing along the direction of flight
and peaks at the \bquark hadron mass if the lost particle is of small mass,
see Fig.~\ref{Fig:mcorr}.
LHCb measured the ratio $|\Vub|/|\Vcb|$ in decays of
\Lb baryons~\cite{LHCb-PAPER-2015-031} and \Bs mesons~\cite{LHCb-PAPER-2020-038},
which are inaccessible to \B factories at the \FourS resonance.
The resulting constraints are shown as diagonal bands in Fig.~\ref{Fig:Vub}.
It is particularly striking that the constraints from
\decay{\Bs}{\Km\mup\neu} in the low and high range of
dilepton masses squared, $\qsq$, are inconsistent, owing
to inconsistent form factors from light-cone sum rules~\cite{Khodjamirian:2017fxg} and
Lattice QCD~\cite{Bazavov:2019aom}, respectively.
The figure also shows the discrepancy between the inclusive
\decay{\bquark}{\quark\ell\neu} and exclusive \decay{\B}{\X\ell\neu}
determinations of $|\Vub|$ and $|\Vcb|$.
This puzzle is still unresolved in spite of multiple
measurements and computations spanning several decades.
Let us hope we won't have to wait for an \epem collider to run at the \W\W threshold
to resolve the issue.

The situation in the measurements of the angles is somewhat clearer.
The angle \ckmbeta is obtained from final states that are \CP eigenstates
reached via interference of 
\decay{\bquark}{\cquark\cquarkbar\squark} and their charged-conjugated processes
after \Bd to \Bdb mixing~\cite{Bigi:1983cj}. 
The angle is known with sub-degree precision:
LHCb now holds the most precise value of $\sin2\ckmbeta=0.717 \pm 0.013 \pm 0.008$,
from a combination of the decay modes \decay{\Bd}{\jpsi\KS} 
and \psitwos\KS~\cite{LHCb-PAPER-2023-013}. 

The angle \ckmgamma is often referred to as a standard candle of the Standard Model ---
to the extent that one is certain that no new physics affects tree-dominated decays, which
is a bold assumption~\cite{Gligorov:2023mji}.
The \CP asymmetry is generated from interference of \decay{\bquark}{\cquark\uquarkbar\quark}
and \decay{\bquark}{\uquark\cquarkbar\quark} where the \D mesons formed by
\cquark\uquarkbar (\cquarkbar\uquark) decay
to a common final state.
If it is assumed that such tree-level processes are free
from new physics contributions, then
the determination of the angle \ckmgamma will be dominated by the experimental
resolution in the foreseeable future and beyond~\cite{Brod:2013sga}.

\begin{table}[t]
\caption{Used integrated luminosities and references of measurements with sensitivity to \protect\ckmgamma split by final states.
  Entries in red are not yet included in the combination~\cite{LHCb-CONF-2022-003}.}\label{Tab:gamma}
\def\nine{{\bf\boldmath$9\invfb$}\xspace}\centering
    \resizebox{\textwidth}{!}{\begin{tabular}{|l|c|c|c|c|c|}
    \hline
    & \multicolumn{5}{c|}{\bf Time-integrated measurements}\\ \hline
    Decays                        & \decay{\Bp}{\D h^+} & \decay{\Bp}{\Dstarz h^+} & \decay{\Bp}{\D\KS\pip} & \decay{\Bz}{\D\Kp\pim} & \decay{\Bp}{\D\Kp\pim\pim} \\ \hline
    \decay{\D}{h^+h^-}            & \nine \cite{LHCb-PAPER-2020-036}        &  \nine \cite{LHCb-PAPER-2020-036}          & 5\invfb \cite{LHCb-PAPER-2017-030}           & 5\invfb \cite{LHCb-PAPER-2019-021}           & 3\invfb \cite{LHCb-PAPER-2015-020}                 \\
    \decay{\D}{h^+\pim\pip\pim}      & 3\invfb \cite{LHCb-PAPER-2016-003}   & \nine \cite{LHCb-PAPER-2019-021}                    & 5\invfb \cite{LHCb-PAPER-2017-030}           & 5\invfb \cite{LHCb-PAPER-2019-021}           &       \\
    \decay{\D}{\Kpm\pimp\pip\pim}      & \nine \cite{LHCb-PAPER-2022-017}  &                        &                        &                        & \\
    \decay{\D}{h^+h^-\pip\pim}      & {\color{red} \nine \cite{LHCb-PAPER-2022-037}}  &                        &                        &                        & \\
    \decay{\D}{h^+h^-\piz}        & \nine \cite{LHCb-PAPER-2021-036}        &                        &                        &                        & \\
    \decay{\D}{\KS h^+h^-}        & \nine \cite{LHCb-PAPER-2020-019}        &  {\color{red}\nine \cite{LHCb-PAPER-2023-012}}  &                        & {\color{red}\nine \cite{LHCb-PAPER-2023-009}}  &  \\
    \decay{\D}{\KS \Kp\pim}       & \nine \cite{LHCb-PAPER-2019-044}        &    &                        &                        & \\ \hline  \hline
  \end{tabular}}
   \vskip 2ex   
    \begin{tabular}{|l|c|c|c|}
    \hline
    & \multicolumn{3}{c|}{\bf Time-dependent measurements}\\ \cline{1-4}  
    Decays                         & \decay{\Bz}{\Dmp\pipm} & \decay{\Bs}{\Dsmp\Kpm} & \decay{\Bs}{\Dsmp\Kpm\pip\pim}  \\ \cline{1-4}
    \decay{\Dpm}{\Kpm h^+h^-}      & 3\invfb \cite{LHCb-PAPER-2018-009}           &   N/A                     &  N/A\\
    \decay{\Dspm}{h^\pm h^\mp\pipm}& N/A                    & {\color{red}\nine} \cite{LHCb-PAPER-2017-047,LHCb-CONF-2023-004}                       & \nine \cite{LHCb-PAPER-2020-030}    \\ \cline{1-4}
  \end{tabular}

\end{table}

However, unlike with \decay{\Bd}{\jpsi\KS} for $\sin2\ckmbeta$ there is no single
process that yields a good resolution on its own.
Multiple \B meson (and potentially \bquark-baryon) decays can be used,
as well as many \D decays. The ultimate precision will be reached
once all combinations are analysed with the full data set.
The present status is shown in Table~\ref{Tab:gamma}, where multiple
possible analyses are still missing. It is however unlikely that this
table will ever be fully filled, as some combination of \B and \D decay modes have
marginal sensitivities.
In addition to the interference pattern above, \ckmgamma can also
be measured from interference of mixing (both \Bd and \Bs) and decay.
These processes are however potentially affected by new physics
entering the  \B mixing loop. A comparison of values of \ckmgamma
obtained by from the two tables in Tab.~\ref{Tab:gamma} thus constitutes
an additional test of the Standard Model.

A precise value of $\ckmgamma=(63.8\aerr{3.5}{3.7})\degrees$ is
obtained from a combination the analyses
reported in Ref.~\cite{LHCb-CONF-2022-003}, which does not yet include some of the latest
LHCb results~\cite{LHCb-PAPER-2022-037,LHCb-PAPER-2023-012,LHCb-PAPER-2023-009}.
Presently, the time-dependent measurements yield
$\ckmgamma=(79\aerr{21}{23})\degrees$, compatible with the
value from time-integrated analyses.

The values reported in this combination also make use of the
charm mixing~\cite{LHCb-PAPER-2022-020} and \CP violation~\cite{LHCb-PAPER-2019-006,LHCb-PAPER-2022-024} measurements
(that slightly affect the \ckmgamma determination for \decay{\B}{\D} modes). 
The first observation of \CP violation in charm~\cite{LHCb-PAPER-2019-006} ---
with a precision in units of $10^{-4}$ --- is a major achievement
but its direct understanding in terms of \CKM matrix elements is not yet within
theoretical reach.

The third angle, \ckmalpha, is obtained from interference in \decay{\bquark}{\uquark} transitions.
LHCb's main contribution is the most precise time-dependent measurement of \CP asymmetries in
\decay{\Bd}{\pip\pim}~\cite{LHCb-PAPER-2020-029}.
A full determination of \ckmalpha needs an isospin analysis of several decay modes
involving \piz mesons~\cite{Gronau:1990ka}, which are hard to reconstruct at
LHCb.\footnote{See however Ref.~\cite{LHCb-CONF-2015-001}.}

The combination of these two sides and three angles yields the over-constrained
triangle depicted in Fig.~\ref{Fig:UT}~\cite{CKMfitter2005}.
This is however only one of several possible triangles, dominated by \B decay modes.

\mywrapfig{0.6\textwidth}{
  \includegraphics[width=\textwidth]{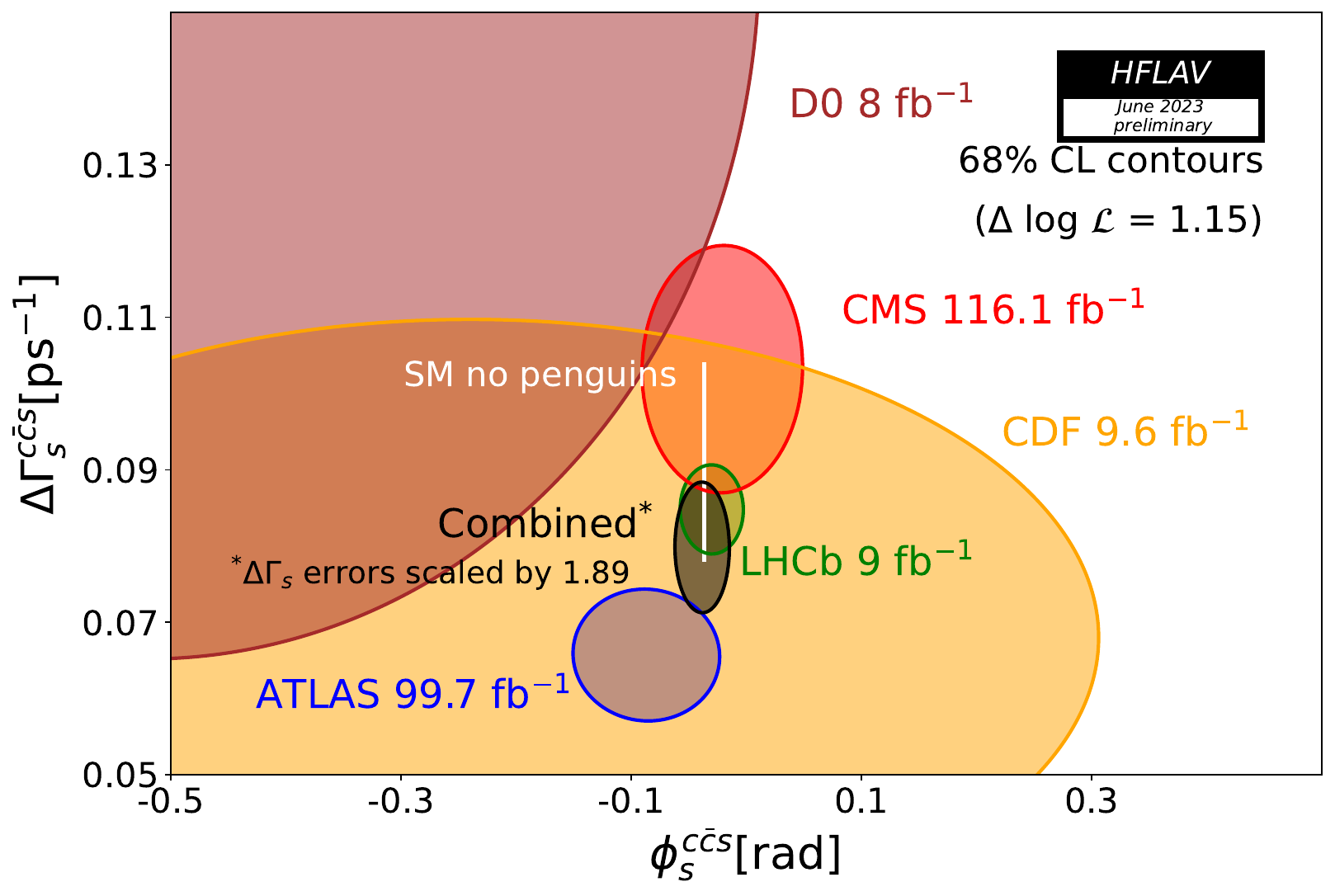}\quad
  \caption{Preliminary \phis combination by HFlav~\cite{HFLAV21}.}\label{Fig:phis}
}
Another triangle, dubbed the \Bs triangle, features the angle \phis.
In analogy to \ckmbeta, it is obtained from \Bs oscillation followed
by a \Bs decay to a \cquark\cquarkbar\squark\squarkbar\ \CP eigenstate,
\eg\ \jpsi{}\Pphi.
It used to be said that the SM expectation is close to vanishing,
making any measurement of a non-zero value a sign of new physics.
We are no longer in this regime: 
LHCb recently released their legacy measurement with
the full 9\invfb dataset and obtain
$\phis= -0.039 \pm 0.022 \pm 0.006\rad$~\cite{LHCb-PAPER-2023-016} from \decay{\Bs}{\jpsi\Kp\Km}
with the \Kp\Km mass in the vicinity of the \Pphi meson.
This value is combined with LHCb measurements using previous datasets or other
decays~\cite{LHCb-PAPER-2019-003,LHCb-PAPER-2017-008,LHCb-PAPER-2016-027,LHCb-PAPER-2014-051}
as well as results from Tevatron~\cite{CDF:2012nqr,D0:2011ymu} and other LHC experiments~\cite{ATLAS:2020lbz,CMS:2020efq}.
The result is $\phis= - 0.050 \pm 0.016\rad$, which is now significantly deviating from zero.
The expectation from \CKM fits is $\phis=-0.037\pm0.001$~\cite{CKMfitter2005,UTfit-UT}, with which
the world average is well compatible, as shown in Fig.~\ref{Fig:phis}.

There is some disagreement in the determination of the total decay width of the \Bs meson $\Gamma_s$
and of the difference of the decays widths of
the heavy and light \Bs mass eigenstates, $\Delta\Gamma_s$, which are by-products of these analyses.
The update of ATLAS with the full Run 2 sample is eagerly awaited, hopefully resolving the issue.

In 2003 there was some excitement due to inconsistent values of
$\sin2\ckmbeta$ obtained from \decay{\bquark}{\cquark\cquarkbar\squark} processes
and from \decay{\bquark}{\squark\squarkbar\squark}, as \decay{\Bd}{\phi\KS}~\cite{Belle:2003vik}.
The latter are also sensitive to the same \CKM phases, but proceed via penguin diagrams and are thus
more likely affected by heavy degrees of freedom in the loop.
The issue turned out to be a fluke but the penguin decay modes remained on
the radar.

A similar test can be done by measuring the time-dependent \CP-violating phase in
from \decay{\Bs}{\Pphi\Pphi}, that is strictly zero in the SM. 
LHCb obtain $\phi_s^{\squark\squarkbar\squark}=-0.18\pm0.09$ in a combination of all data
samples~\cite{LHCb-PAPER-2023-001}, which is compatible with the SM expectation.

\begin{figure}[b]\centering
  \includegraphics[width=0.8\textwidth]{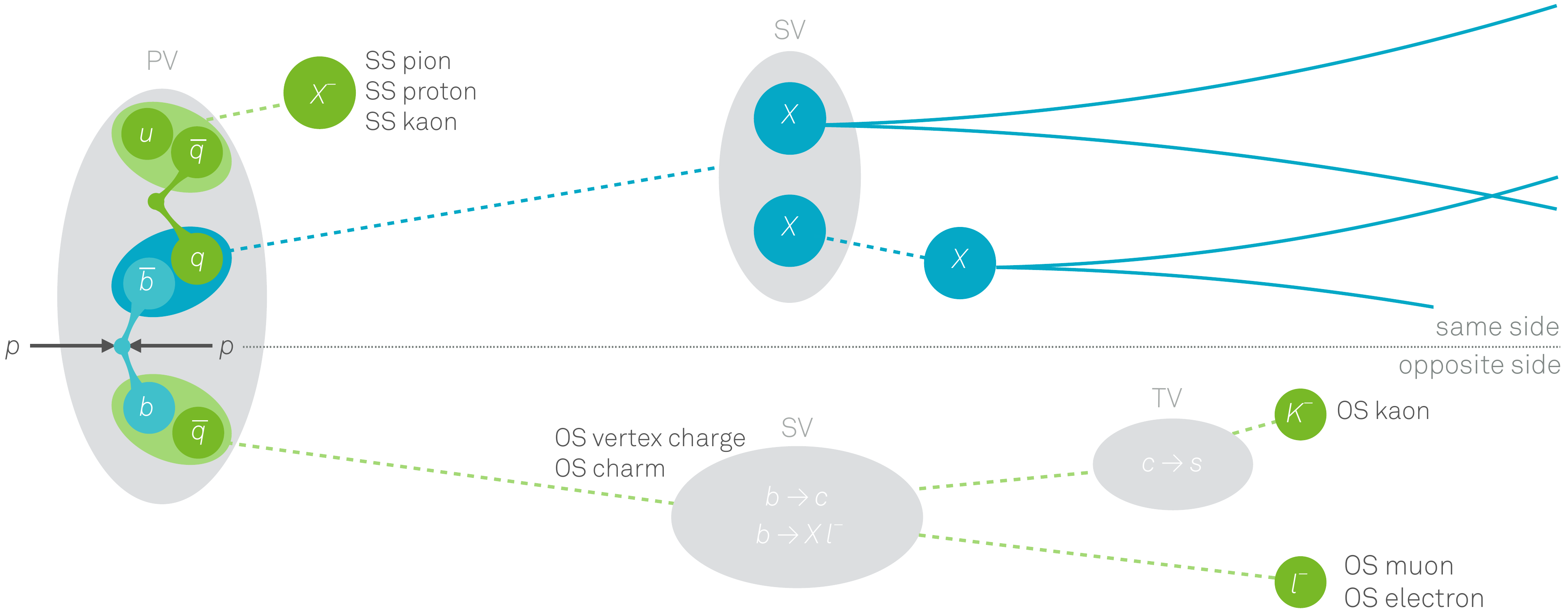}
  \caption{Schematics of information used in flavour tagging.
    Figure from Ref.~\cite{LHCb-FIGURE-2020-002}.}\label{Fig:tagging}
\end{figure}

Resolution of the above-mentioned 
interference patterns require flavour tagging, i.e. the identification
of the original flavour of the \Bd or \Bs meson. LHCb uses
the flavour of the accompanying \bquark from the \bquark\bquarkbar pair,
using muon, electron and kaon particles, as well as the vertex charge (Fig.~\ref{Fig:tagging}).
In addition, accompanying pions, kaons and protons from fragmentation
are used to determine the \B meson flavour directly~\cite{LHCb-PAPER-2016-039}.

Improved understanding of proton-proton collisions at LHC energies has allowed continuous increase of the
effective tagging power from less than 2\% in 2011~\cite{LHCb-PAPER-2011-021} to
more than 6\% for selected modes
nowadays~\cite{LHCb-PAPER-2023-001,LHCb-PAPER-2023-013}.

A (somewhat outdated) comparison of several decays modes is shown in Fig.~\ref{Fig:tagging2}.
Note however that the tagging performance anti-correlates with the hardware trigger efficiency:
Modes with muons profit from low trigger thresholds while those
with hadrons or electrons have lower efficiencies.

\mywrapfig{0.5\textwidth}{
  \includegraphics[width=\textwidth]{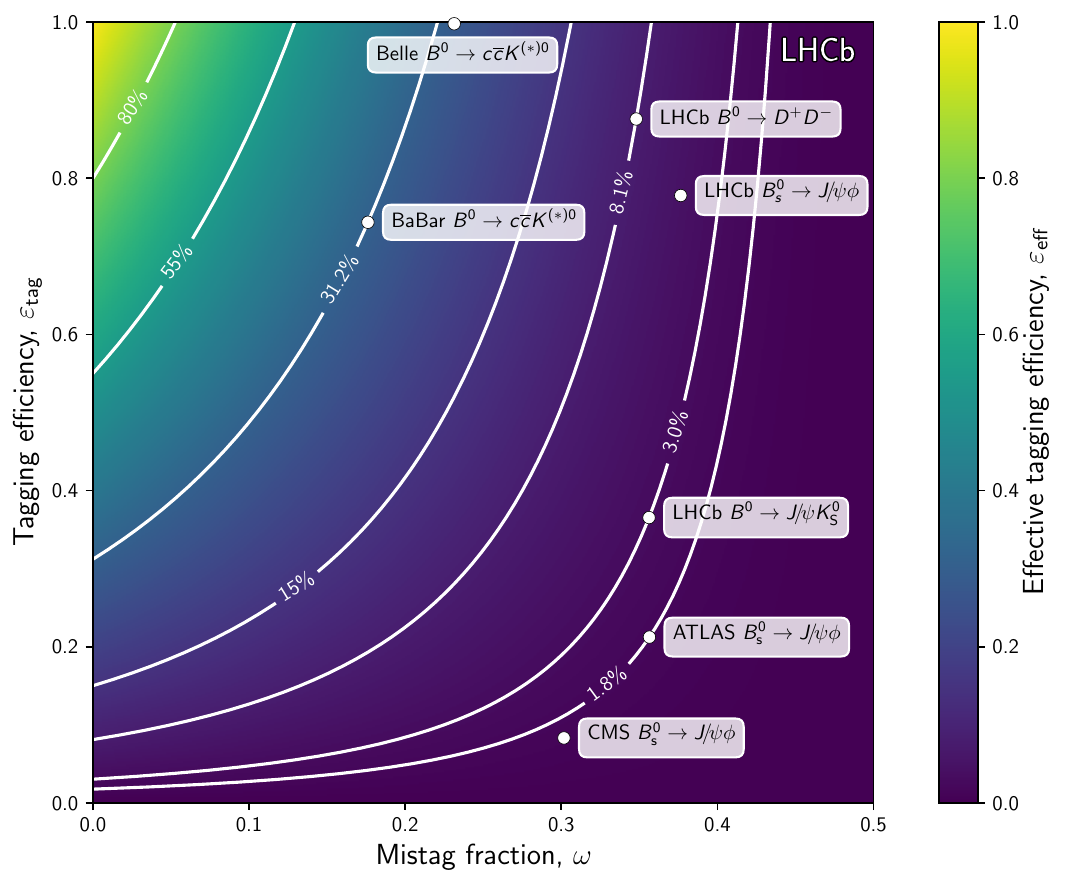}
  \caption{Tagging performance for selected analyses.
    The lines indicate constant values of the effective tagging power $\epsilon_\text{tag}(1-2\omega_\text{tag})^2$.
    Figure from Ref.~\cite{LHCb-FIGURE-2020-002}.}\label{Fig:tagging2}
  \vskip -6ex
}

These thresholds
however bias the transverse momentum of the accompanying \B and the tagging
particles from fragmentation, which improves the tagging performance.
As an example, the best tagging performance at the LHC is obtained by CMS~\cite{CMS:2020efq}
thanks for their triggering on the tagging muon: Trigger efficiency is traded off for
better tagging.

\section{Anomalies coming and going}\label{Sec:RD}
\begin{figure}[tb]\centering
  \includegraphics[width=0.9\textwidth]{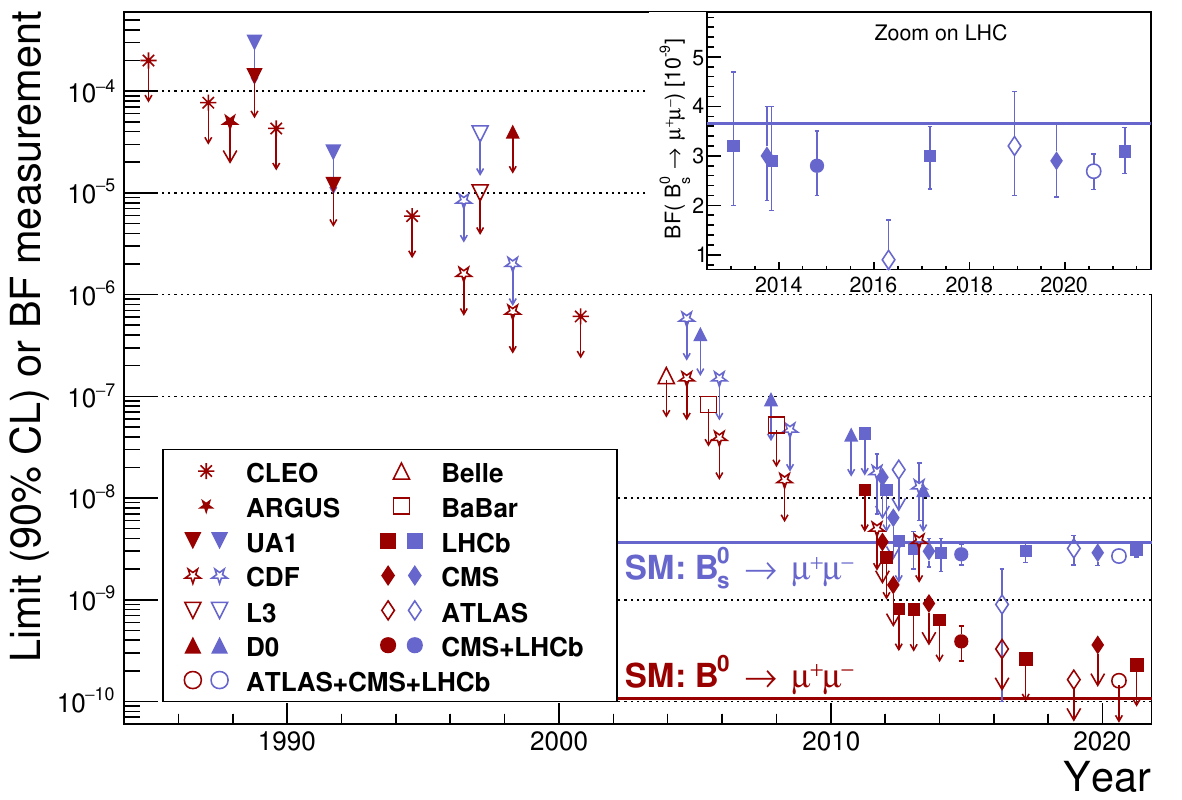}
  \caption{(left) Summary of all limits and measurements of \Bsmm and \Bdmm branching fractions.
    Figure courtesy F.~Dettori adapted from Ref.~\cite{LHCb-PAPER-2014-049}.}
  \label{Fig:Bmm}
\end{figure}

After 30 years of unsuccessful searches at multiple experiments,
the observation of \Bsmm was one of the benchmark goals at the LHC.
This loop-induced, GIM and helicity-suppressed process is very rare in the SM
--- its branching fraction is 3 in a billion~\cite{Beneke:2019slt,DeBruyn:2012wj} ---
which makes it sensitive to any new physics contribution at this or higher level.
In supersymmetric extensions its rate is enhanced proportionally to the sixth power of $\tan\beta$ --- 
the ratio of the two vacuum expectations of the two neutral Higgs bosons.
Its non-observation increasingly shattered hopes for large-$\tan\beta$ supersymmetry as
the limit on the branching fraction decreased (Fig.~\ref{Fig:Bmm}).
The first evidence was then finally reported by LHCb in 2013~\cite{LHCb-PAPER-2012-043},
the first observation was achieved via from a joint fit to LHCb and CMS data in
2014~\cite{LHCb-PAPER-2014-049}\footnote{The fit to a joint dataset of two experiments
  was a major enterprise. It required first to align all definitions and treatments of backgrounds
  in the two experiments. Once this was done the complex simultaneous fit to multiple datasets
  returned the same result as one would have obtained from combining the LHCb and CMS
  likelihoods~\cite{LHCb-PAPER-2014-049}.
  This teaches a lesson on the relative importance of agreeing between experiments versus developing complex
fitting frameworks.}
and LHCb reported the first single-experiment observation in 2017~\cite{LHCb-PAPER-2017-001},
soon followed by CMS~\cite{CMS:2019bbr}. ATLAS are just a bit short of an observation~\cite{ATLAS:2018cur}.
The LHCb and CMS results were in the meantime updated
using the full Run 1--2 data sets~\cite{LHCb-PAPER-2021-007,LHCb-PAPER-2021-008,CMS:2022mgd}.
The average of the \Bsmm measurements~\cite{Allanach:2022iod} is consistent with the Standard Model
expectation, which sets strong constraints on new physics affecting $\bquark\squarkbar\ell\overline{\ell}$ operators.

The further suppressed \Bdmm mode still escapes clear detection,
but one can nevertheless determine a ratio of branching fractions
which is precisely predicted in the SM as it is essentially a measurement
of \CKM matrix elements:
\eq[eq:Rmumu]{
{\cal R}_\mumu  \equiv \frac{\BF(\Bdmm)}{\BF(\Bsmm)}  
\stackrel{\text{SM}}{=} \frac{\Gamma_s^\text{H}}{\Gamma_d}\left(\frac{f_\Bd}{f_\Bs}\right)^2
\frac{|\Vtd|^2}{|\Vts|^2}\frac{\sqrt{m^2_\Bd-4m^2_\mu}}{\sqrt{m^2_\Bs-4m^2_\mu}},
}
and is measured as~\cite{LHCb-PAPER-2021-007,LHCb-PAPER-2021-008}
\eq{
{\cal R}_\mumu^\text{exp}=\left(3.9\aerr{3.0}{2.4}\aerr{0.6}{0.4}\right)\times10^{-2}.
}

\begin{figure}[t]%
\includegraphics[width=0.30\textwidth]{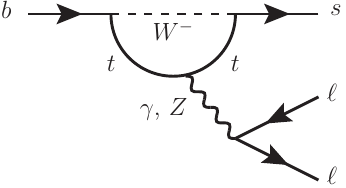}\hskip 0.05\textwidth%
\includegraphics[width=0.30\textwidth]{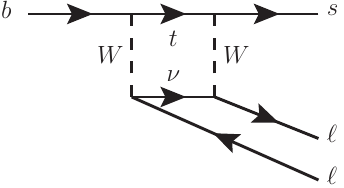}\hskip 0.05\textwidth%
\includegraphics[width=0.30\textwidth]{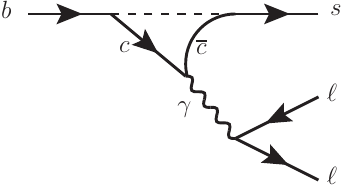}%
\caption{Feynman diagrams of the dominant Standard Model contributions to \blls: %
(left) electroweak loop, (centre) box, %
(right) \cquark{}\cquarkbar loop diagram.}\label{Fig:blls}%
\end{figure}

Another decay mode sensitive to new physics in these operators is \BllKs.
While \Bsmm is essentially probing axial currents, \BmmKs also probes vector currents,
and their interference, as shown in Fig.~\ref{Fig:blls}.

The exclusive decay \decay{\Bz}{\Kstarz\ellp\ellm}, with 
\decay{\Kstarz}{\Kp\pim}, provides a rich set 
of observables with different sensitivities to new physics, and for which theoretical 
predictions are  available. 
This process is complicated by a dependence on $q^2$, 
the dilepton mass squared.
At low $q^2$, \BdllKs behaves like \BgKsz, with a slightly off-shell photon 
decaying to two leptons. 
At higher \qsq values, there is an interference of the amplitudes controlled by the
${\cal O}_{9}$ and ${\cal O}_{10}$ operators~\cite{Buchalla:1995vs},  related to the \Z loop and \W  box diagrams, respectively. 
This ``low-$q^2$'' region between 1 and 6\:\gevgevcccc is the most 
interesting and theoretically cleanest.
The observation of high mass resonances above the \psitwos meson
by the LHCb collaboration~\cite{LHCb-PAPER-2013-039} is an indication that a lot of
care is needed when interpreting the high-$q^2$ region.

\begin{figure}[b]
\centering
\includegraphics[width=0.49\textwidth]{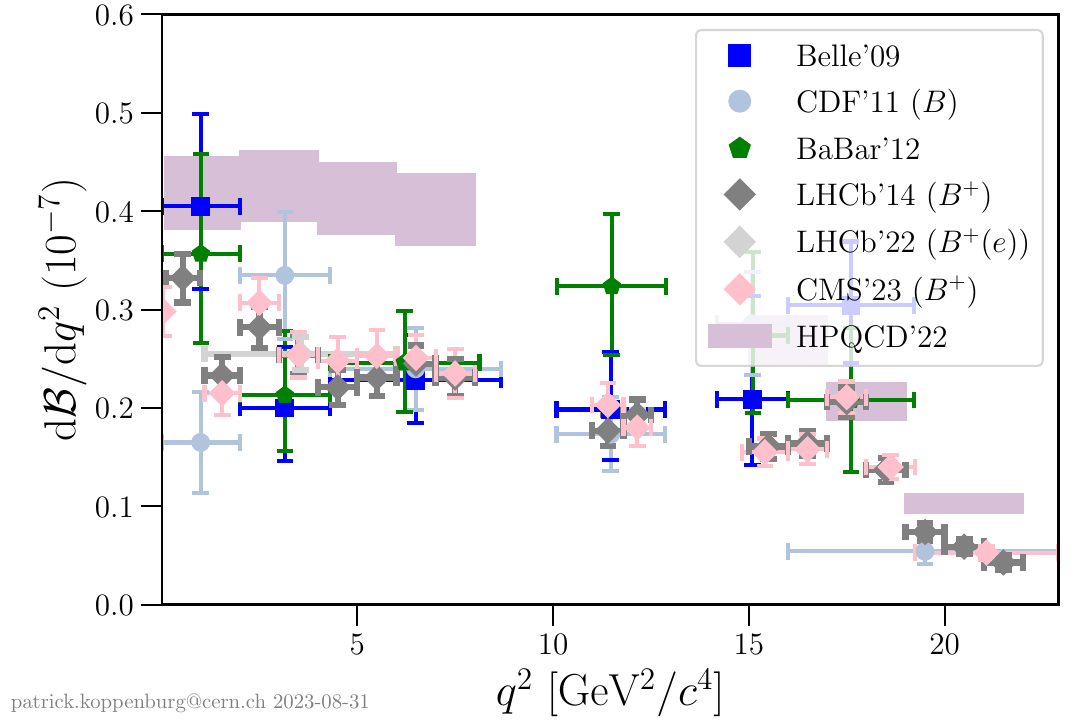}\hskip 0.01\textwidth
\includegraphics[width=0.49\textwidth]{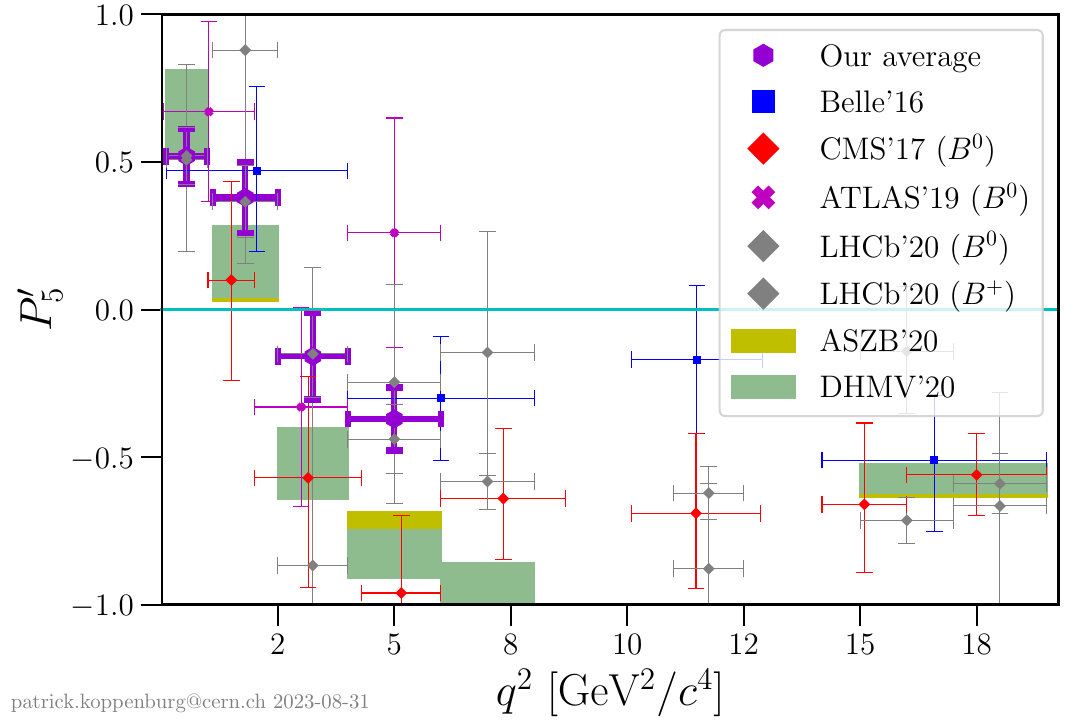}
\caption{Experimental results on the (left) \BllK differential decay rate~\cite{Lees:2012tva,Wei:2009zv,Aaltonen:2011qs,LHCb-PAPER-2012-011,LHCb-PAPER-2022-045,LHCb-PAPER-2022-046} compared to prediction of Ref.~\cite{Parrott:2022zte} and (right) $P_5'$~\cite{Lees:2015ymt,Wehle:2016yoi,Khachatryan:2015isa,Sirunyan:2017dhj,Aaboud:2018krd,LHCb-PAPER-2020-002,LHCb-PAPER-2020-041} asymmetry compared to predictions from Refs.~\cite{Straub:2015ica,Altmannshofer:2014rta} (sea green) and Refs.~\cite{Descotes-Genon:2014uoa,Khodjamirian:2010vf} (green).}
\label{Fig:PAPER-2020-002}
\end{figure}

Branching fraction predictions are affected by hadronic uncertainties (see also below),
but selected ratios of observables benefit from cancellations of 
uncertainties, thus providing a cleaner test of the 
Standard Model~\cite{Eilam:1986fs,Ali:1999mm,Kruger:2005ep,Altmannshofer:2008dz,Egede:2008uy,Bobeth:2008ij,Descotes-Genon:2013vna}.
The observable $P_5'$~\cite{LHCb-PAPER-2020-002} for instance is in tension
with the theoretical prediction, as seen in Fig.~\ref{Fig:PAPER-2020-002}, but the
jury is still out on determining what the cause is.
In order to address this question, 
LHCb recently performed an amplitude analysis in the \BdmmKs decay
in which the short-distance Wilson coefficients and long-distance nuisance parameters
are determined from the data~\cite{LHCb-PAPER-2023-033}.
The overall level of discrepancy with the Standard Model is at the level of $2\sigma$.

Until recently there was excitement about the evidence of lepton-universality breaking in
the decays \BeeK and \BmmK~\cite{LHCb-PAPER-2021-004,LHCb-PAPER-2021-038}, 
and similarly but less significantly in
\BeeKs and \BmmKs~\cite{LHCb-PAPER-2017-013,LHCb-PAPER-2021-038}.
With these decays one defines the ratio~\cite{Hiller:2003js}
\begin{equation}
  R_{\X} = \frac{%
            \int\limits_{q_\text{min}^2}^{q_\text{max}^2}\rmd\qsq\frac{d\Gamma\left(\decay{\B}{X\mumu}\right)}{\rmd\qsq}%
          }{%
            \int\limits_{q_\text{min}^2}^{q_\text{max}^2}\rmd\qsq\frac{d\Gamma\left(\decay{\B}{X\epem}\right)}{\rmd\qsq}}
\end{equation}
in a well chosen range of \qsq (usually 1 to 6\gevgevcccc) in order to avoid charmonium resonances and photon poles.
These ratios should be identical to unity at a level of precision well below experimental resolution~\cite{Isidori:2020acz}.

Initially Hiller and Kr\"uger~\cite{Hiller:2003js} had introduced these ratios as they are linearly correlated with the branching fraction
of \Bsmm (under some reasonable assumptions, notably minimal flavour violation). In 2003 the hope was thus
that Belle and BaBar could see sign of new physics in these ratios before the LHC would observe \Bsmm.
At a high-luminosity workshop at SLAC in 2003, Hiller convinced me to check whether LHCb could
measure this ratio. I showed that LHCb could so~\cite{Koppenburg:1027442}
thereby demonstrating that LHCb had physics potential with electrons.\footnote{Inspection of old
  reports to the LHCC~\cite{Dijkstra:691698,CERN-LHCC-98-004,:2009ny,LHCb-TDR-009} shows that the use of electrons was essentially for flavour tagging,
  and adding some data to \phis and $\sin2\ckmbeta$ measurements. }
A first proof-of-concept analysis
was the topic of a master thesis~\cite{Skidmore} that exceeded my expectations
and led to an LHCb measurement that was
``compatible with the SM prediction within 2.6 standard deviations''~\cite{LHCb-PAPER-2014-024}.\footnote{The
  result was submitted to PRL after the editor had expressed their interest in
  this result following a CERN news update.
}
The result was {\it below} unity,\footnote{Since Hiller and Kr\"uger expected $R_X\ge1$ they defined
  it as muon/electrons. This is an annoyance for experiments as the uncertainty on $R_X$ is dominated by the
  electron modes, which leads to asymmetric uncertainties when expressed in the ratio.
  In hindsight one should have defined $R_X^{-1}$, which is what is reported in Ref.~\cite{LHCb-PAPER-2019-040}.}
meaning we saw a deficit of muons; 
a case that had not been foreseen by Hiller in the original publication.

Three years later the analysis of \BdllKs yielded a similar deviation
from unity. However this result already contained a hint that something was
not quite right: the ratio \RKstar in the low \qsq bin should have been unity,
as this region is dominated by the photon pole and lepton-universality is known to be
respected in electromagnetic decays. In spite of lengthy investigations nothing could be found
that explained the discrepancy and the result was published as is~\cite{LHCb-PAPER-2017-013}.
The value of \RK in \BllKp was updated twice with Run~2 data~\cite{LHCb-PAPER-2019-009,LHCb-PAPER-2021-004},
the latter of which exhibited a $3\sigma$ evidence.
This was exciting news for model builders. Indeed, no hadronic effects can mimic
values of \RK different from unity, unlike other deviations,
which can be generated by stretching some not-so-certain QCD predictions.
It was known that the discrepancy with the SM was driven by
\bmms channels, for which all rates are measured below the SM
expectation~\cite{LHCb-PAPER-2012-011,LHCb-PAPER-2021-014}, while
the electrons were thought to be SM-like.

The anomaly disappeared after a reappraisal
of the LHCb result~\cite{LHCb-PAPER-2022-045,LHCb-PAPER-2022-046}.
More stringent particle identification requirements reduced the amount of
backgrounds with two hadrons mis-identified as electrons, and these and similar
backgrounds were determined from data and incorporated in the mass fit.
The measurements of two bins in \qsq (below and above 1\gevgevcccc)
in \BullK and \BdllKs are now all compatible with unity.

\mywrapfig{0.5\textwidth}{
  \includegraphics[width=\textwidth]{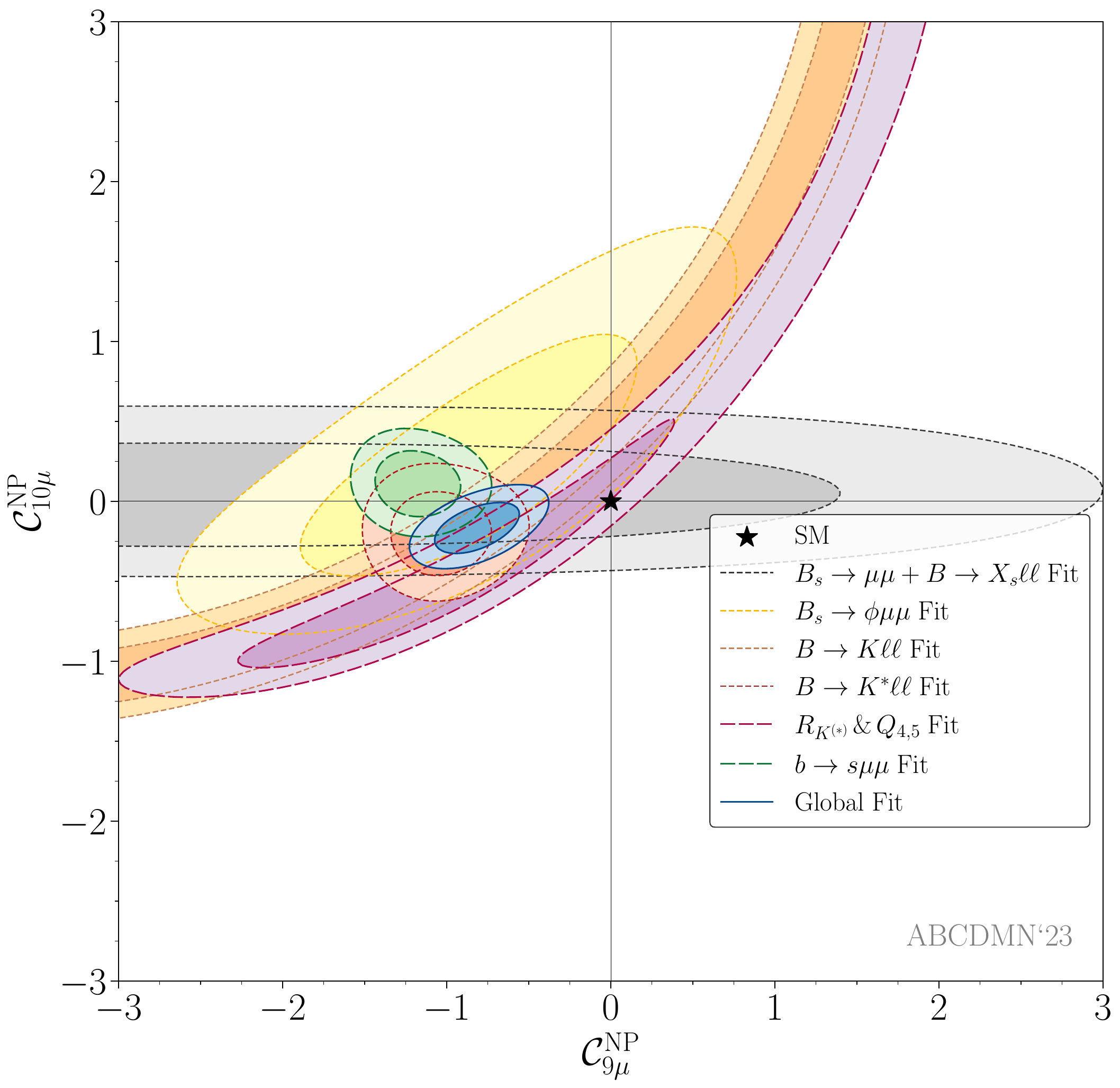}
\caption{Fits to \cnine and \cten Wilson coefficients from \bsll modes.
    Figure from Ref.~\cite{Alguero:2023jeh}.}\label{Fig:Fits}%
}%
However, the demise of \RK replaced one anomaly with another.
After the correction, the electron modes also have measured decay rates below the SM expectation.
Some lattice QCD groups report discrepancies in excess of $4\sigma$~\cite{Parrott:2022zte}
for the muon modes, while the electron and muon modes are experimentally compatible.
Other groups however determine form factors from the data and get much lower
tensions~\cite{Gubernari:2022hxn}.


Input from all \blls, \Bmm and \bsg modes is used to perform fits to Wilson coefficients \cnine and \cten
corresponding to the vector and axial operators, respectively.
Multiple groups~\cite{DeBlas:2019ehy,Isidori:2021vtc,Hurth:2021nsi,Altmannshofer:2021qrr,Geng:2021nhg,Angelescu:2021lln,Ciuchini:2022wbq,Greljo:2022jac,Gubernari:2022hxn,Alguero:2023jeh} consistently hint at 
a modified vector $b\squarkbar\ell\bar{\ell}$ operator --- with varying significance --- as shown in
in Fig.~\ref{Fig:Fits}~\cite{Alguero:2023jeh}.

The elephant in the room are non-local \cquark{}\cquarkbar contributions
(Fig.~\ref{Fig:blls}, right).
Unlike the local form factors that are in principle calculable, the charm loops
are nonlocal --- i.e. the \bquark\squarkbar{}\cquark{}\cquarkbar and \cquark{}\cquarkbar{}$\ell{}\bar{\ell}$ operators
appear at different points in spacetime --- which requires involved integrals. 
In the \BmmK channel, light-cone-sum-rule (LCSR) methods predict small nonlocal
effects~\cite{Khodjamirian:2012rm},
and attempts to constrain charm loops from data~\cite{Gubernari:2022hxn}
reach the conclusion that they are too small to explain the observed
discrepancies. Are the discrepancies due to form factors?
The latest evidence by Belle~II for the \decay{\Bp}{\Kp\neu\neub} decay~\cite{B2Knunu},
at a rate above the SM predicted branching fraction,
tends to indicate that the \decay{\B}{\kaon} form factor is not low
compared to the SM expectation.
In \BmmKs decays the form factors and nonlocal effects are less well controlled~\cite{Khodjamirian:2010vf}.


\mywrapfig{0.6\textwidth}{
  \includegraphics[width=\textwidth]{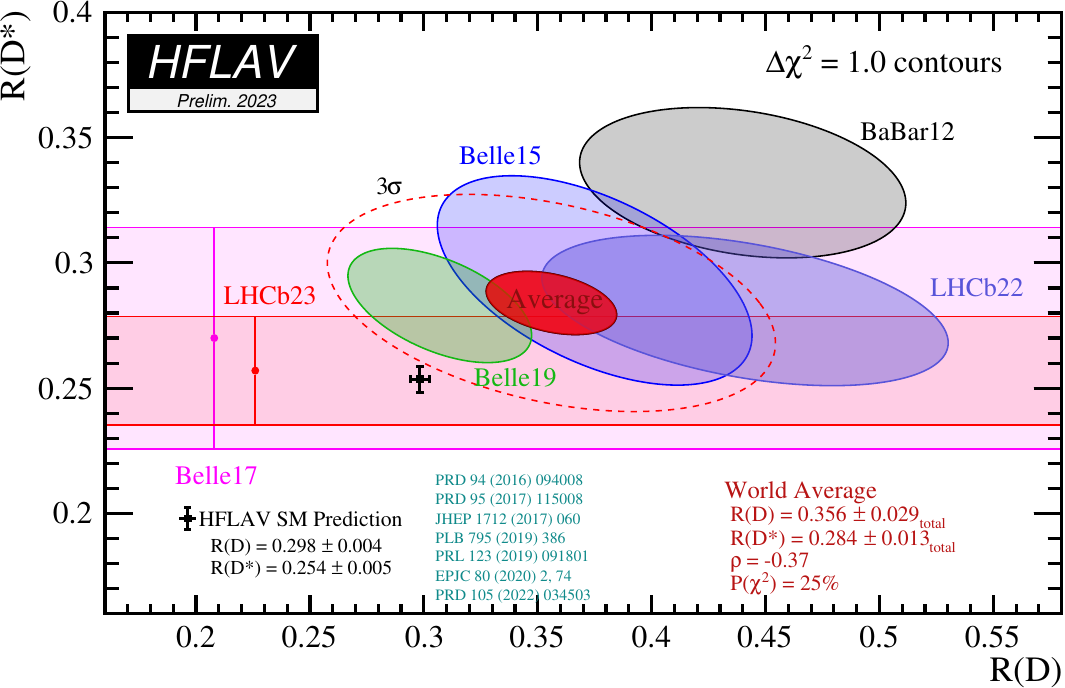}%
\caption{Experimental constraints on \RD and \RDstar. Figure from Ref.~\cite{HFLAV21}.}\label{Fig:RD}%
}

Another test of lepton universality in tree decays is done with
semileptonic decays to \Ptau leptons. The ratio \RD 
of the \decay{\B}{\D\Ptau\neu} to \decay{\B}{\D\muon\neu} decay rates
--- and mutatis mutandis for \RDstar\ --- are not unity because of phase-space factors,
but well predicted in the SM. There has been a long-standing discrepancy
at the level of $3\sigma$, mostly driven by a measurement from BaBar
\cite{BaBar:2012obs,BaBar:2013mob}, that pulls the experimental average away
from the SM, see Fig.~\ref{Fig:RD}.
In the spirit of ``things LHCb cannot do (but still does)'',
LHCb is contributing to this programme: in spite of
the missing neutrino and the multiplicity of backgrounds affecting these
decays, LHCb is able to reconstruct the missing-mass, \qsq and muon energy in
the \B frame, owing to the missing-momentum correction coming from
the \B pointing requirement. 
The LHCb results are compatible with the SM in \RDstar, but show a slight tension
in \RD~\cite{LHCb-PAPER-2022-039,LHCb-PAPER-2022-052}. The full legacy dataset is
not yet exploited, so more updates will be coming, while waiting for
the first results from Belle~II.

\begin{figure}[tb]\centering
  \includegraphics[width=0.8\textwidth]{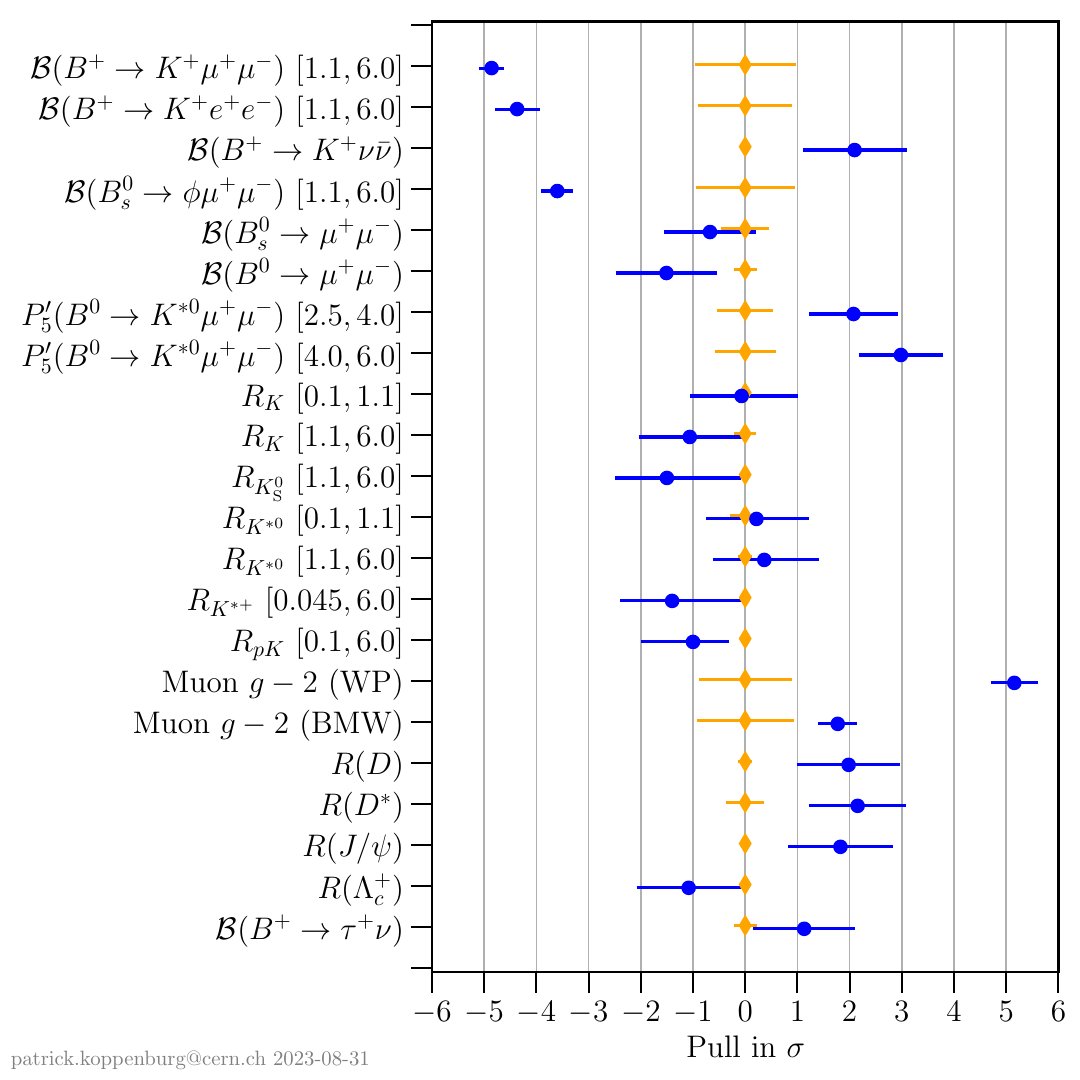}\quad
  \caption{
    Cherry-picked selection of measurements in flavour physics. The SM prediction is placed at zero.
    The experimental value is then offset its deviation from the SM in units of standard deviations.
    The quadratic sum of the uncertainties is unity by construction.
    Figure from ref.~\cite{PKanomalies}.
    Values from Refs.~\cite{LHCb-PAPER-2014-006,LHCb-PAPER-2022-045,LHCb-PAPER-2022-046,Parrott:2022zte,LHCb-PAPER-2021-014,Horgan:2015vla,Allanach:2022iod,LHCb-PAPER-2021-007,LHCb-PAPER-2021-008,ATLAS:2018cur,CMS:2022mgd,HFLAV21,Muong-2:2021ojo,Aoyama:2020ynm,LHCb-PAPER-2017-035,Harrison:2020nrv,LHCb-PAPER-2021-044,Bernlochner:2018bfn,PDG2022,UTfit:2006vpt,LHCb-PAPER-2021-005,DiLuzio:2019jyq,ALEPH:2005ab}.
    Figure from Ref.~\cite{Alguero:2023jeh}.
  }
  \label{Fig:anomalies}
\end{figure}
Figure~\ref{Fig:anomalies} shows a cherry-picked selection of measurements which are of particular interest.
For each measurements, the SM prediction is placed at zero.
The experimental value is then offset its deviation from the SM in units of standard deviations.
The quadratic sum of the two uncertainties is therefore unity by construction.
This presentation shows which observables have an uncertainty dominated by
experiment (\RD, $\BF(\decay{\Bp}{\taup\neu}$ stick out) or by theory ($\BF(\decay{\Bp}{\Kp\mumu}$),
which tells where efforts are needed to reduce uncertainties.\footnote{Additionally there
  are anomalies like the muon $g-2$ value where multiple theory determinations do not agree
  (we report Ref.~\cite{Aoyama:2020ynm}) or where there is disagreement between experiments
  (as the \W mass, which is not reported here).}

\section{Spectroscopy}\label{Sec:spectro}
\begin{figure}[t]\centering
  \includegraphics[width=\textwidth]{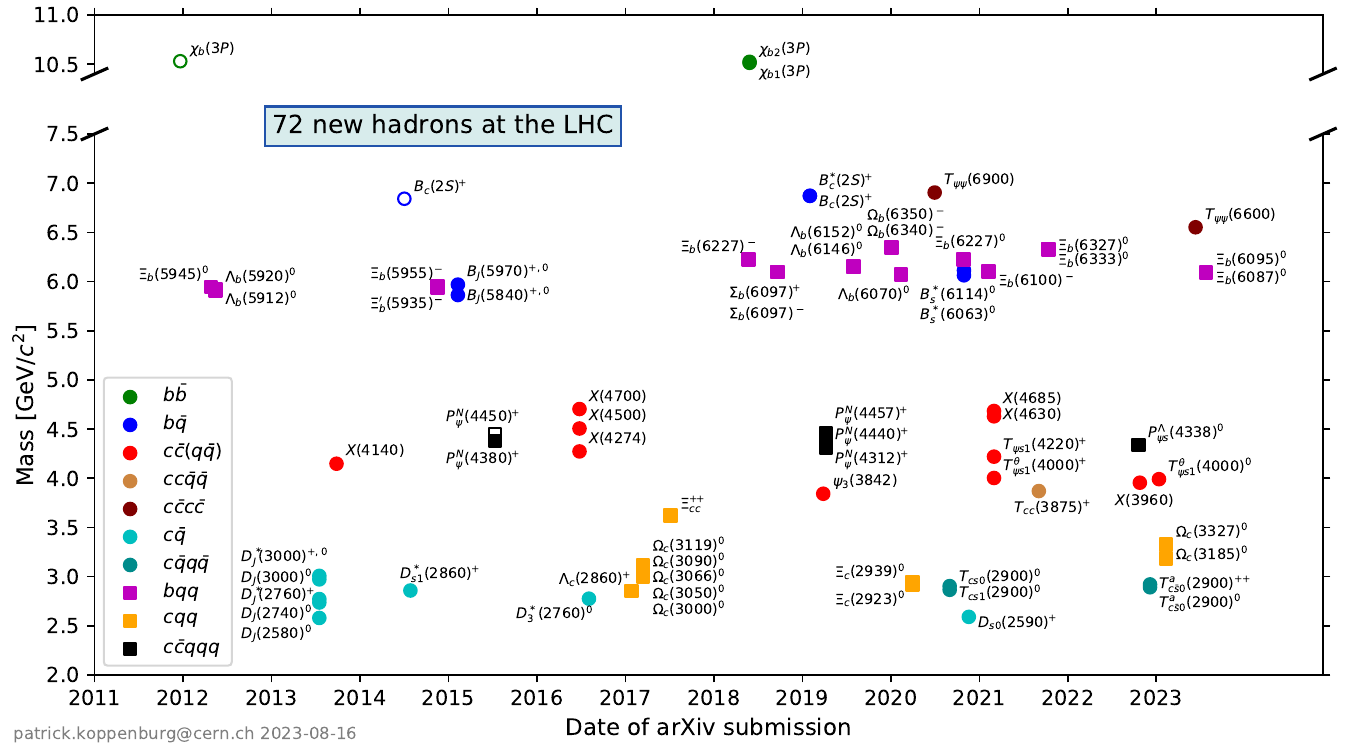}
  \caption{Masses and date of {\tt arxiv} submission for all states observed at the LHC.
  Figure from Ref.~\cite{LHCb-FIGURE-2021-001}.}\label{Fig:hadrons}
\end{figure}

The LHC has become a hadron discovery machine~\cite{ConversationHadrons}
with 72 states observed so far, 64 of which were discovered by LHCb.
A timeline is shown in Figure~\ref{Fig:hadrons} \cite{LHCb-FIGURE-2021-001}.

Many of these new states are exotic; but let us rewind history to 2003.
I remember well the internal discussions about a bump
in the \jpsi\pip\pim mass that was seen in Belle data of the \decay{\Bp}{\jpsi\Kp\pip\pim}
decay, an considered it likely to be yet another charmonium state.
Instead, this accidental observation~\cite{Choi:2003ue}
of what turned out to be the first tetraquark, called the $X(3872)$ and then the \theX meson,
started the whole new field of exotic spectroscopy.
Multiple similar states --- some of which charged --- have then been
discovered by the \B factories; see Ref.~\cite{Lebed:2016hpi} for a history.
We now know that \theX is an isospin-singlet that proceeds mostly in an isospin-violating
decay to \Dz\Dzb\piz or \Dstarz\Dzb~\cite{Belle:2006olv}, or to \jpsi\pip\pim
via a \rhoz~\cite{Belle:2011vlx}, but with a sizeable \Pomega
component~\cite{LHCb-PAPER-2021-045}.\footnote{This \Pomega component has been
underestimated for a long time because of a bug in the \evtgen model
which made the generated \pip\pim mass from pure \decay{\theX}{\jpsi\rhoz} decays look
like what was seen in data; see \eg mass plots in
Refs.~\cite{CMS:2013fpt,ATLAS:2016kwu,LHCb-PAPER-2015-015}.}
LHCb showed that \theX meson has quantum numbers $J^{PC}=1^{++}$~\cite{LHCb-PAPER-2013-001,LHCb-PAPER-2015-015}
and is consistent with \Dstarz\Dzb bound state with 24\kev
binding energy and a width of $1.39\pm0.26\mev$~\cite{LHCb-PAPER-2020-008}
(see also Ref.~\cite{Belle:2023zxm}). This makes it look like a molecule,
while its production mode is that of charmonium~\cite{LHCb-PAPER-2021-026}.
If both possibilities exist, reality must be a superposition of the two~\cite{Weinberg:1965zz,LHCb-PAPER-2020-008}.

In spite of some early papers (notably Ref.~\cite{LHCb-PAPER-2011-033}),
most members of the LHCb experiment only realised the potential of exotic spectroscopy
when we stumbled over the \jpsi\proton pentaquarks with minimal quark content
\cquark\cquarkbar\uquark\uquark\dquark~\cite{LHCb-PAPER-2015-029}.
These states were first seen in \decay{\Lb}{\jpsi\proton\Km} decays,\footnote{simultaneously by multiple people
  including a CERN summer student supervised by the author.} and established in
an amplitude analysis. The 3\invfb data set was best fit with a wide $P_\psi^N(4380)^+$ and a narrow $P_\psi^N(4450)^+$
state (using the naming scheme of Ref.~\cite{Gershon:2022xnn}).
A subsequent simplified analysis~\cite{LHCb-PAPER-2022-031} of the 9\invfb legacy data showed that the latter state is split in
two states, $P_\psi^N(4440)^+$ and $P_\psi^N(4457)^+$, and that another is needed at $4312\mev$.
The full amplitude analysis is underway.

\begin{table}[t]
  \caption{Selected tetraquark states listed quark content, beyond \cquark{}\cquarkbar{}\quark{}\quarkbar~\cite{LHCb-PAPER-2020-011,CMS:2013jru,LHCb-PAPER-2016-019,CMS:2023owd,ATLAS:2023bft,LHCb-PAPER-2022-040,LHCb-PAPER-2020-044,LHCb-PAPER-2022-026,LHCb-PAPER-2022-027,LHCb-PAPER-2020-025,LHCb-PAPER-2021-031,LHCb-PAPER-2021-032,Choi:2003ue}.}\label{Tab:TQ}
    \begin{tabular}{L||L|L|L|L|L|L}
        \rowcolor{gray} & \multicolumn{1}{c|}{\cquark\cquarkbar} & \multicolumn{1}{c|}{\cquark\dquarkbar} & \multicolumn{1}{c|}{\squark\squarkbar} & \multicolumn{1}{c|}{\squarkbar\uquark} &  \multicolumn{1}{c|}{\squarkbar\dquark} & \multicolumn{1}{c}{\squark\uquarkbar} \\
        \hline\hline
        \cellcolor{gray}  & T_{\psi\psi}(6900) &  & X(4140) & T^\theta_{\psi s1}(4000)^-  & T^\theta_{\psi s1}(4000)^0 & \\
        \multirow{-2}{*}{\cellcolor{gray}\cquark\cquarkbar} &  T_{\psi\psi}(6600) &  & \text{\bf+5 more}& T_{\psi s1}(4220)^- & &\\
        \hline
        \cellcolor{gray}\cquark\uquarkbar & & T_{cc}(3875)^+ & & & T^a_{c\squarkbar0}(2900)^0 & \\
        \hline
        \cellcolor{gray} &  & & & & T^a_{c\squarkbar0}(2900)^{++} & T_{c\squarkbar0}(2900)^{0}\\
        \multirow{-2}{*}{\cellcolor{gray}\cquark\dquarkbar}  & &  & & & & T_{c\squarkbar1}(2900)^{0}\\
        \hline
        \cellcolor{gray}\quark\quarkbar & \theX\dots & & & & \\
  \end{tabular}

\end{table}
With the full legacy data sample, LHCb also discovered doubly charmed states \Tcc~\cite{LHCb-PAPER-2021-031,LHCb-PAPER-2021-032}, which differ from previously discovered tetraquarks in that they have two charm quarks and two light anti-quarks,
the $T^a_{cs0,1}(2900)$ states with a single charm and a strange quark~\cite{LHCb-PAPER-2022-026,LHCb-PAPER-2022-027},
and the heavy $T_{\psi\psi}$ with two charm and two anti-charm quarks~\cite{LHCb-PAPER-2020-011,ATLAS:2023bft,CMS:2023owd}.
Table~\ref{Tab:TQ} attempts to classify these states by quark content.

Mapping out existing (and eventually non-existing) quark contents helps understanding
the internal structure of exotic hadrons. The dispute between proponents of the
molecular picture and those of the compact tetra- and pentaquarks has calmed down
lately. Most likely there are representatives of both kinds.
Let's look at states with two heavy quarks $Q$, and $Q$ or $\overline{Q}$.
If $Q\overline{Q}$ is in a colour-singlet configuration, it will immediately hadronise
into quarkonium. If the quarks are in different colourless hadrons, they may form a molecule.
On the other hand $QQ$ can never be in a singlet configuration. A $QQ\quarkbar\quarkbar$
state may thus be compact. A $Q\overline{Q}\quark\quarkbar$ may not~\cite{Karliner}.
It is thus likely that all kinds of structures exist in nature: hadronic molecules,
compact multi-quark objects, superposition of those, as well as rescattering effects.
Sorting them out will be an enterprise for the next decades~\cite{Brambilla:2022ura}.

\begin{figure}[tb]\centering
  \def\ww{0.3\textwidth}
  \includegraphics[width=\ww]{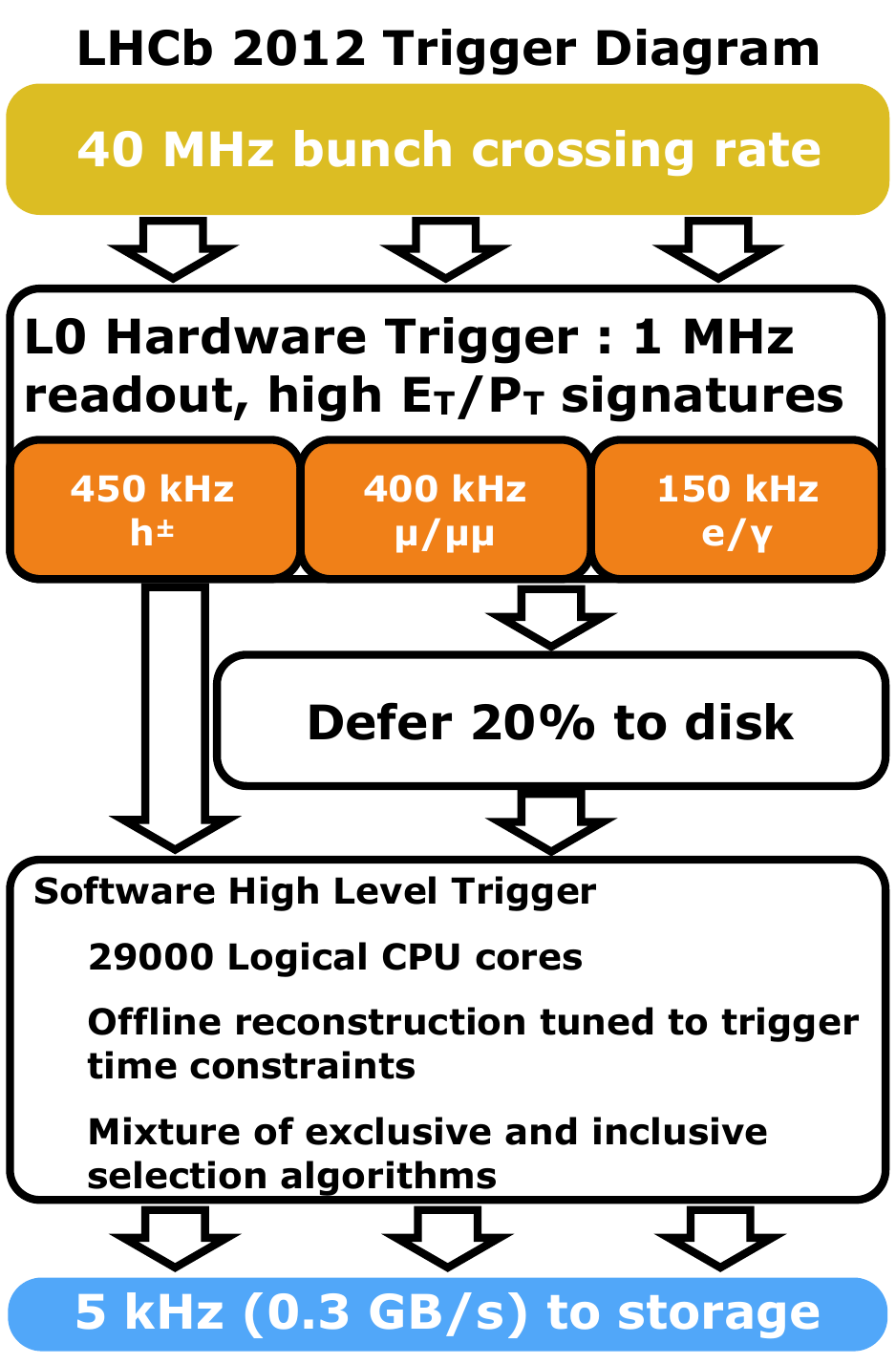}\quad
  \includegraphics[width=\ww]{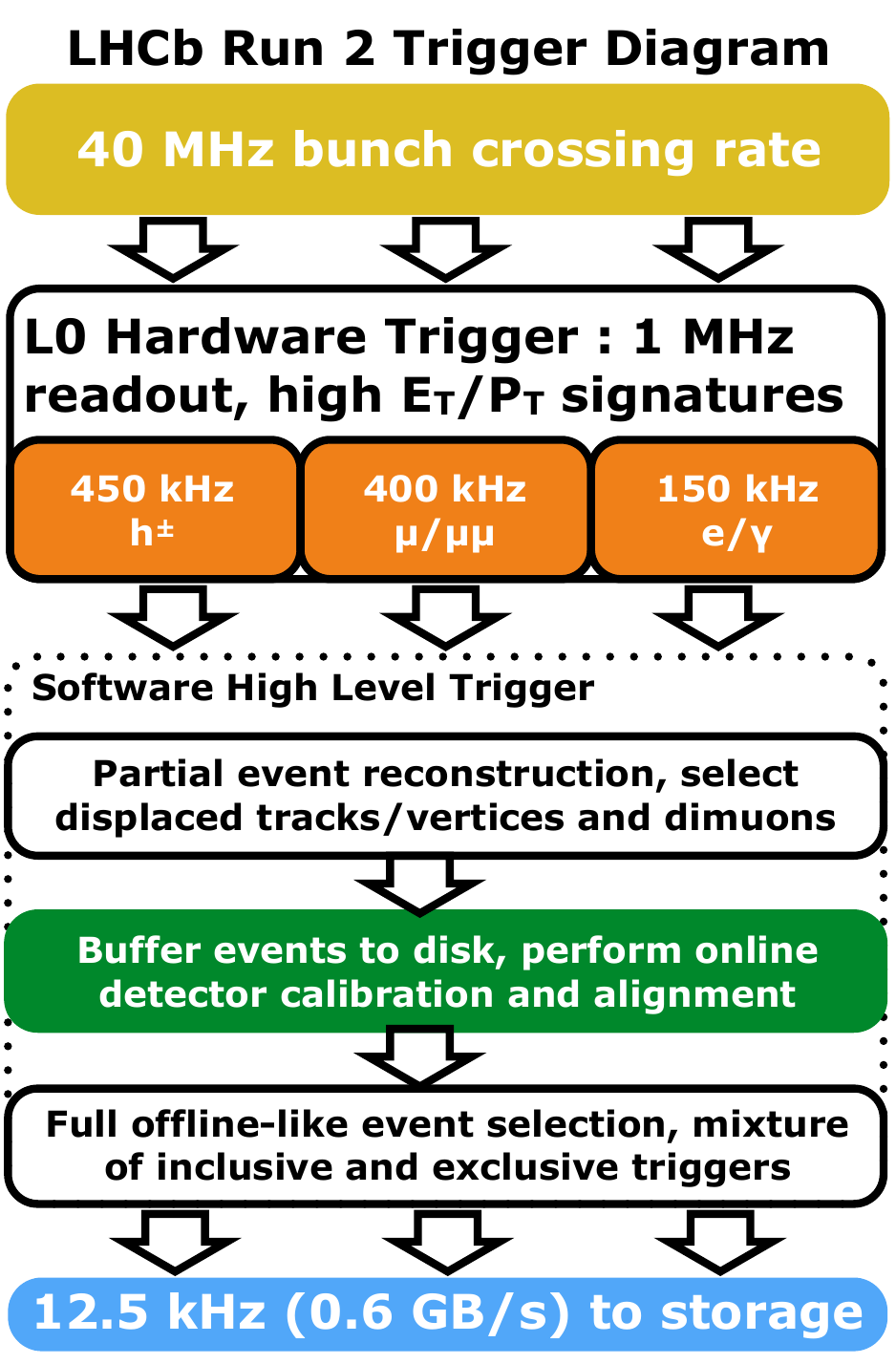}\quad
  \includegraphics[width=\ww]{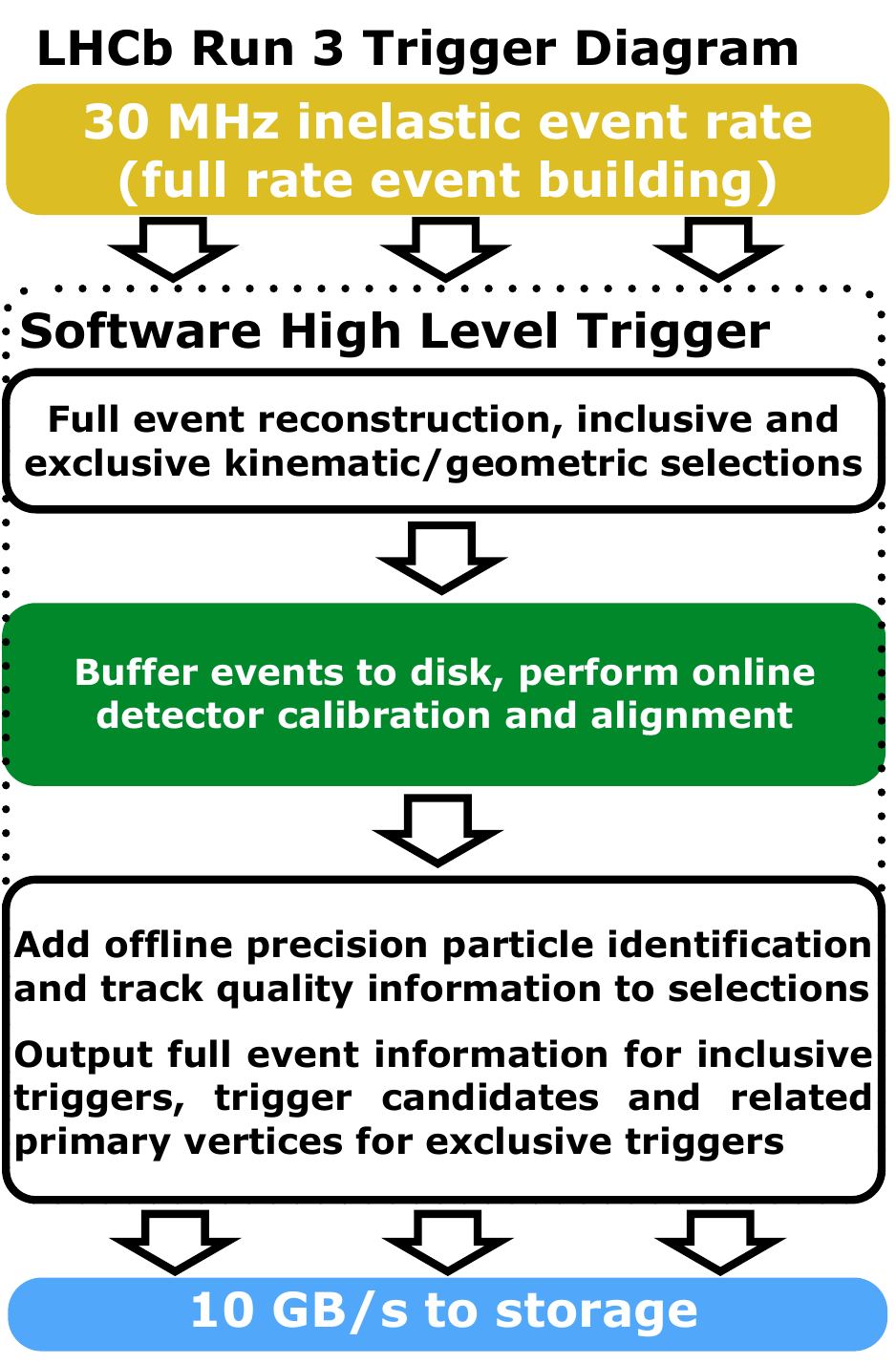}
  \caption{Evolution of LHCb trigger schemes in Run 1, 2 and 3.
    Figures from Ref.~\cite{LHCb-FIGURE-2020-016}.}\label{Fig:Trigger}
\end{figure}
\section{LHCb status and prospects}\label{Sec:U1}
The LHCb experiment has just undergone a major upgrade~\cite{LHCb-DP-2022-002}. The
detector layout is hardly changed --- actually the image in Fig.~\ref{Fig:LHCb} is
that of the new detector --- but many components have been changed.
The goal of the upgrade was to allow for an increased instantaneous luminosity,
in the $10^{33}\lumiunit$ range. In order to achieve that, the hardware trigger
needed to be removed: Meeting a 1\mhz bottleneck would require \pt thresholds of several
\gev, which starts to remove as much \B signal as backgrounds. This in turn
required to change all readout electronics of the detectors not included in
the hardware trigger, namely the vertex detector, the tracking system, and the RICH.
The silicon strip vertex detector was replaced by a pixel detector and the
trackers consisting of silicon strips near the beam and straw tubes elsewhere
were replaced by scintillating fibres.

All this detector is read out at 40\mhz, of which 30\mhz are non-empty events.
These data are sent to a first trigger farm of graphical processing units.
The selected events
are then buffered on disk while a calibration and alignment procedure is run.
Only when all calibration constants are available are the data processed
by the second trigger level. The final selected data thus have the full
offline-level quality and no further reconstruction is required.
Most of the events are saved partially, only keeping the objects of interest
for the analysis.
The evolution of the trigger scheme is shown in Fig.~\ref{Fig:Trigger}.

\begin{figure}[tb]\centering
  \includegraphics[width=0.9\textwidth]{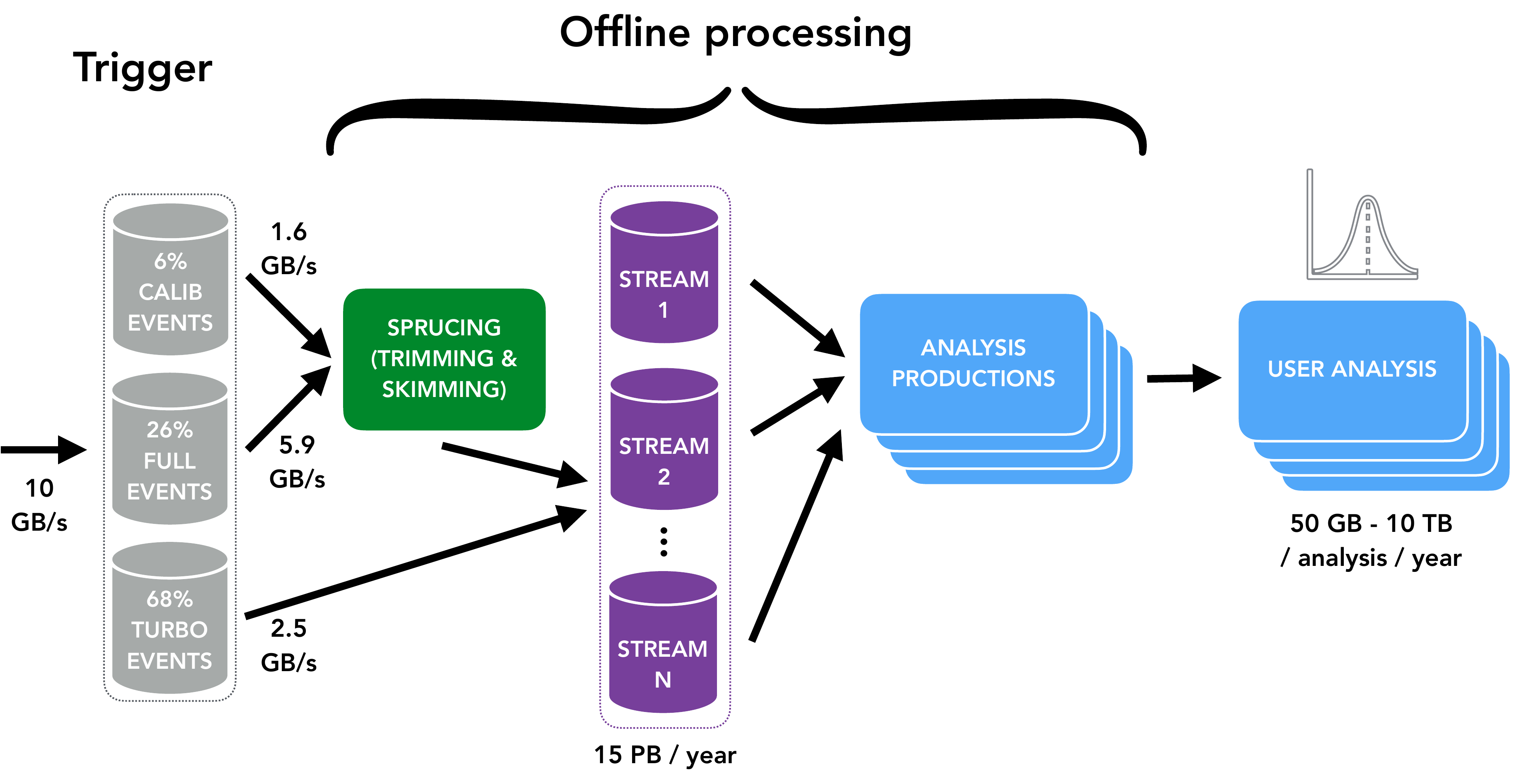}
  \caption{LHCb offline dataflow. Figure from Ref.~\cite{LHCb-FIGURE-2020-016}}\label{Fig:DPA}
\end{figure}

The data stored by the trigger are massaged by a ``sprucing'' process,
which may for instance add neighbouring tracks to a selected \B candidate to form
a potential excited state to be used in spectroscopy studies.
These neighbouring tracks have to be duly requested by the
relevant trigger selection as they may otherwise be lost.
The spruced candidates are stored in data streams that
are analysed by centrally managed user analysis productions,
as shown in Fig.~\ref{Fig:DPA}.

The detector is now fully installed and commissioning is ongoing.
There was a recent set-back as the vacuum inside the vertex detector
exceeded the specifications due to a faulty sensor. The pressure difference
with the vacuum of the LHC deformed the thin RF foil that separates the two volumes.
Luckily no sensors were affected, but the vertex detector cannot fully close
to its nominal position until the foil is replaced during the 2023--24 winter shutdown.
In the meantime the experiment can run with a partially open vertex detector
and a consequently degraded resolution. Here degraded means a resolution
similar to that of the decommissioned Run 1--2 vertex detector. The main
issue is rather the lack of simulation perfectly matching this sub-optimal
situation and thus determining the geometrical acceptance of the partially open detector.
It should all be recovered in 2024.

\begin{table}[tb]\centering
  \caption{Expected sensitivities for selected observables.
    Table adapted from Refs.~\cite{LHCb-PII-Physics,LHCbCollaboration:2806113}.}\label{Tab:U2}
  \ifdefined\talk
\else
\def\talk{0}
\fi
\ifdefined\tmpcite
\else
\newcommand{\tmpcite}[1]{
  \IF{\talk}{1}{\LHCbShortCitation{#1}}{\cite{#1}}}
\fi
\begin{tabular}{L|Rl|C|C|LLLL}
                 & \multicolumn{2}{c|}{\text{Legacy}} & 2026 & \text{U2} \\ 
  \text{Observable} & \multicolumn{2}{c|}{(9\invfb)} & (23\invfb) & (300\invfb) \\ 
  \hline
\sin 2 \ckmbeta\text{, with \decay{\Bd}{\jpsi\KS}} & 0.015 & \tmpcite{LHCb-PAPER-2023-013} & 0.011  & 0.003 \\
\phis\text{, with \decay{\Bs}{\jpsi\Kp\Km}}\ [\!\mrad] & 23 & \tmpcite{LHCb-PAPER-2014-059}  & 14  & 4  \\
\phi_s^{s{\bar{s}s}}\text{, with \decay{\Bs}{\phi\phi}}\ [\!\mrad]  & 80 & \tmpcite{LHCb-PAPER-2023-001} & 39 & 11  \\
\ckmgamma & 4^\circ & \tmpcite{LHCb-CONF-2022-003} & 1.5^\circ  & 0.35^\circ \\
|\Vub|/|\Vcb|  & 6\% & \tmpcite{LHCb-PAPER-2015-013}  & 3\%    &  1\%  \\
\hline
{\cal R}_\mumu & 90\% & \tmpcite{LHCb-PAPER-2021-007}  & 34\%  &  10\%  \\
R_K\ (1<q^2<6\gevgevcccc)            & 0.1 & \tmpcite{LHCb-PAPER-2022-045}  &  0.025    &   0.007 \\
R_{K^\ast}\ (1<q^2<6\gevgevcccc)  & 0.1 & \tmpcite{LHCb-PAPER-2022-045}  &  0.031   &   0.008 \\
\hline
R(D^\ast) & 0.022 & \tmpcite{LHCb-PAPER-2022-052}  & 0.0072  & 0.002 \\
R(J/\psi) & 0.24 & \tmpcite{LHCb-PAPER-2017-035} & 0.071 & 0.02  \\
\hline
\Delta A_{\CP}(KK-\pi\pi)\ [10^{-5}] &  85 & \tmpcite{LHCb-PAPER-2015-055}  & 17 &  3.0 \\
\end{tabular}

\end{table}
\section{Future prospects}\label{sec:Future}
After Runs 3 and 4, LHCb plans to upgrade again the detector in order to keep up
with the requirements of the High Luminosity LHC. A luminosity
in the vicinity of $10^{34}\lumiunit$ will however generate several tens of
$pp$ collisions per bunch crossings. Fishing out \bquark and \cquark hadrons
from such busy events will required 4D tracking including timing~\cite{LHCb-TDR-023}.
With such a detector one could achieve unprecedented sensitivities,
some of which are listed in Table~\ref{Tab:U2}.

\begin{figure}[b]\centering
  \includegraphics[height=0.23\textheight]{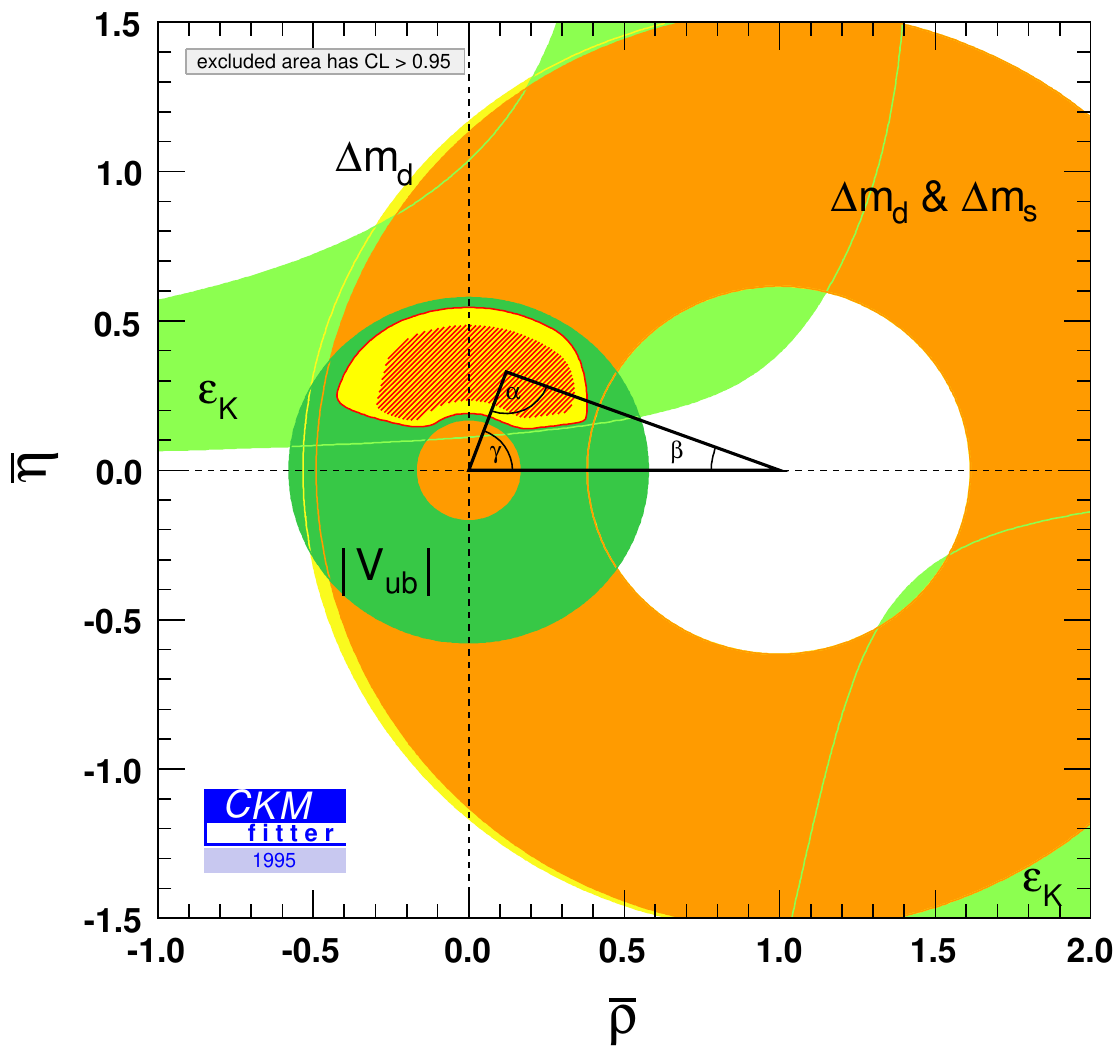}%
  \includegraphics[height=0.25\textheight]{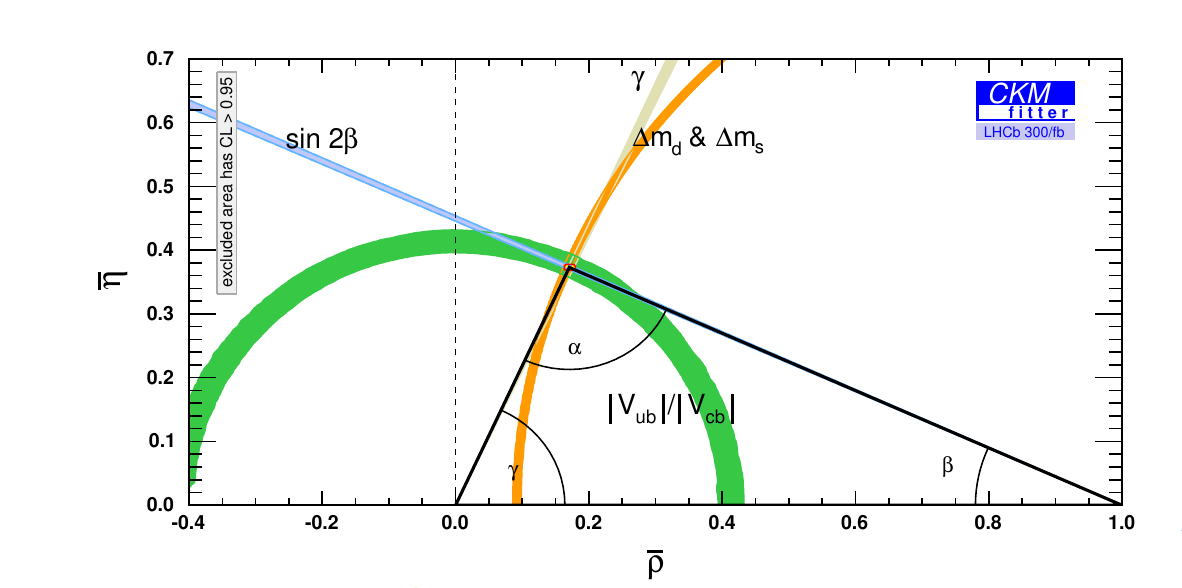}
  \caption{(left) The unitarity triangle in 1995~\cite{CKMfitter2005}.
    (right) Possible status after Run 5 of the LHC~\cite{LHCb-PII-Physics}.}\label{Fig:UT300}\label{Fig:UT95}
\end{figure}

\section{Conclusion}
Such fantastic accuracies will require enormous efforts on
understanding the detector, the backgrounds, the simulation.
The prize will be an extremely precise knowledge of the flavour sector,
including of course the \CKM matrix at the origin of this meeting.
Figures~\ref{Fig:UT95} (left) and~\ref{Fig:UT}
show the progress between the early
dates of \CKM metrology and now,
and Fig.~\ref{Fig:UT300} (right)
shows how much there is still to do.

Precision metrology should not be dismissed. It is about the best possible
understanding of Nature and the best possible exploitation of the machines
we have built. It is a prerequisite to discovery of new physics.
Who knows if at this level of precision the triangle will still close?
Either way, it is a journey worth taking.

\section*{Acknowledgements}
I would like to thank Hashimoto-san for the kind invitation to the workshop. It
was a pleasure to visit KEK after 20 years of absence.
Many thanks for Carla G\"obel for commenting on the manuscript.

\addcontentsline{toc}{section}{References}
\bibliographystyle{LHCb}
\bibliography{exp,theory,standard,LHCb-PAPER,LHCb-CONF,LHCb-DP,LHCb-TDR,lhcbnotes}

\ifx\mcitethebibliography\mciteundefinedmacro
\PackageError{LHCb.bst}{mciteplus.sty has not been loaded}
{This bibstyle requires the use of the mciteplus package.}\fi
\providecommand{\href}[2]{#2}
\begin{mcitethebibliography}{100}
\mciteSetBstSublistMode{n}
\mciteSetBstMaxWidthForm{subitem}{\alph{mcitesubitemcount})}
\mciteSetBstSublistLabelBeginEnd{\mcitemaxwidthsubitemform\space}
{\relax}{\relax}

\bibitem{Kobayashi:1973fv}
M.~Kobayashi and T.~Maskawa,
  \ifthenelse{\boolean{articletitles}}{\emph{{\CP-violation in the
  renormalizable theory of weak interaction}},
  }{}\href{https://doi.org/10.1143/PTP.49.652}{Prog.\ Theor.\ Phys.\
  \textbf{49} (1973) 652}\relax
\mciteBstWouldAddEndPuncttrue
\mciteSetBstMidEndSepPunct{\mcitedefaultmidpunct}
{\mcitedefaultendpunct}{\mcitedefaultseppunct}\relax
\EndOfBibitem
\bibitem{Christenson:1964fg}
J.~H. Christenson, J.~W. Cronin, V.~L. Fitch, and R.~Turlay,
  \ifthenelse{\boolean{articletitles}}{\emph{{Evidence for the $2\pion$ decay
  of the $K_2^0$ meson}},
  }{}\href{https://doi.org/10.1103/PhysRevLett.13.138}{Phys.\ Rev.\ Lett.\
  \textbf{13} (1964) 138}\relax
\mciteBstWouldAddEndPuncttrue
\mciteSetBstMidEndSepPunct{\mcitedefaultmidpunct}
{\mcitedefaultendpunct}{\mcitedefaultseppunct}\relax
\EndOfBibitem
\bibitem{Wu:1957}
C.~S. Wu {\em et~al.}, \ifthenelse{\boolean{articletitles}}{\emph{{Experimental
  test of parity conservation in beta decay}},
  }{}\href{https://doi.org/10.1103/PhysRev.105.1413}{Phys.\ Rev.\  \textbf{105}
  (1957) 1413}\relax
\mciteBstWouldAddEndPuncttrue
\mciteSetBstMidEndSepPunct{\mcitedefaultmidpunct}
{\mcitedefaultendpunct}{\mcitedefaultseppunct}\relax
\EndOfBibitem
\bibitem{Aubert:2001nu}
BaBar collaboration, B.~Aubert {\em et~al.},
  \ifthenelse{\boolean{articletitles}}{\emph{{Observation of \CP violation in
  the $B^0$ meson system}},
  }{}\href{https://doi.org/10.1103/PhysRevLett.87.091801}{Phys.\ Rev.\ Lett.\
  \textbf{87} (2001) 091801},
  \href{http://arxiv.org/abs/hep-ex/0107013}{{\normalfont\ttfamily
  arXiv:hep-ex/0107013}}\relax
\mciteBstWouldAddEndPuncttrue
\mciteSetBstMidEndSepPunct{\mcitedefaultmidpunct}
{\mcitedefaultendpunct}{\mcitedefaultseppunct}\relax
\EndOfBibitem
\bibitem{Abe:2001xe}
Belle collaboration, K.~Abe {\em et~al.},
  \ifthenelse{\boolean{articletitles}}{\emph{{Observation of large \CP
  violation in the neutral $B$ meson system}},
  }{}\href{https://doi.org/10.1103/PhysRevLett.87.091802}{Phys.\ Rev.\ Lett.\
  \textbf{87} (2001) 091802},
  \href{http://arxiv.org/abs/hep-ex/0107061}{{\normalfont\ttfamily
  arXiv:hep-ex/0107061}}\relax
\mciteBstWouldAddEndPuncttrue
\mciteSetBstMidEndSepPunct{\mcitedefaultmidpunct}
{\mcitedefaultendpunct}{\mcitedefaultseppunct}\relax
\EndOfBibitem
\bibitem{LHCb-PAPER-2013-018}
LHCb collaboration, R.~Aaij {\em et~al.},
  \ifthenelse{\boolean{articletitles}}{\emph{{First observation of \CP
  violation in the decays of \Bs mesons}},
  }{}\href{https://doi.org/10.1103/PhysRevLett.110.221601}{Phys.\ Rev.\ Lett.\
  \textbf{110} (2013) 221601},
  \href{http://arxiv.org/abs/1304.6173}{{\normalfont\ttfamily
  arXiv:1304.6173}}\relax
\mciteBstWouldAddEndPuncttrue
\mciteSetBstMidEndSepPunct{\mcitedefaultmidpunct}
{\mcitedefaultendpunct}{\mcitedefaultseppunct}\relax
\EndOfBibitem
\bibitem{LHCb-PAPER-2019-006}
LHCb collaboration, R.~Aaij {\em et~al.},
  \ifthenelse{\boolean{articletitles}}{\emph{{Observation of \CP violation in
  charm decays}},
  }{}\href{https://doi.org/10.1103/PhysRevLett.122.211803}{Phys.\ Rev.\ Lett.\
  \textbf{122} (2019) 211803},
  \href{http://arxiv.org/abs/1903.08726}{{\normalfont\ttfamily
  arXiv:1903.08726}}\relax
\mciteBstWouldAddEndPuncttrue
\mciteSetBstMidEndSepPunct{\mcitedefaultmidpunct}
{\mcitedefaultendpunct}{\mcitedefaultseppunct}\relax
\EndOfBibitem
\bibitem{KTeV:1999kad}
KTeV collaboration, A.~Alavi-Harati {\em et~al.},
  \ifthenelse{\boolean{articletitles}}{\emph{{Observation of direct CP
  violation in $K_{S,L} \to \pi \pi$ decays}},
  }{}\href{https://doi.org/10.1103/PhysRevLett.83.22}{Phys.\ Rev.\ Lett.\
  \textbf{83} (1999) 22},
  \href{http://arxiv.org/abs/hep-ex/9905060}{{\normalfont\ttfamily
  arXiv:hep-ex/9905060}}\relax
\mciteBstWouldAddEndPuncttrue
\mciteSetBstMidEndSepPunct{\mcitedefaultmidpunct}
{\mcitedefaultendpunct}{\mcitedefaultseppunct}\relax
\EndOfBibitem
\bibitem{NA48:2002tmj}
NA48 collaboration, J.~R. Batley {\em et~al.},
  \ifthenelse{\boolean{articletitles}}{\emph{{A Precision measurement of direct
  \CP violation in the decay of neutral kaons into two pions}},
  }{}\href{https://doi.org/10.1016/S0370-2693(02)02476-0}{Phys.\ Lett.\
  \textbf{B544} (2002) 97},
  \href{http://arxiv.org/abs/hep-ex/0208009}{{\normalfont\ttfamily
  arXiv:hep-ex/0208009}}\relax
\mciteBstWouldAddEndPuncttrue
\mciteSetBstMidEndSepPunct{\mcitedefaultmidpunct}
{\mcitedefaultendpunct}{\mcitedefaultseppunct}\relax
\EndOfBibitem
\bibitem{BaBar:2004gyj}
BaBar collaboration, B.~Aubert {\em et~al.},
  \ifthenelse{\boolean{articletitles}}{\emph{{Observation of direct CP
  violation in $B^0 \to K^+ \pi^-$ decays}},
  }{}\href{https://doi.org/10.1103/PhysRevLett.93.131801}{Phys.\ Rev.\ Lett.\
  \textbf{93} (2004) 131801},
  \href{http://arxiv.org/abs/hep-ex/0407057}{{\normalfont\ttfamily
  arXiv:hep-ex/0407057}}\relax
\mciteBstWouldAddEndPuncttrue
\mciteSetBstMidEndSepPunct{\mcitedefaultmidpunct}
{\mcitedefaultendpunct}{\mcitedefaultseppunct}\relax
\EndOfBibitem
\bibitem{Chao:2004mn}
Belle collaboration, Y.~Chao {\em et~al.},
  \ifthenelse{\boolean{articletitles}}{\emph{{Evidence for direct \CP violation
  in $\Bz\to \Kp \pim$ decays}}, }{}Phys.\ Rev.\ Lett.\  \textbf{93} (2004)
  191802, \href{http://arxiv.org/abs/hep-ex/0408100}{{\normalfont\ttfamily
  arXiv:hep-ex/0408100}}\relax
\mciteBstWouldAddEndPuncttrue
\mciteSetBstMidEndSepPunct{\mcitedefaultmidpunct}
{\mcitedefaultendpunct}{\mcitedefaultseppunct}\relax
\EndOfBibitem
\bibitem{LHCb-PAPER-2020-029}
LHCb collaboration, R.~Aaij {\em et~al.},
  \ifthenelse{\boolean{articletitles}}{\emph{{Observation of \CP violation in
  two-body \mbox{$\BdorBs$-meson} decays to charged pions and kaons}},
  }{}\href{https://doi.org/10.1007/JHEP03(2021)075}{JHEP \textbf{03} (2021)
  075}, \href{http://arxiv.org/abs/2012.05319}{{\normalfont\ttfamily
  arXiv:2012.05319}}\relax
\mciteBstWouldAddEndPuncttrue
\mciteSetBstMidEndSepPunct{\mcitedefaultmidpunct}
{\mcitedefaultendpunct}{\mcitedefaultseppunct}\relax
\EndOfBibitem
\bibitem{LHCb-PAPER-2012-001}
LHCb collaboration, R.~Aaij {\em et~al.},
  \ifthenelse{\boolean{articletitles}}{\emph{{Observation of \CP violation in
  \mbox{\decay{\Bpm}{\D\Kpm}} decays}},
  }{}\href{https://doi.org/10.1016/j.physletb.2012.04.060}{Phys.\ Lett.\
  \textbf{B712} (2012) 203}, Erratum
  \href{https://doi.org/10.1016/j.physletb.2012.05.060}{ibid.\   \textbf{B713}
  (2012) 351}, \href{http://arxiv.org/abs/1203.3662}{{\normalfont\ttfamily
  arXiv:1203.3662}}\relax
\mciteBstWouldAddEndPuncttrue
\mciteSetBstMidEndSepPunct{\mcitedefaultmidpunct}
{\mcitedefaultendpunct}{\mcitedefaultseppunct}\relax
\EndOfBibitem
\bibitem{LHCb-DP-2022-002}
LHCb collaboration, R.~Aaij {\em et~al.},
  \ifthenelse{\boolean{articletitles}}{\emph{{The LHCb Upgrade I}},
  }{}\href{http://arxiv.org/abs/2305.10515}{{\normalfont\ttfamily
  arXiv:2305.10515}}, {to appear in JINST}\relax
\mciteBstWouldAddEndPuncttrue
\mciteSetBstMidEndSepPunct{\mcitedefaultmidpunct}
{\mcitedefaultendpunct}{\mcitedefaultseppunct}\relax
\EndOfBibitem
\bibitem{Schmidtler:1991tv}
M.~Schmidtler and K.~R. Schubert,
  \ifthenelse{\boolean{articletitles}}{\emph{{Experimental constraints on the
  phase in the Cabibbo-Kobayashi-Maskawa matrix}},
  }{}\href{https://doi.org/10.1007/BF01597574}{Z.\ Phys.\  \textbf{C53} (1992)
  347}\relax
\mciteBstWouldAddEndPuncttrue
\mciteSetBstMidEndSepPunct{\mcitedefaultmidpunct}
{\mcitedefaultendpunct}{\mcitedefaultseppunct}\relax
\EndOfBibitem
\bibitem{Anonymous}
{Anonymous LHCb physicist}. circa 1998\relax
\mciteBstWouldAddEndPuncttrue
\mciteSetBstMidEndSepPunct{\mcitedefaultmidpunct}
{\mcitedefaultendpunct}{\mcitedefaultseppunct}\relax
\EndOfBibitem
\bibitem{Abulencia:2006ze}
CDF collaboration, A.~Abulencia {\em et~al.},
  \ifthenelse{\boolean{articletitles}}{\emph{{Observation of \Bs--\Bsb
  oscillations}},
  }{}\href{https://doi.org/10.1103/PhysRevLett.97.242003}{Phys.\ Rev.\ Lett.\
  \textbf{97} (2006) 242003},
  \href{http://arxiv.org/abs/hep-ex/0609040}{{\normalfont\ttfamily
  arXiv:hep-ex/0609040}}\relax
\mciteBstWouldAddEndPuncttrue
\mciteSetBstMidEndSepPunct{\mcitedefaultmidpunct}
{\mcitedefaultendpunct}{\mcitedefaultseppunct}\relax
\EndOfBibitem
\bibitem{LHCb-DP-2008-001}
LHCb collaboration, A.~A. Alves~Jr.\ {\em et~al.},
  \ifthenelse{\boolean{articletitles}}{\emph{{The \lhcb detector at the LHC}},
  }{}\href{https://doi.org/10.1088/1748-0221/3/08/S08005}{JINST \textbf{3}
  (2008) S08005}\relax
\mciteBstWouldAddEndPuncttrue
\mciteSetBstMidEndSepPunct{\mcitedefaultmidpunct}
{\mcitedefaultendpunct}{\mcitedefaultseppunct}\relax
\EndOfBibitem
\bibitem{LHCb-DP-2014-002}
LHCb collaboration, R.~Aaij {\em et~al.},
  \ifthenelse{\boolean{articletitles}}{\emph{{LHCb detector performance}},
  }{}\href{https://doi.org/10.1142/S0217751X15300227}{Int.\ J.\ Mod.\ Phys.\
  \textbf{A30} (2015) 1530022},
  \href{http://arxiv.org/abs/1412.6352}{{\normalfont\ttfamily
  arXiv:1412.6352}}\relax
\mciteBstWouldAddEndPuncttrue
\mciteSetBstMidEndSepPunct{\mcitedefaultmidpunct}
{\mcitedefaultendpunct}{\mcitedefaultseppunct}\relax
\EndOfBibitem
\bibitem{LHCC2010}
P.~Koppenburg, \ifthenelse{\boolean{articletitles}}{\emph{{LHCb status
  report}}, }{}
  \href{https://indico.cern.ch/event/92525/#2-lhcb-status-report}{LHCC open
  session}\relax
\mciteBstWouldAddEndPuncttrue
\mciteSetBstMidEndSepPunct{\mcitedefaultmidpunct}
{\mcitedefaultendpunct}{\mcitedefaultseppunct}\relax
\EndOfBibitem
\bibitem{LHCb-PAPER-2020-046}
LHCb collaboration, R.~Aaij {\em et~al.},
  \ifthenelse{\boolean{articletitles}}{\emph{{Precise measurement of the
  $f_s/f_d$ ratio of fragmentation fractions and of $B^0_s$ decay branching
  fractions}}, }{}\href{https://doi.org/10.1103/PhysRevD.104.032005}{Phys.\
  Rev.\  \textbf{D104} (2021) 032005},
  \href{http://arxiv.org/abs/2103.06810}{{\normalfont\ttfamily
  arXiv:2103.06810}}\relax
\mciteBstWouldAddEndPuncttrue
\mciteSetBstMidEndSepPunct{\mcitedefaultmidpunct}
{\mcitedefaultendpunct}{\mcitedefaultseppunct}\relax
\EndOfBibitem
\bibitem{CKMfitter2005}
CKMfitter group, J.~Charles {\em et~al.},
  \ifthenelse{\boolean{articletitles}}{\emph{{\CP violation and the CKM matrix:
  Assessing the impact of the asymmetric $B$ factories}},
  }{}\href{https://doi.org/10.1140/epjc/s2005-02169-1}{Eur.\ Phys.\ J.\
  \textbf{C41} (2005) 1},
  \href{http://arxiv.org/abs/hep-ph/0406184}{{\normalfont\ttfamily
  arXiv:hep-ph/0406184}}, {updated results and plots available at
  \href{http://ckmfitter.in2p3.fr/}{{\texttt{http://ckmfitter.in2p3.fr/}}}}\relax
\mciteBstWouldAddEndPuncttrue
\mciteSetBstMidEndSepPunct{\mcitedefaultmidpunct}
{\mcitedefaultendpunct}{\mcitedefaultseppunct}\relax
\EndOfBibitem
\bibitem{LHCb-PAPER-2021-005}
LHCb collaboration, R.~Aaij {\em et~al.},
  \ifthenelse{\boolean{articletitles}}{\emph{{Precise determination of the
  $\Bs$-$\Bsb$ oscillation frequency}},
  }{}\href{https://doi.org/10.1038/s41567-021-01394-x}{Nature Physics
  \textbf{18} (2022) 1},
  \href{http://arxiv.org/abs/2104.04421}{{\normalfont\ttfamily
  arXiv:2104.04421}}\relax
\mciteBstWouldAddEndPuncttrue
\mciteSetBstMidEndSepPunct{\mcitedefaultmidpunct}
{\mcitedefaultendpunct}{\mcitedefaultseppunct}\relax
\EndOfBibitem
\bibitem{LHCb-PAPER-2015-031}
LHCb collaboration, R.~Aaij {\em et~al.},
  \ifthenelse{\boolean{articletitles}}{\emph{{A precise measurement of the \Bz
  meson oscillation frequency}},
  }{}\href{https://doi.org/10.1140/epjc/s10052-016-4250-2}{Eur.\ Phys.\ J.\
  \textbf{C76} (2016) 412},
  \href{http://arxiv.org/abs/1604.03475}{{\normalfont\ttfamily
  arXiv:1604.03475}}\relax
\mciteBstWouldAddEndPuncttrue
\mciteSetBstMidEndSepPunct{\mcitedefaultmidpunct}
{\mcitedefaultendpunct}{\mcitedefaultseppunct}\relax
\EndOfBibitem
\bibitem{LHCb-PAPER-2020-038}
LHCb collaboration, R.~Aaij {\em et~al.},
  \ifthenelse{\boolean{articletitles}}{\emph{{First observation of the decay
  $B^0_s \to K^- \mu^+ \nu_{\mu}$ and measurement of $|V_{ub}| / |V_{cb}| $}},
  }{}\href{https://doi.org/10.1103/PhysRevLett.126.081804}{Phys.\ Rev.\ Lett.\
  \textbf{126} (2021) 081804},
  \href{http://arxiv.org/abs/2012.05143}{{\normalfont\ttfamily
  arXiv:2012.05143}}\relax
\mciteBstWouldAddEndPuncttrue
\mciteSetBstMidEndSepPunct{\mcitedefaultmidpunct}
{\mcitedefaultendpunct}{\mcitedefaultseppunct}\relax
\EndOfBibitem
\bibitem{Khodjamirian:2017fxg}
A.~Khodjamirian and A.~V. Rusov,
  \ifthenelse{\boolean{articletitles}}{\emph{{$B_{s}\to K \ell \nu_\ell$ and
  $B_{(s)} \to \pi (K) \ell^+\ell^-$ decays at large recoil and CKM matrix
  elements}}, }{}\href{https://doi.org/10.1007/JHEP08(2017)112}{JHEP
  \textbf{08} (2017) 112},
  \href{http://arxiv.org/abs/1703.04765}{{\normalfont\ttfamily
  arXiv:1703.04765}}\relax
\mciteBstWouldAddEndPuncttrue
\mciteSetBstMidEndSepPunct{\mcitedefaultmidpunct}
{\mcitedefaultendpunct}{\mcitedefaultseppunct}\relax
\EndOfBibitem
\bibitem{Bazavov:2019aom}
Fermilab Lattice, MILC collaborations, A.~Bazavov {\em et~al.},
  \ifthenelse{\boolean{articletitles}}{\emph{{$B_s\to K\ell\nu$ decay from
  lattice QCD}}, }{}\href{https://doi.org/10.1103/PhysRevD.100.034501}{Phys.\
  Rev.\  \textbf{D100} (2019) 034501},
  \href{http://arxiv.org/abs/1901.02561}{{\normalfont\ttfamily
  arXiv:1901.02561}}\relax
\mciteBstWouldAddEndPuncttrue
\mciteSetBstMidEndSepPunct{\mcitedefaultmidpunct}
{\mcitedefaultendpunct}{\mcitedefaultseppunct}\relax
\EndOfBibitem
\bibitem{Bigi:1983cj}
I.~I.~Y. Bigi and A.~I. Sanda, \ifthenelse{\boolean{articletitles}}{\emph{{On
  \Bz--\Bzb mixing and violations of \CP symmetry}},
  }{}\href{https://doi.org/10.1103/PhysRevD.29.1393}{Phys.\ Rev.\  \textbf{D29}
  (1984) 1393}\relax
\mciteBstWouldAddEndPuncttrue
\mciteSetBstMidEndSepPunct{\mcitedefaultmidpunct}
{\mcitedefaultendpunct}{\mcitedefaultseppunct}\relax
\EndOfBibitem
\bibitem{LHCb-PAPER-2023-013}
LHCb collaboration, R.~Aaij {\em et~al.},
  \ifthenelse{\boolean{articletitles}}{\emph{{Measurement of $\CP$ violation in
  \mbox{$\Bz \rightarrow \psires (\to \ellell) \KS(\to \pip \pim)$} decays}},
  }{}\href{http://arxiv.org/abs/2309.09728}{{\normalfont\ttfamily
  arXiv:2309.09728}}, {submitted to Phys.~Rev.~Lett.}\relax
\mciteBstWouldAddEndPunctfalse
\mciteSetBstMidEndSepPunct{\mcitedefaultmidpunct}
{}{\mcitedefaultseppunct}\relax
\EndOfBibitem
\bibitem{Gligorov:2023mji}
V.~V. Gligorov, \ifthenelse{\boolean{articletitles}}{\emph{{Quark flavour
  physics: status and future prospects}},
  }{}\href{http://arxiv.org/abs/2306.12728}{{\normalfont\ttfamily
  arXiv:2306.12728}}\relax
\mciteBstWouldAddEndPuncttrue
\mciteSetBstMidEndSepPunct{\mcitedefaultmidpunct}
{\mcitedefaultendpunct}{\mcitedefaultseppunct}\relax
\EndOfBibitem
\bibitem{Brod:2013sga}
J.~Brod and J.~Zupan, \ifthenelse{\boolean{articletitles}}{\emph{{The ultimate
  theoretical error on $\gamma$ from $B \to DK$ decays}},
  }{}\href{https://doi.org/10.1007/JHEP01(2014)051}{JHEP \textbf{01} (2014)
  051}, \href{http://arxiv.org/abs/1308.5663}{{\normalfont\ttfamily
  arXiv:1308.5663}}\relax
\mciteBstWouldAddEndPuncttrue
\mciteSetBstMidEndSepPunct{\mcitedefaultmidpunct}
{\mcitedefaultendpunct}{\mcitedefaultseppunct}\relax
\EndOfBibitem
\bibitem{LHCb-CONF-2022-003}
{LHCb collaboration}, \ifthenelse{\boolean{articletitles}}{\emph{{Simultaneous
  determination of the CKM angle $\gamma$ and parameters related to mixing and
  CP violation in the charm sector}}, }{}
  \href{http://cdsweb.cern.ch/search?p=LHCb-CONF-2022-003&f=reportnumber&action_search=Search&c=LHCb+Conference+Contributions}
  {LHCb-CONF-2022-003}, {2022}\relax
\mciteBstWouldAddEndPuncttrue
\mciteSetBstMidEndSepPunct{\mcitedefaultmidpunct}
{\mcitedefaultendpunct}{\mcitedefaultseppunct}\relax
\EndOfBibitem
\bibitem{LHCb-PAPER-2020-036}
LHCb collaboration, R.~Aaij {\em et~al.},
  \ifthenelse{\boolean{articletitles}}{\emph{{Measurement of \CP observables in
  $B^\pm \to D^{(*)} K^{\pm}$ and $B^\pm \to D^{(*)} \pi^{\pm} $ decays using
  two-body $D$ final states}},
  }{}\href{https://doi.org/10.1007/JHEP04(2021)081}{JHEP \textbf{04} (2021)
  081}, \href{http://arxiv.org/abs/2012.09903}{{\normalfont\ttfamily
  arXiv:2012.09903}}\relax
\mciteBstWouldAddEndPuncttrue
\mciteSetBstMidEndSepPunct{\mcitedefaultmidpunct}
{\mcitedefaultendpunct}{\mcitedefaultseppunct}\relax
\EndOfBibitem
\bibitem{LHCb-PAPER-2017-030}
LHCb collaboration, R.~Aaij {\em et~al.},
  \ifthenelse{\boolean{articletitles}}{\emph{{Measurement of \CP observables in
  \mbox{\decay{\Bpm}{DK^{\ast \pm}}} decays using two- and four-body $D$-meson
  final states}}, }{}\href{https://doi.org/10.1007/JHEP11(2017)156}{JHEP
  \textbf{11} (2017) 156}, Erratum
  \href{https://doi.org/10.1007/JHEP05(2018)067}{ibid.\   \textbf{05} (2018)
  067}, \href{http://arxiv.org/abs/1709.05855}{{\normalfont\ttfamily
  arXiv:1709.05855}}\relax
\mciteBstWouldAddEndPuncttrue
\mciteSetBstMidEndSepPunct{\mcitedefaultmidpunct}
{\mcitedefaultendpunct}{\mcitedefaultseppunct}\relax
\EndOfBibitem
\bibitem{LHCb-PAPER-2019-021}
LHCb collaboration, R.~Aaij {\em et~al.},
  \ifthenelse{\boolean{articletitles}}{\emph{{Measurement of \CP observables in
  the process \mbox{\decay{\Bz}{\D\Kstarz}} with two- and four-body \D
  decays}}, }{}\href{https://doi.org/10.1007/JHEP08(2019)041}{JHEP \textbf{08}
  (2019) 041}, \href{http://arxiv.org/abs/1906.08297}{{\normalfont\ttfamily
  arXiv:1906.08297}}\relax
\mciteBstWouldAddEndPuncttrue
\mciteSetBstMidEndSepPunct{\mcitedefaultmidpunct}
{\mcitedefaultendpunct}{\mcitedefaultseppunct}\relax
\EndOfBibitem
\bibitem{LHCb-PAPER-2015-020}
LHCb collaboration, R.~Aaij {\em et~al.},
  \ifthenelse{\boolean{articletitles}}{\emph{{Study of
  \mbox{\decay{\Bm}{\D\Km\pip\pim}} and \mbox{\decay{\Bm}{\D\pim\pip\pim}}
  decays and determination of the CKM angle $\gamma$}},
  }{}\href{https://doi.org/10.1103/PhysRevD.92.112005}{Phys.\ Rev.\
  \textbf{D92} (2015) 112005},
  \href{http://arxiv.org/abs/1505.07044}{{\normalfont\ttfamily
  arXiv:1505.07044}}\relax
\mciteBstWouldAddEndPuncttrue
\mciteSetBstMidEndSepPunct{\mcitedefaultmidpunct}
{\mcitedefaultendpunct}{\mcitedefaultseppunct}\relax
\EndOfBibitem
\bibitem{LHCb-PAPER-2016-003}
LHCb collaboration, R.~Aaij {\em et~al.},
  \ifthenelse{\boolean{articletitles}}{\emph{{Measurement of \CP observables in
  \mbox{\decay{\Bpm}{\D \Kpm}} and \mbox{\decay{\Bpm}{\D\pipm}} with two- and
  four-body \D decays}},
  }{}\href{https://doi.org/10.1016/j.physletb.2016.06.022}{Phys.\ Lett.\
  \textbf{B760} (2016) 117},
  \href{http://arxiv.org/abs/1603.08993}{{\normalfont\ttfamily
  arXiv:1603.08993}}\relax
\mciteBstWouldAddEndPuncttrue
\mciteSetBstMidEndSepPunct{\mcitedefaultmidpunct}
{\mcitedefaultendpunct}{\mcitedefaultseppunct}\relax
\EndOfBibitem
\bibitem{LHCb-PAPER-2022-017}
LHCb collaboration, R.~Aaij {\em et~al.},
  \ifthenelse{\boolean{articletitles}}{\emph{{Measurement of the CKM angle
  $\gamma$ with \mbox{$\Bmp \rightarrow D[\Kpm\pimp\pimp\pipm] h^{\mp}$ decays
  using a binned phase-space approach}}},
  }{}\href{https://doi.org/10.1007/JHEP07(2023)138}{JHEP \textbf{07} (2023)
  138}, \href{http://arxiv.org/abs/2209.03692}{{\normalfont\ttfamily
  arXiv:2209.03692}}\relax
\mciteBstWouldAddEndPuncttrue
\mciteSetBstMidEndSepPunct{\mcitedefaultmidpunct}
{\mcitedefaultendpunct}{\mcitedefaultseppunct}\relax
\EndOfBibitem
\bibitem{LHCb-PAPER-2022-037}
LHCb collaboration, R.~Aaij {\em et~al.},
  \ifthenelse{\boolean{articletitles}}{\emph{{A study of CP violation in the
  decays \mbox{$\Bpm \rightarrow [\Kp\Km\pip\pim]_D \hadron^{\pm}$}
  \mbox{($\hadron=\kaon,\pion$)} and \mbox{$\Bpm \rightarrow
  [\pip\pim\pip\pim]_D \hadron^{\pm}$}}},
  }{}\href{https://doi.org/10.1140/epjc/s10052-023-11560-5}{Eur.\ Phys.\ J.\
  \textbf{C84} (2023) 547},
  \href{http://arxiv.org/abs/2301.10328}{{\normalfont\ttfamily
  arXiv:2301.10328}}\relax
\mciteBstWouldAddEndPuncttrue
\mciteSetBstMidEndSepPunct{\mcitedefaultmidpunct}
{\mcitedefaultendpunct}{\mcitedefaultseppunct}\relax
\EndOfBibitem
\bibitem{LHCb-PAPER-2021-036}
LHCb collaboration, R.~Aaij {\em et~al.},
  \ifthenelse{\boolean{articletitles}}{\emph{{Constraints on the CKM angle \g
  from $\Bpm \to \D \hadron^\pm$ decays using $D \to \hadron^\pm
  \hadron^{\prime\mp} \piz$ final states}}, }{}JHEP \textbf{07} (2022) 099,
  \href{http://arxiv.org/abs/2112.10617}{{\normalfont\ttfamily
  arXiv:2112.10617}}\relax
\mciteBstWouldAddEndPuncttrue
\mciteSetBstMidEndSepPunct{\mcitedefaultmidpunct}
{\mcitedefaultendpunct}{\mcitedefaultseppunct}\relax
\EndOfBibitem
\bibitem{LHCb-PAPER-2020-019}
LHCb collaboration, R.~Aaij {\em et~al.},
  \ifthenelse{\boolean{articletitles}}{\emph{{Measurement of the CKM angle
  $\gamma$ in \mbox{$B^{\pm} \to D K^{\pm}$ and $B^{\pm} \to D \pi^{\pm}$}
  decays with $D \to \KS h^+h^-$}},
  }{}\href{https://doi.org/10.1007/JHEP02(2021)169}{JHEP \textbf{02} (2021)
  0169}, \href{http://arxiv.org/abs/2010.08483}{{\normalfont\ttfamily
  arXiv:2010.08483}}\relax
\mciteBstWouldAddEndPuncttrue
\mciteSetBstMidEndSepPunct{\mcitedefaultmidpunct}
{\mcitedefaultendpunct}{\mcitedefaultseppunct}\relax
\EndOfBibitem
\bibitem{LHCb-PAPER-2023-012}
LHCb collaboration, R.~Aaij {\em et~al.},
  \ifthenelse{\boolean{articletitles}}{\emph{{Measurement of the CKM angle
  $\gamma$ using the $\Bpm\to \Dstar \hadron^\pm$ channels}},
  }{}\href{http://arxiv.org/abs/2310.04277}{{\normalfont\ttfamily
  arXiv:2310.04277}}, {submitted to JHEP}\relax
\mciteBstWouldAddEndPuncttrue
\mciteSetBstMidEndSepPunct{\mcitedefaultmidpunct}
{\mcitedefaultendpunct}{\mcitedefaultseppunct}\relax
\EndOfBibitem
\bibitem{LHCb-PAPER-2023-009}
LHCb collaboration, R.~Aaij {\em et~al.},
  \ifthenelse{\boolean{articletitles}}{\emph{{Measurement of the CKM angle
  $\gamma$ in the \mbox{$\Bz \to \Dz \Kstarz$} channel using self-conjugate
  $\Dz \to \KS \hadron^+\hadron^-$ decays}},
  }{}\href{http://arxiv.org/abs/2309.05514}{{\normalfont\ttfamily
  arXiv:2309.05514}}, {submitted to Eur. Phys. J. C}\relax
\mciteBstWouldAddEndPuncttrue
\mciteSetBstMidEndSepPunct{\mcitedefaultmidpunct}
{\mcitedefaultendpunct}{\mcitedefaultseppunct}\relax
\EndOfBibitem
\bibitem{LHCb-PAPER-2019-044}
LHCb collaboration, R.~Aaij {\em et~al.},
  \ifthenelse{\boolean{articletitles}}{\emph{{Measurement of \CP observables in
  \mbox{\decay{\Bpm}{D\Kpm}} and \mbox{\decay{\Bpm}{D\pipm}} with
  \mbox{\decay{D}{\KS\Kpm\pimp}} decays}},
  }{}\href{https://doi.org/10.1007/JHEP06(2020)058}{JHEP \textbf{06} (2020)
  58}, \href{http://arxiv.org/abs/2002.08858}{{\normalfont\ttfamily
  arXiv:2002.08858}}\relax
\mciteBstWouldAddEndPuncttrue
\mciteSetBstMidEndSepPunct{\mcitedefaultmidpunct}
{\mcitedefaultendpunct}{\mcitedefaultseppunct}\relax
\EndOfBibitem
\bibitem{LHCb-PAPER-2018-009}
LHCb collaboration, R.~Aaij {\em et~al.},
  \ifthenelse{\boolean{articletitles}}{\emph{{Measurement of \CP violation in
  \mbox{\decay{\Bz}{D^\pm \pimp}} decays}},
  }{}\href{https://doi.org/10.1007/JHEP06(2018)084}{JHEP \textbf{06} (2018)
  084}, \href{http://arxiv.org/abs/1805.03448}{{\normalfont\ttfamily
  arXiv:1805.03448}}\relax
\mciteBstWouldAddEndPuncttrue
\mciteSetBstMidEndSepPunct{\mcitedefaultmidpunct}
{\mcitedefaultendpunct}{\mcitedefaultseppunct}\relax
\EndOfBibitem
\bibitem{LHCb-PAPER-2017-047}
LHCb collaboration, R.~Aaij {\em et~al.},
  \ifthenelse{\boolean{articletitles}}{\emph{{Measurement of \CP asymmetry in
  \mbox{\decay{\Bs}{D_s^\mp \Kpm}} decays}},
  }{}\href{https://doi.org/10.1007/JHEP03(2018)059}{JHEP \textbf{03} (2018)
  059}, \href{http://arxiv.org/abs/1712.07428}{{\normalfont\ttfamily
  arXiv:1712.07428}}\relax
\mciteBstWouldAddEndPuncttrue
\mciteSetBstMidEndSepPunct{\mcitedefaultmidpunct}
{\mcitedefaultendpunct}{\mcitedefaultseppunct}\relax
\EndOfBibitem
\bibitem{LHCb-CONF-2023-004}
{LHCb collaboration}, \ifthenelse{\boolean{articletitles}}{\emph{{Measurement
  of CP asymmetry in $\Bs \to \Dsmp \Kpm$ decays}}, }{}
  \href{http://cdsweb.cern.ch/search?p=LHCb-CONF-2023-004&f=reportnumber&action_search=Search&c=LHCb+Conference+Contributions}
  {LHCb-CONF-2023-004}, {2023}\relax
\mciteBstWouldAddEndPuncttrue
\mciteSetBstMidEndSepPunct{\mcitedefaultmidpunct}
{\mcitedefaultendpunct}{\mcitedefaultseppunct}\relax
\EndOfBibitem
\bibitem{LHCb-PAPER-2020-030}
LHCb collaboration, R.~Aaij {\em et~al.},
  \ifthenelse{\boolean{articletitles}}{\emph{{Measurement of the CKM angle
  $\gamma$ and \mbox{\Bs-\Bsb} mixing frequency with $\Bs \to \Dsmp h^\pm \pipm
  \pimp$ decays}}, }{}\href{https://doi.org/10.1007/JHEP03(2021)137}{JHEP
  \textbf{03} (2021) 137},
  \href{http://arxiv.org/abs/2011.12041}{{\normalfont\ttfamily
  arXiv:2011.12041}}\relax
\mciteBstWouldAddEndPuncttrue
\mciteSetBstMidEndSepPunct{\mcitedefaultmidpunct}
{\mcitedefaultendpunct}{\mcitedefaultseppunct}\relax
\EndOfBibitem
\bibitem{LHCb-PAPER-2022-020}
LHCb collaboration, R.~Aaij {\em et~al.},
  \ifthenelse{\boolean{articletitles}}{\emph{{Model-indepedent measurement of
  charm mixing parameters in $\Bbar \to \Dz (\to \Kstarz \pip \pim) \mun \neumb
  X$ decays}}, }{}\href{https://doi.org/10.1103/PhysRevD.108.052005}{Phys.\
  Rev.\  \textbf{D108} (2023) 052005},
  \href{http://arxiv.org/abs/2208.06512}{{\normalfont\ttfamily
  arXiv:2208.06512}}\relax
\mciteBstWouldAddEndPuncttrue
\mciteSetBstMidEndSepPunct{\mcitedefaultmidpunct}
{\mcitedefaultendpunct}{\mcitedefaultseppunct}\relax
\EndOfBibitem
\bibitem{LHCb-PAPER-2022-024}
LHCb collaboration, R.~Aaij {\em et~al.},
  \ifthenelse{\boolean{articletitles}}{\emph{{Measurement of the
  time-integrated \CP asymmetry in $\Dz \to \Km \Kp$ decays}},
  }{}\href{https://doi.org/10.1103/PhysRevLett.131.091802}{Phys.\ Rev.\ Lett.\
  \textbf{131} (2023) 091802},
  \href{http://arxiv.org/abs/2209.03179}{{\normalfont\ttfamily
  arXiv:2209.03179}}\relax
\mciteBstWouldAddEndPuncttrue
\mciteSetBstMidEndSepPunct{\mcitedefaultmidpunct}
{\mcitedefaultendpunct}{\mcitedefaultseppunct}\relax
\EndOfBibitem
\bibitem{Gronau:1990ka}
M.~Gronau and D.~London, \ifthenelse{\boolean{articletitles}}{\emph{{Isospin
  analysis of \CP asymmetries in \B decays}},
  }{}\href{https://doi.org/10.1103/PhysRevLett.65.3381}{Phys.\ Rev.\ Lett.\
  \textbf{65} (1990) 3381}\relax
\mciteBstWouldAddEndPuncttrue
\mciteSetBstMidEndSepPunct{\mcitedefaultmidpunct}
{\mcitedefaultendpunct}{\mcitedefaultseppunct}\relax
\EndOfBibitem
\bibitem{LHCb-CONF-2015-001}
{LHCb collaboration}, \ifthenelse{\boolean{articletitles}}{\emph{{Study of the
  decay $B^+\to K^+\pi^0$ at LHCb}}, }{}
  \href{http://cdsweb.cern.ch/search?p=LHCb-CONF-2015-001&f=reportnumber&action_search=Search&c=LHCb+Conference+Contributions}
  {LHCb-CONF-2015-001}, {2015}\relax
\mciteBstWouldAddEndPuncttrue
\mciteSetBstMidEndSepPunct{\mcitedefaultmidpunct}
{\mcitedefaultendpunct}{\mcitedefaultseppunct}\relax
\EndOfBibitem
\bibitem{HFLAV21}
Y.~Amhis {\em et~al.}, \ifthenelse{\boolean{articletitles}}{\emph{{Averages of
  $b$-hadron, $c$-hadron, and $\tau$-lepton properties as of 2021}},
  }{}\href{https://doi.org/10.1103/PhysRevD.107.052008}{Phys.\ Rev.\
  \textbf{D107} (2023) 052008},
  \href{http://arxiv.org/abs/2206.07501}{{\normalfont\ttfamily
  arXiv:2206.07501}}, {updated results and plots available at
  \href{https://hflav.web.cern.ch}{{\texttt{https://hflav.web.cern.ch}}}}\relax
\mciteBstWouldAddEndPuncttrue
\mciteSetBstMidEndSepPunct{\mcitedefaultmidpunct}
{\mcitedefaultendpunct}{\mcitedefaultseppunct}\relax
\EndOfBibitem
\bibitem{LHCb-PAPER-2023-016}
LHCb collaboration, R.~Aaij {\em et~al.},
  \ifthenelse{\boolean{articletitles}}{\emph{{Improved measurement of CP
  violation parameters in $\Bs\to \jpsi \Kp\Km$ decays in the vicinty of the
  $\phiz(1020)$ resonance}},
  }{}\href{http://arxiv.org/abs/2308.01468}{{\normalfont\ttfamily
  arXiv:2308.01468}}, {submitted to Phys. Rev. Lett.}\relax
\mciteBstWouldAddEndPunctfalse
\mciteSetBstMidEndSepPunct{\mcitedefaultmidpunct}
{}{\mcitedefaultseppunct}\relax
\EndOfBibitem
\bibitem{LHCb-PAPER-2019-003}
LHCb collaboration, R.~Aaij {\em et~al.},
  \ifthenelse{\boolean{articletitles}}{\emph{{Measurement of the \CP-violating
  phase \phis from \mbox{\decay{\Bs}{\jpsi\pip\pim}} decays in 13\tev\
  \proton\proton collisions}},
  }{}\href{https://doi.org/10.1016/j.physletb.2019.07.036}{Phys.\ Lett.\
  \textbf{B797} (2019) 134789},
  \href{http://arxiv.org/abs/1903.05530}{{\normalfont\ttfamily
  arXiv:1903.05530}}\relax
\mciteBstWouldAddEndPuncttrue
\mciteSetBstMidEndSepPunct{\mcitedefaultmidpunct}
{\mcitedefaultendpunct}{\mcitedefaultseppunct}\relax
\EndOfBibitem
\bibitem{LHCb-PAPER-2017-008}
LHCb collaboration, R.~Aaij {\em et~al.},
  \ifthenelse{\boolean{articletitles}}{\emph{{Resonances and \CP-violation in
  \Bs and \mbox{\decay{\Bsb}{\jpsi\Kp\Km}} decays in the mass region above the
  $\phiz(1020)$}}, }{}\href{https://doi.org/10.1007/JHEP08(2017)037}{JHEP
  \textbf{08} (2017) 037},
  \href{http://arxiv.org/abs/1704.08217}{{\normalfont\ttfamily
  arXiv:1704.08217}}\relax
\mciteBstWouldAddEndPuncttrue
\mciteSetBstMidEndSepPunct{\mcitedefaultmidpunct}
{\mcitedefaultendpunct}{\mcitedefaultseppunct}\relax
\EndOfBibitem
\bibitem{LHCb-PAPER-2016-027}
LHCb collaboration, R.~Aaij {\em et~al.},
  \ifthenelse{\boolean{articletitles}}{\emph{{Measurement of the \CP violating
  phase and decay-width difference in \mbox{\decay{\Bs}{\psitwos\phi}}
  decays}}, }{}\href{https://doi.org/10.1016/j.physletb.2016.09.028}{Phys.\
  Lett.\  \textbf{B762} (2016) 253},
  \href{http://arxiv.org/abs/1608.04855}{{\normalfont\ttfamily
  arXiv:1608.04855}}\relax
\mciteBstWouldAddEndPuncttrue
\mciteSetBstMidEndSepPunct{\mcitedefaultmidpunct}
{\mcitedefaultendpunct}{\mcitedefaultseppunct}\relax
\EndOfBibitem
\bibitem{LHCb-PAPER-2014-051}
LHCb collaboration, R.~Aaij {\em et~al.},
  \ifthenelse{\boolean{articletitles}}{\emph{{Measurement of the \CP-violating
  phase \phis in \mbox{\decay{\Bsb}{\Dsp\Dsm}} decays}},
  }{}\href{https://doi.org/10.1103/PhysRevLett.113.211801}{Phys.\ Rev.\ Lett.\
  \textbf{113} (2014) 211801},
  \href{http://arxiv.org/abs/1409.4619}{{\normalfont\ttfamily
  arXiv:1409.4619}}\relax
\mciteBstWouldAddEndPuncttrue
\mciteSetBstMidEndSepPunct{\mcitedefaultmidpunct}
{\mcitedefaultendpunct}{\mcitedefaultseppunct}\relax
\EndOfBibitem
\bibitem{CDF:2012nqr}
CDF collaboration, T.~Aaltonen {\em et~al.},
  \ifthenelse{\boolean{articletitles}}{\emph{{Measurement of the bottom-strange
  meson mixing phase in the full CDF data set}},
  }{}\href{https://doi.org/10.1103/PhysRevLett.109.171802}{Phys.\ Rev.\ Lett.\
  \textbf{109} (2012) 171802},
  \href{http://arxiv.org/abs/1208.2967}{{\normalfont\ttfamily
  arXiv:1208.2967}}\relax
\mciteBstWouldAddEndPuncttrue
\mciteSetBstMidEndSepPunct{\mcitedefaultmidpunct}
{\mcitedefaultendpunct}{\mcitedefaultseppunct}\relax
\EndOfBibitem
\bibitem{D0:2011ymu}
D0 collaboration, V.~M. Abazov {\em et~al.},
  \ifthenelse{\boolean{articletitles}}{\emph{{Measurement of the CP-violating
  phase $\phi_s^{\jpsi \phi}$ using the flavor-tagged decay $B_s^0 \rightarrow
  \jpsi \phi$ in 8 fb$^{-1}$ of $p \bar p$ collisions}},
  }{}\href{https://doi.org/10.1103/PhysRevD.85.032006}{Phys.\ Rev.\
  \textbf{D85} (2012) 032006},
  \href{http://arxiv.org/abs/1109.3166}{{\normalfont\ttfamily
  arXiv:1109.3166}}\relax
\mciteBstWouldAddEndPuncttrue
\mciteSetBstMidEndSepPunct{\mcitedefaultmidpunct}
{\mcitedefaultendpunct}{\mcitedefaultseppunct}\relax
\EndOfBibitem
\bibitem{ATLAS:2020lbz}
ATLAS collaboration, G.~Aad {\em et~al.},
  \ifthenelse{\boolean{articletitles}}{\emph{{Measurement of the $CP$-violating
  phase $\phi_s$ in $B^0_s \to \jpsi\phi$ decays in ATLAS at 13 TeV}},
  }{}\href{https://doi.org/10.1140/epjc/s10052-021-09011-0}{Eur.\ Phys.\ J.\
  \textbf{C81} (2021) 342},
  \href{http://arxiv.org/abs/2001.07115}{{\normalfont\ttfamily
  arXiv:2001.07115}}\relax
\mciteBstWouldAddEndPuncttrue
\mciteSetBstMidEndSepPunct{\mcitedefaultmidpunct}
{\mcitedefaultendpunct}{\mcitedefaultseppunct}\relax
\EndOfBibitem
\bibitem{CMS:2020efq}
CMS collaboration, A.~M. Sirunyan {\em et~al.},
  \ifthenelse{\boolean{articletitles}}{\emph{{Measurement of the $CP$-violating
  phase $\phi_\mathrm{s}$ in the B$^0_\mathrm{s}\to$ J$/\psi\, \phi$(1020) $\to
  \mu^+\mu^-$K$^+$K$^-$ channel in proton-proton collisions at $\sqrt{s} =$ 13
  TeV}}, }{}\href{https://doi.org/10.1016/j.physletb.2021.136188}{Phys.\ Lett.\
   \textbf{B816} (2021) 136188},
  \href{http://arxiv.org/abs/2007.02434}{{\normalfont\ttfamily
  arXiv:2007.02434}}\relax
\mciteBstWouldAddEndPuncttrue
\mciteSetBstMidEndSepPunct{\mcitedefaultmidpunct}
{\mcitedefaultendpunct}{\mcitedefaultseppunct}\relax
\EndOfBibitem
\bibitem{UTfit-UT}
UTfit collaboration, M.~Bona {\em et~al.},
  \ifthenelse{\boolean{articletitles}}{\emph{{The unitarity triangle fit in the
  standard model and hadronic parameters from lattice QCD: A reappraisal after
  the measurements of $\Delta m_{s}$ and $BR(B\to\tau\nu_{\tau})$}},
  }{}\href{https://doi.org/10.1088/1126-6708/2006/10/081}{JHEP \textbf{10}
  (2006) 081}, \href{http://arxiv.org/abs/hep-ph/0606167}{{\normalfont\ttfamily
  arXiv:hep-ph/0606167}}, {updated results and plots available at
  \href{http://www.utfit.org/}{{\texttt{http://www.utfit.org/}}}}\relax
\mciteBstWouldAddEndPuncttrue
\mciteSetBstMidEndSepPunct{\mcitedefaultmidpunct}
{\mcitedefaultendpunct}{\mcitedefaultseppunct}\relax
\EndOfBibitem
\bibitem{Belle:2003vik}
Belle collaboration, K.~Abe {\em et~al.},
  \ifthenelse{\boolean{articletitles}}{\emph{{Measurement of time-dependent
  \CP-violating asymmetries in \Bz\to\Pphi{}\KS, \Kp{}\Km{}\KS, and $\eta'$\KS
  decays}}, }{}Phys.\ Rev.\ Lett.\  \textbf{91} (2003) 261602,
  \href{http://arxiv.org/abs/hep-ex/0308035}{{\normalfont\ttfamily
  arXiv:hep-ex/0308035}}\relax
\mciteBstWouldAddEndPuncttrue
\mciteSetBstMidEndSepPunct{\mcitedefaultmidpunct}
{\mcitedefaultendpunct}{\mcitedefaultseppunct}\relax
\EndOfBibitem
\bibitem{LHCb-PAPER-2023-001}
LHCb collaboration, R.~Aaij {\em et~al.},
  \ifthenelse{\boolean{articletitles}}{\emph{{Precision measurement of \CP
  violation in the penguin-mediated decay \decay{\Bs}{\phi\phi}}},
  }{}\href{http://arxiv.org/abs/2304.06198}{{\normalfont\ttfamily
  arXiv:2304.06198}}, {to appear in Phys. Rev. Lett.}\relax
\mciteBstWouldAddEndPunctfalse
\mciteSetBstMidEndSepPunct{\mcitedefaultmidpunct}
{}{\mcitedefaultseppunct}\relax
\EndOfBibitem
\bibitem{LHCb-FIGURE-2020-002}
LHCb collaboration, \ifthenelse{\boolean{articletitles}}{\emph{{Comparison of
  flavour tagging performances displayed in the
  $\omega$-$\varepsilon_\mathrm{tag}$-plane}}, }{}
  \href{http://cdsweb.cern.ch/search?p=LHCb-FIGURE-2020-002&f=reportnumber&action_search=Search&c=LHCb+Figures}
  {LHCb-FIGURE-2020-002}, 2020\relax
\mciteBstWouldAddEndPuncttrue
\mciteSetBstMidEndSepPunct{\mcitedefaultmidpunct}
{\mcitedefaultendpunct}{\mcitedefaultseppunct}\relax
\EndOfBibitem
\bibitem{LHCb-PAPER-2016-039}
LHCb collaboration, R.~Aaij {\em et~al.},
  \ifthenelse{\boolean{articletitles}}{\emph{{New algorithms for identifying
  the flavour of \Bz mesons using pions and protons}},
  }{}\href{https://doi.org/10.1140/epjc/s10052-017-4731-y}{Eur.\ Phys.\ J.\
  \textbf{C77} (2017) 238},
  \href{http://arxiv.org/abs/1610.06019}{{\normalfont\ttfamily
  arXiv:1610.06019}}\relax
\mciteBstWouldAddEndPuncttrue
\mciteSetBstMidEndSepPunct{\mcitedefaultmidpunct}
{\mcitedefaultendpunct}{\mcitedefaultseppunct}\relax
\EndOfBibitem
\bibitem{LHCb-PAPER-2011-021}
LHCb collaboration, R.~Aaij {\em et~al.},
  \ifthenelse{\boolean{articletitles}}{\emph{{Measurement of the \CP-violating
  phase \phis in the decay \mbox{\decay{\Bs}{\jpsi\phiz}}}},
  }{}\href{https://doi.org/10.1103/PhysRevLett.108.101803}{Phys.\ Rev.\ Lett.\
  \textbf{108} (2012) 101803},
  \href{http://arxiv.org/abs/1112.3183}{{\normalfont\ttfamily
  arXiv:1112.3183}}\relax
\mciteBstWouldAddEndPuncttrue
\mciteSetBstMidEndSepPunct{\mcitedefaultmidpunct}
{\mcitedefaultendpunct}{\mcitedefaultseppunct}\relax
\EndOfBibitem
\bibitem{Beneke:2019slt}
M.~Beneke, C.~Bobeth, and R.~Szafron,
  \ifthenelse{\boolean{articletitles}}{\emph{{Power-enhanced
  leading-logarithmic QED corrections to $B_q \to \mu^+\mu^-$}},
  }{}\href{https://doi.org/10.1007/JHEP10(2019)232}{JHEP \textbf{10} (2019)
  232}, \href{http://arxiv.org/abs/1908.07011}{{\normalfont\ttfamily
  arXiv:1908.07011}}, [Erratum: JHEP 11, 099 (2022)]\relax
\mciteBstWouldAddEndPuncttrue
\mciteSetBstMidEndSepPunct{\mcitedefaultmidpunct}
{\mcitedefaultendpunct}{\mcitedefaultseppunct}\relax
\EndOfBibitem
\bibitem{DeBruyn:2012wj}
K.~De~Bruyn {\em et~al.}, \ifthenelse{\boolean{articletitles}}{\emph{{Branching
  ratio measurements of $B_s$ decays}},
  }{}\href{https://doi.org/10.1103/PhysRevD.86.014027}{Phys.\ Rev.\
  \textbf{D86} (2012) 014027},
  \href{http://arxiv.org/abs/1204.1735}{{\normalfont\ttfamily
  arXiv:1204.1735}}\relax
\mciteBstWouldAddEndPuncttrue
\mciteSetBstMidEndSepPunct{\mcitedefaultmidpunct}
{\mcitedefaultendpunct}{\mcitedefaultseppunct}\relax
\EndOfBibitem
\bibitem{LHCb-PAPER-2012-043}
LHCb collaboration, R.~Aaij {\em et~al.},
  \ifthenelse{\boolean{articletitles}}{\emph{{First evidence for the decay
  \mbox{\decay{\Bs}{\mumu}}}},
  }{}\href{https://doi.org/10.1103/PhysRevLett.110.021801}{Phys.\ Rev.\ Lett.\
  \textbf{110} (2013) 021801},
  \href{http://arxiv.org/abs/1211.2674}{{\normalfont\ttfamily
  arXiv:1211.2674}}\relax
\mciteBstWouldAddEndPuncttrue
\mciteSetBstMidEndSepPunct{\mcitedefaultmidpunct}
{\mcitedefaultendpunct}{\mcitedefaultseppunct}\relax
\EndOfBibitem
\bibitem{LHCb-PAPER-2014-049}
CMS and LHCb collaborations, V.~Khachatryan {\em et~al.},
  \ifthenelse{\boolean{articletitles}}{\emph{{Observation of the rare
  \mbox{\decay{\Bs}{\mumu}} decay from the combined analysis of CMS and LHCb
  data}}, }{}\href{https://doi.org/10.1038/nature14474}{Nature \textbf{522}
  (2015) 68}, \href{http://arxiv.org/abs/1411.4413}{{\normalfont\ttfamily
  arXiv:1411.4413}}\relax
\mciteBstWouldAddEndPuncttrue
\mciteSetBstMidEndSepPunct{\mcitedefaultmidpunct}
{\mcitedefaultendpunct}{\mcitedefaultseppunct}\relax
\EndOfBibitem
\bibitem{LHCb-PAPER-2017-001}
LHCb collaboration, R.~Aaij {\em et~al.},
  \ifthenelse{\boolean{articletitles}}{\emph{{Measurement of the
  \mbox{\decay{\Bs}{\mumu}} branching fraction and effective lifetime and
  search for \mbox{\decay{\Bz}{\mumu}} decays}},
  }{}\href{https://doi.org/10.1103/PhysRevLett.118.191801}{Phys.\ Rev.\ Lett.\
  \textbf{118} (2017) 191801},
  \href{http://arxiv.org/abs/1703.05747}{{\normalfont\ttfamily
  arXiv:1703.05747}}\relax
\mciteBstWouldAddEndPuncttrue
\mciteSetBstMidEndSepPunct{\mcitedefaultmidpunct}
{\mcitedefaultendpunct}{\mcitedefaultseppunct}\relax
\EndOfBibitem
\bibitem{CMS:2019bbr}
CMS collaboration, A.~M. Sirunyan {\em et~al.},
  \ifthenelse{\boolean{articletitles}}{\emph{{Measurement of properties of
  \Bsmm decays and search for \Bdmm with the CMS experiment}},
  }{}\href{https://doi.org/10.1007/JHEP04(2020)188}{JHEP \textbf{04} (2020)
  188}, \href{http://arxiv.org/abs/1910.12127}{{\normalfont\ttfamily
  arXiv:1910.12127}}\relax
\mciteBstWouldAddEndPuncttrue
\mciteSetBstMidEndSepPunct{\mcitedefaultmidpunct}
{\mcitedefaultendpunct}{\mcitedefaultseppunct}\relax
\EndOfBibitem
\bibitem{ATLAS:2018cur}
ATLAS collaboration, M.~Aaboud {\em et~al.},
  \ifthenelse{\boolean{articletitles}}{\emph{{Study of the rare decays of
  $B^0_s$ and $B^0$ mesons into muon pairs using data collected during 2015 and
  2016 with the ATLAS detector}},
  }{}\href{https://doi.org/10.1007/JHEP04(2019)098}{JHEP \textbf{04} (2019)
  098}, \href{http://arxiv.org/abs/1812.03017}{{\normalfont\ttfamily
  arXiv:1812.03017}}\relax
\mciteBstWouldAddEndPuncttrue
\mciteSetBstMidEndSepPunct{\mcitedefaultmidpunct}
{\mcitedefaultendpunct}{\mcitedefaultseppunct}\relax
\EndOfBibitem
\bibitem{LHCb-PAPER-2021-007}
LHCb collaboration, R.~Aaij {\em et~al.},
  \ifthenelse{\boolean{articletitles}}{\emph{{Analysis of neutral $B$-meson
  decays into two muons}},
  }{}\href{https://doi.org/10.1103/PhysRevLett.128.041801}{Phys.\ Rev.\ Lett.\
  \textbf{128} (2022) 041801},
  \href{http://arxiv.org/abs/2108.09284}{{\normalfont\ttfamily
  arXiv:2108.09284}}\relax
\mciteBstWouldAddEndPuncttrue
\mciteSetBstMidEndSepPunct{\mcitedefaultmidpunct}
{\mcitedefaultendpunct}{\mcitedefaultseppunct}\relax
\EndOfBibitem
\bibitem{LHCb-PAPER-2021-008}
LHCb collaboration, R.~Aaij {\em et~al.},
  \ifthenelse{\boolean{articletitles}}{\emph{{Measurement of the $\Bs \to \mup
  \mun$ decay properties and search for the $\Bd \to \mup \mun$ and $\Bs \to
  \mup \mun \gamma$ decays}},
  }{}\href{https://doi.org/10.1103/PhysRevD.105.012010}{Phys.\ Rev.\
  \textbf{D105} (2022) 012010},
  \href{http://arxiv.org/abs/2108.09283}{{\normalfont\ttfamily
  arXiv:2108.09283}}\relax
\mciteBstWouldAddEndPuncttrue
\mciteSetBstMidEndSepPunct{\mcitedefaultmidpunct}
{\mcitedefaultendpunct}{\mcitedefaultseppunct}\relax
\EndOfBibitem
\bibitem{CMS:2022mgd}
CMS collaboration, A.~Tumasyan {\em et~al.},
  \ifthenelse{\boolean{articletitles}}{\emph{{Measurement of the \Bsmm decay
  properties and search for the \Bdmm decay in proton-proton collisions at
  $\sqs=13\tev$}},
  }{}\href{https://doi.org/10.1016/j.physletb.2023.137955}{Phys.\ Lett.\
  \textbf{B842} (2023) 137955},
  \href{http://arxiv.org/abs/2212.10311}{{\normalfont\ttfamily
  arXiv:2212.10311}}\relax
\mciteBstWouldAddEndPuncttrue
\mciteSetBstMidEndSepPunct{\mcitedefaultmidpunct}
{\mcitedefaultendpunct}{\mcitedefaultseppunct}\relax
\EndOfBibitem
\bibitem{Allanach:2022iod}
B.~Allanach and J.~Davighi, \ifthenelse{\boolean{articletitles}}{\emph{{The
  Rumble in the Meson: a leptoquark versus a $Z'$ to fit \decay{b}{s\mumu}
  anomalies including 2022 LHCb $ {R}_{K^{\left(\ast \right)}} $
  measurements}}, }{}\href{https://doi.org/10.1007/JHEP04(2023)033}{JHEP
  \textbf{04} (2023) 033},
  \href{http://arxiv.org/abs/2211.11766}{{\normalfont\ttfamily
  arXiv:2211.11766}}\relax
\mciteBstWouldAddEndPuncttrue
\mciteSetBstMidEndSepPunct{\mcitedefaultmidpunct}
{\mcitedefaultendpunct}{\mcitedefaultseppunct}\relax
\EndOfBibitem
\bibitem{Buchalla:1995vs}
G.~Buchalla, A.~J. Buras, and M.~E. Lautenbacher,
  \ifthenelse{\boolean{articletitles}}{\emph{{Weak decays beyond leading
  logarithms}}, }{}\href{https://doi.org/10.1103/RevModPhys.68.1125}{Rev.\
  Mod.\ Phys.\  \textbf{68} (1996) 1125},
  \href{http://arxiv.org/abs/hep-ph/9512380}{{\normalfont\ttfamily
  arXiv:hep-ph/9512380}}\relax
\mciteBstWouldAddEndPuncttrue
\mciteSetBstMidEndSepPunct{\mcitedefaultmidpunct}
{\mcitedefaultendpunct}{\mcitedefaultseppunct}\relax
\EndOfBibitem
\bibitem{LHCb-PAPER-2013-039}
LHCb collaboration, R.~Aaij {\em et~al.},
  \ifthenelse{\boolean{articletitles}}{\emph{{Observation of a resonance in
  \mbox{\decay{\Bp}{\Kp\mumu}} decays at low recoil}},
  }{}\href{https://doi.org/10.1103/PhysRevLett.111.112003}{Phys.\ Rev.\ Lett.\
  \textbf{111} (2013) 112003},
  \href{http://arxiv.org/abs/1307.7595}{{\normalfont\ttfamily
  arXiv:1307.7595}}\relax
\mciteBstWouldAddEndPuncttrue
\mciteSetBstMidEndSepPunct{\mcitedefaultmidpunct}
{\mcitedefaultendpunct}{\mcitedefaultseppunct}\relax
\EndOfBibitem
\bibitem{Eilam:1986fs}
G.~Eilam, J.~L. Hewett, and T.~G. Rizzo,
  \ifthenelse{\boolean{articletitles}}{\emph{{$B \to K \ell^+ \ell^-$ with four
  generations: rates and {\CP} violation}},
  }{}\href{https://doi.org/10.1103/PhysRevD.34.2773}{Phys.\ Rev.\  \textbf{D34}
  (1986) 2773}\relax
\mciteBstWouldAddEndPuncttrue
\mciteSetBstMidEndSepPunct{\mcitedefaultmidpunct}
{\mcitedefaultendpunct}{\mcitedefaultseppunct}\relax
\EndOfBibitem
\bibitem{Ali:1999mm}
A.~Ali, P.~Ball, L.~Handoko, and G.~Hiller,
  \ifthenelse{\boolean{articletitles}}{\emph{{A Comparative Study of the Decays
  \mbox{\decay{\B}{(\kaon,\Kstar)\ellell}} in the Standard Model and
  Supersymmetric Theories}},
  }{}\href{https://doi.org/10.1103/PhysRevD.61.074024}{Phys.\ Rev.\
  \textbf{D61} (2000) 074024},
  \href{http://arxiv.org/abs/hep-ph/9910221}{{\normalfont\ttfamily
  arXiv:hep-ph/9910221}}\relax
\mciteBstWouldAddEndPuncttrue
\mciteSetBstMidEndSepPunct{\mcitedefaultmidpunct}
{\mcitedefaultendpunct}{\mcitedefaultseppunct}\relax
\EndOfBibitem
\bibitem{Kruger:2005ep}
F.~{Kr\"uger} and J.~Matias,
  \ifthenelse{\boolean{articletitles}}{\emph{{Probing new physics via the
  transverse amplitudes of \mbox{\decay{\Bz}{\Kstarz(\Kp\pip)\ellell}} at large
  recoil}}, }{}\href{https://doi.org/10.1103/PhysRevD.71.094009}{Phys.\ Rev.\
  \textbf{D71} (2005) 094009},
  \href{http://arxiv.org/abs/hep-ph/0502060}{{\normalfont\ttfamily
  arXiv:hep-ph/0502060}}\relax
\mciteBstWouldAddEndPuncttrue
\mciteSetBstMidEndSepPunct{\mcitedefaultmidpunct}
{\mcitedefaultendpunct}{\mcitedefaultseppunct}\relax
\EndOfBibitem
\bibitem{Altmannshofer:2008dz}
W.~Altmannshofer {\em et~al.},
  \ifthenelse{\boolean{articletitles}}{\emph{{Symmetries and asymmetries of $B
  \to K^{*} \mu^{+} \mu^{-}$ decays in the Standard Model and beyond}},
  }{}\href{https://doi.org/10.1088/1126-6708/2009/01/019}{JHEP \textbf{0901}
  (2009) 019}, \href{http://arxiv.org/abs/0811.1214}{{\normalfont\ttfamily
  arXiv:0811.1214}}\relax
\mciteBstWouldAddEndPuncttrue
\mciteSetBstMidEndSepPunct{\mcitedefaultmidpunct}
{\mcitedefaultendpunct}{\mcitedefaultseppunct}\relax
\EndOfBibitem
\bibitem{Egede:2008uy}
U.~Egede {\em et~al.}, \ifthenelse{\boolean{articletitles}}{\emph{{New
  observables in the decay mode \Bbar\to\Kstarb{}\ellp{}\ellm}},
  }{}\href{https://doi.org/10.1088/1126-6708/2008/11/032}{JHEP \textbf{11}
  (2008) 032}, \href{http://arxiv.org/abs/0807.2589}{{\normalfont\ttfamily
  arXiv:0807.2589}}\relax
\mciteBstWouldAddEndPuncttrue
\mciteSetBstMidEndSepPunct{\mcitedefaultmidpunct}
{\mcitedefaultendpunct}{\mcitedefaultseppunct}\relax
\EndOfBibitem
\bibitem{Bobeth:2008ij}
C.~Bobeth, G.~Hiller, and G.~Piranishvili,
  \ifthenelse{\boolean{articletitles}}{\emph{{\CP Asymmetries in $\Bb \to
  \Kstarb (\to \Kb \pi) \bar{\ell} \ell$ and Untagged $\Bb_s$, $B_s \to \phi
  (\to K^{+} K^-) \bar{\ell} \ell$ Decays at NLO}},
  }{}\href{https://doi.org/10.1088/1126-6708/2008/07/106}{JHEP \textbf{0807}
  (2008) 106}, \href{http://arxiv.org/abs/0805.2525}{{\normalfont\ttfamily
  arXiv:0805.2525}}\relax
\mciteBstWouldAddEndPuncttrue
\mciteSetBstMidEndSepPunct{\mcitedefaultmidpunct}
{\mcitedefaultendpunct}{\mcitedefaultseppunct}\relax
\EndOfBibitem
\bibitem{Descotes-Genon:2013vna}
S.~Descotes-Genon, T.~Hurth, J.~Matias, and J.~Virto,
  \ifthenelse{\boolean{articletitles}}{\emph{{Optimizing the basis of ${B} \to
  {K}^{*}\ell^+ \ell^-$ observables in the full kinematic range}},
  }{}\href{https://doi.org/10.1007/JHEP05(2013)137}{JHEP \textbf{1305} (2013)
  137}, \href{http://arxiv.org/abs/1303.5794}{{\normalfont\ttfamily
  arXiv:1303.5794}}\relax
\mciteBstWouldAddEndPuncttrue
\mciteSetBstMidEndSepPunct{\mcitedefaultmidpunct}
{\mcitedefaultendpunct}{\mcitedefaultseppunct}\relax
\EndOfBibitem
\bibitem{LHCb-PAPER-2020-002}
LHCb collaboration, R.~Aaij {\em et~al.},
  \ifthenelse{\boolean{articletitles}}{\emph{{Measurement of \CP-averaged
  observables in the \mbox{\decay{\Bz}{\Kstarz\mumu}} decay}},
  }{}\href{https://doi.org/10.1103/PhysRevLett.125.011802}{Phys.\ Rev.\ Lett.\
  \textbf{125} (2020) 011802},
  \href{http://arxiv.org/abs/2003.04831}{{\normalfont\ttfamily
  arXiv:2003.04831}}\relax
\mciteBstWouldAddEndPuncttrue
\mciteSetBstMidEndSepPunct{\mcitedefaultmidpunct}
{\mcitedefaultendpunct}{\mcitedefaultseppunct}\relax
\EndOfBibitem
\bibitem{LHCb-PAPER-2023-033}
LHCb collaboration, R.~Aaij {\em et~al.},
  \ifthenelse{\boolean{articletitles}}{\emph{{Direct determination of Wilson
  coefficients $C_9^{(')}$ and $C_{10}^{(')}$ via amplitude analysis of $\Bz
  \to \Kstarz\mup\mun$ decays}}, }{} {LHCb-PAPER-2023-033}, {in
  preparation}\relax
\mciteBstWouldAddEndPuncttrue
\mciteSetBstMidEndSepPunct{\mcitedefaultmidpunct}
{\mcitedefaultendpunct}{\mcitedefaultseppunct}\relax
\EndOfBibitem
\bibitem{Lees:2012tva}
BaBar collaboration, J.~P. Lees {\em et~al.},
  \ifthenelse{\boolean{articletitles}}{\emph{{Measurement of branching
  fractions and rate asymmetries in the rare decays $B \to K^{(*)} l^+ l^-$}},
  }{}\href{https://doi.org/10.1103/PhysRevD.86.032012}{Phys.\ Rev.\
  \textbf{D86} (2012) 032012},
  \href{http://arxiv.org/abs/1204.3933}{{\normalfont\ttfamily
  arXiv:1204.3933}}\relax
\mciteBstWouldAddEndPuncttrue
\mciteSetBstMidEndSepPunct{\mcitedefaultmidpunct}
{\mcitedefaultendpunct}{\mcitedefaultseppunct}\relax
\EndOfBibitem
\bibitem{Wei:2009zv}
Belle collaboration, J.-T. Wei {\em et~al.},
  \ifthenelse{\boolean{articletitles}}{\emph{{Measurement of the Differential
  Branching Fraction and Forward-Backward Asymmetry for $B\to\Kstar\ellell$}},
  }{}\href{https://doi.org/10.1103/PhysRevLett.103.171801}{Phys.\ Rev.\ Lett.\
  \textbf{103} (2009) 171801},
  \href{http://arxiv.org/abs/0904.0770}{{\normalfont\ttfamily
  arXiv:0904.0770}}\relax
\mciteBstWouldAddEndPuncttrue
\mciteSetBstMidEndSepPunct{\mcitedefaultmidpunct}
{\mcitedefaultendpunct}{\mcitedefaultseppunct}\relax
\EndOfBibitem
\bibitem{Aaltonen:2011qs}
CDF collaboration, T.~Aaltonen {\em et~al.},
  \ifthenelse{\boolean{articletitles}}{\emph{{Observation of the baryonic
  flavor-changing neutral current decay $\Lambdares_{b} \to \Lambdares \mu^{+}
  \mu^{-}$}}, }{}\href{https://doi.org/10.1103/PhysRevLett.107.201802}{Phys.\
  Rev.\ Lett.\  \textbf{107} (2011) 201802},
  \href{http://arxiv.org/abs/1107.3753}{{\normalfont\ttfamily
  arXiv:1107.3753}}\relax
\mciteBstWouldAddEndPuncttrue
\mciteSetBstMidEndSepPunct{\mcitedefaultmidpunct}
{\mcitedefaultendpunct}{\mcitedefaultseppunct}\relax
\EndOfBibitem
\bibitem{LHCb-PAPER-2012-011}
LHCb collaboration, R.~Aaij {\em et~al.},
  \ifthenelse{\boolean{articletitles}}{\emph{{Measurement of the isospin
  asymmetry in \mbox{\decay{\B}{\Kstar\mumu}} decays}},
  }{}\href{https://doi.org/10.1007/JHEP07(2012)133}{JHEP \textbf{07} (2012)
  133}, \href{http://arxiv.org/abs/1205.3422}{{\normalfont\ttfamily
  arXiv:1205.3422}}\relax
\mciteBstWouldAddEndPuncttrue
\mciteSetBstMidEndSepPunct{\mcitedefaultmidpunct}
{\mcitedefaultendpunct}{\mcitedefaultseppunct}\relax
\EndOfBibitem
\bibitem{LHCb-PAPER-2022-045}
LHCb collaboration, R.~Aaij {\em et~al.},
  \ifthenelse{\boolean{articletitles}}{\emph{{Measurement of lepton
  universality parameters in $\Bp\to\Kp\ellp\ellm$ and
  $\Bz\to\Kstarz\ellp\ellm$ decays}},
  }{}\href{https://doi.org/10.1103/PhysRevD.108.032002}{Phys.\ Rev.\
  \textbf{D108} (2023) 032002},
  \href{http://arxiv.org/abs/2212.09153}{{\normalfont\ttfamily
  arXiv:2212.09153}}\relax
\mciteBstWouldAddEndPuncttrue
\mciteSetBstMidEndSepPunct{\mcitedefaultmidpunct}
{\mcitedefaultendpunct}{\mcitedefaultseppunct}\relax
\EndOfBibitem
\bibitem{LHCb-PAPER-2022-046}
LHCb collaboration, R.~Aaij {\em et~al.},
  \ifthenelse{\boolean{articletitles}}{\emph{{Test of lepton universality in
  $\bquark \to \squark \ellp\ellm$ decays}},
  }{}\href{https://doi.org/10.1103/PhysRevLett.131.051803}{Phys.\ Rev.\ Lett.\
  \textbf{131} (2023) 051803},
  \href{http://arxiv.org/abs/2212.09152}{{\normalfont\ttfamily
  arXiv:2212.09152}}\relax
\mciteBstWouldAddEndPuncttrue
\mciteSetBstMidEndSepPunct{\mcitedefaultmidpunct}
{\mcitedefaultendpunct}{\mcitedefaultseppunct}\relax
\EndOfBibitem
\bibitem{Parrott:2022zte}
HPQCD collaboration, W.~G. Parrott, C.~Bouchard, and C.~T.~H. Davies,
  \ifthenelse{\boolean{articletitles}}{\emph{{Standard Model predictions for
  \BllK, \decay{\B}{K\ell_1^-\ell_2^+} and \decay{\B}{kaon\neu\neub} using form
  factors from $N_f=2+1+1$ lattice QCD}},
  }{}\href{https://doi.org/10.1103/PhysRevD.107.014511}{Phys.\ Rev.\
  \textbf{D107} (2023) 014511}, Erratum \href{https://doi.org/}{ibid.\
  \textbf{107} (2023) 119903},
  \href{http://arxiv.org/abs/2207.13371}{{\normalfont\ttfamily
  arXiv:2207.13371}}, [Erratum: Phys.Rev.D 107, 119903 (2023)]\relax
\mciteBstWouldAddEndPuncttrue
\mciteSetBstMidEndSepPunct{\mcitedefaultmidpunct}
{\mcitedefaultendpunct}{\mcitedefaultseppunct}\relax
\EndOfBibitem
\bibitem{Lees:2015ymt}
BaBar collaboration, J.~P. Lees {\em et~al.},
  \ifthenelse{\boolean{articletitles}}{\emph{{Measurement of angular
  asymmetries in the decays $B \to K^*\ell^+\ell^-$}},
  }{}\href{https://doi.org/10.1103/PhysRevD.93.052015}{Phys.\ Rev.\
  \textbf{93} (2016) 052015},
  \href{http://arxiv.org/abs/1508.07960}{{\normalfont\ttfamily
  arXiv:1508.07960}}\relax
\mciteBstWouldAddEndPuncttrue
\mciteSetBstMidEndSepPunct{\mcitedefaultmidpunct}
{\mcitedefaultendpunct}{\mcitedefaultseppunct}\relax
\EndOfBibitem
\bibitem{Wehle:2016yoi}
Belle collaboration, S.~Wehle {\em et~al.},
  \ifthenelse{\boolean{articletitles}}{\emph{{Lepton-flavor-dependent angular
  analysis of $B\to K^\ast \ell^+\ell^-$}},
  }{}\href{https://doi.org/10.1103/PhysRevLett.118.111801}{Phys.\ Rev.\ Lett.\
  \textbf{118} (2016) 111801},
  \href{http://arxiv.org/abs/1612.05014}{{\normalfont\ttfamily
  arXiv:1612.05014}}\relax
\mciteBstWouldAddEndPuncttrue
\mciteSetBstMidEndSepPunct{\mcitedefaultmidpunct}
{\mcitedefaultendpunct}{\mcitedefaultseppunct}\relax
\EndOfBibitem
\bibitem{Khachatryan:2015isa}
CMS collaboration, V.~Khachatryan {\em et~al.},
  \ifthenelse{\boolean{articletitles}}{\emph{{Angular analysis of the decay $
  B^0 \to K^{*0} \mu^{+} \mu^{-}$ from pp collisions at $\sqrt{s}=8$ TeV}},
  }{}\href{https://doi.org/10.1016/j.physletb.2015.12.020}{Phys.\ Lett.\
  \textbf{B753} (2016) 424},
  \href{http://arxiv.org/abs/1507.08126}{{\normalfont\ttfamily
  arXiv:1507.08126}}\relax
\mciteBstWouldAddEndPuncttrue
\mciteSetBstMidEndSepPunct{\mcitedefaultmidpunct}
{\mcitedefaultendpunct}{\mcitedefaultseppunct}\relax
\EndOfBibitem
\bibitem{Sirunyan:2017dhj}
CMS collaboration, A.~M. Sirunyan {\em et~al.},
  \ifthenelse{\boolean{articletitles}}{\emph{{Measurement of angular parameters
  from the decay $\Bz \to \Kstarz \mumu$ in proton-proton collisions at
  $\sqrt{s} = $ 8 TeV}},
  }{}\href{https://doi.org/10.1016/j.physletb.2018.04.030}{Phys.\ Lett.\
  \textbf{B781} (2018) 517},
  \href{http://arxiv.org/abs/1710.02846}{{\normalfont\ttfamily
  arXiv:1710.02846}}\relax
\mciteBstWouldAddEndPuncttrue
\mciteSetBstMidEndSepPunct{\mcitedefaultmidpunct}
{\mcitedefaultendpunct}{\mcitedefaultseppunct}\relax
\EndOfBibitem
\bibitem{Aaboud:2018krd}
ATLAS collaboration, M.~Aaboud {\em et~al.},
  \ifthenelse{\boolean{articletitles}}{\emph{{Angular analysis of $B^0_d
  \rightarrow K^{*}\mu^+\mu^-$ decays in $pp$ collisions at $\sqrt{s}= 8$ TeV
  with the ATLAS detector}},
  }{}\href{https://doi.org/10.1007/JHEP10(2018)047}{JHEP \textbf{10} (2018)
  047}, \href{http://arxiv.org/abs/1805.04000}{{\normalfont\ttfamily
  arXiv:1805.04000}}\relax
\mciteBstWouldAddEndPuncttrue
\mciteSetBstMidEndSepPunct{\mcitedefaultmidpunct}
{\mcitedefaultendpunct}{\mcitedefaultseppunct}\relax
\EndOfBibitem
\bibitem{LHCb-PAPER-2020-041}
LHCb collaboration, R.~Aaij {\em et~al.},
  \ifthenelse{\boolean{articletitles}}{\emph{{Angular analysis of the $B^{+}\to
  K^{\ast+}\mu^+\mu^-$ decay}},
  }{}\href{https://doi.org/10.1103/PhysRevLett.126.161802}{Phys.\ Rev.\ Lett.\
  \textbf{126} (2021) 161802},
  \href{http://arxiv.org/abs/2012.13241}{{\normalfont\ttfamily
  arXiv:2012.13241}}\relax
\mciteBstWouldAddEndPuncttrue
\mciteSetBstMidEndSepPunct{\mcitedefaultmidpunct}
{\mcitedefaultendpunct}{\mcitedefaultseppunct}\relax
\EndOfBibitem
\bibitem{Straub:2015ica}
A.~Bharucha, D.~M. Straub, and R.~Zwicky,
  \ifthenelse{\boolean{articletitles}}{\emph{{$B\to V\ell^+\ell^-$ in the
  Standard Model from Light-Cone Sum Rules}},
  }{}\href{https://doi.org/10.1007/JHEP08(2016)098}{JHEP \textbf{08} (2015)
  098}, \href{http://arxiv.org/abs/1503.05534}{{\normalfont\ttfamily
  arXiv:1503.05534}}\relax
\mciteBstWouldAddEndPuncttrue
\mciteSetBstMidEndSepPunct{\mcitedefaultmidpunct}
{\mcitedefaultendpunct}{\mcitedefaultseppunct}\relax
\EndOfBibitem
\bibitem{Altmannshofer:2014rta}
W.~Altmannshofer and D.~M. Straub,
  \ifthenelse{\boolean{articletitles}}{\emph{{New physics in $b\rightarrow s$
  transitions after LHC run 1}},
  }{}\href{https://doi.org/10.1140/epjc/s10052-015-3602-7}{Eur.\ Phys.\ J.\
  \textbf{C75} (2015) 382},
  \href{http://arxiv.org/abs/1411.3161}{{\normalfont\ttfamily
  arXiv:1411.3161}}\relax
\mciteBstWouldAddEndPuncttrue
\mciteSetBstMidEndSepPunct{\mcitedefaultmidpunct}
{\mcitedefaultendpunct}{\mcitedefaultseppunct}\relax
\EndOfBibitem
\bibitem{Descotes-Genon:2014uoa}
S.~Descotes-Genon, L.~Hofer, J.~Matias, and J.~Virto,
  \ifthenelse{\boolean{articletitles}}{\emph{{On the impact of power
  corrections in the prediction of $B \to K^*\mu^+\mu^-$ observables}},
  }{}\href{https://doi.org/10.1007/JHEP12(2014)125}{JHEP \textbf{12} (2014)
  125}, \href{http://arxiv.org/abs/1407.8526}{{\normalfont\ttfamily
  arXiv:1407.8526}}\relax
\mciteBstWouldAddEndPuncttrue
\mciteSetBstMidEndSepPunct{\mcitedefaultmidpunct}
{\mcitedefaultendpunct}{\mcitedefaultseppunct}\relax
\EndOfBibitem
\bibitem{Khodjamirian:2010vf}
A.~Khodjamirian, T.~Mannel, A.~A. Pivovarov, and Y.-M. Wang,
  \ifthenelse{\boolean{articletitles}}{\emph{{Charm-loop effect in $B \to
  K^{(*)} \ell^{+} \ell^{-}$ and $B\to K^*\gamma$}},
  }{}\href{https://doi.org/10.1007/JHEP09(2010)089}{JHEP \textbf{09} (2010)
  089}, \href{http://arxiv.org/abs/1006.4945}{{\normalfont\ttfamily
  arXiv:1006.4945}}\relax
\mciteBstWouldAddEndPuncttrue
\mciteSetBstMidEndSepPunct{\mcitedefaultmidpunct}
{\mcitedefaultendpunct}{\mcitedefaultseppunct}\relax
\EndOfBibitem
\bibitem{LHCb-PAPER-2021-004}
LHCb collaboration, R.~Aaij {\em et~al.},
  \ifthenelse{\boolean{articletitles}}{\emph{{Test of lepton universality in
  beauty-quark decays}},
  }{}\href{https://doi.org/10.1038/s41567-021-01478-8}{Nature Physics
  \textbf{18} (2022) 277},
  \href{http://arxiv.org/abs/2103.11769}{{\normalfont\ttfamily
  arXiv:2103.11769}}\relax
\mciteBstWouldAddEndPuncttrue
\mciteSetBstMidEndSepPunct{\mcitedefaultmidpunct}
{\mcitedefaultendpunct}{\mcitedefaultseppunct}\relax
\EndOfBibitem
\bibitem{LHCb-PAPER-2021-038}
LHCb collaboration, R.~Aaij {\em et~al.},
  \ifthenelse{\boolean{articletitles}}{\emph{{Tests of lepton universality
  using \mbox{$B^0 \to \KS \ell^+\ell^-$} and \mbox{$B^+\to K^{*+}\ell^+
  \ell^-$} decays}},
  }{}\href{https://doi.org/10.1103/PhysRevLett.128.191802}{Phys.\ Rev.\ Lett.\
  \textbf{128} (2022) 191802},
  \href{http://arxiv.org/abs/2110.09501}{{\normalfont\ttfamily
  arXiv:2110.09501}}\relax
\mciteBstWouldAddEndPuncttrue
\mciteSetBstMidEndSepPunct{\mcitedefaultmidpunct}
{\mcitedefaultendpunct}{\mcitedefaultseppunct}\relax
\EndOfBibitem
\bibitem{LHCb-PAPER-2017-013}
LHCb collaboration, R.~Aaij {\em et~al.},
  \ifthenelse{\boolean{articletitles}}{\emph{{Test of lepton universality with
  \mbox{\decay{\Bz}{\Kstarz\ellell}} decays}},
  }{}\href{https://doi.org/10.1007/JHEP08(2017)055}{JHEP \textbf{08} (2017)
  055}, \href{http://arxiv.org/abs/1705.05802}{{\normalfont\ttfamily
  arXiv:1705.05802}}\relax
\mciteBstWouldAddEndPuncttrue
\mciteSetBstMidEndSepPunct{\mcitedefaultmidpunct}
{\mcitedefaultendpunct}{\mcitedefaultseppunct}\relax
\EndOfBibitem
\bibitem{Hiller:2003js}
G.~Hiller and F.~Kr{\"u}ger, \ifthenelse{\boolean{articletitles}}{\emph{More
  model-independent analysis of \bquark\to\squark processes},
  }{}\href{https://doi.org/10.1103/PhysRevD.69.074020}{Phys.\ Rev.\
  \textbf{D69} (2004) 074020},
  \href{http://arxiv.org/abs/hep-ph/0310219}{{\normalfont\ttfamily
  arXiv:hep-ph/0310219}}\relax
\mciteBstWouldAddEndPuncttrue
\mciteSetBstMidEndSepPunct{\mcitedefaultmidpunct}
{\mcitedefaultendpunct}{\mcitedefaultseppunct}\relax
\EndOfBibitem
\bibitem{Isidori:2020acz}
G.~Isidori, S.~Nabeebaccus, and R.~Zwicky,
  \ifthenelse{\boolean{articletitles}}{\emph{{QED corrections in $ \Bb\to
  \overline{K}{\mathrm{\ell}}^{+}{\mathrm{\ell}}^{-} $ at the
  double-differential level}},
  }{}\href{https://doi.org/10.1007/JHEP12(2020)104}{JHEP \textbf{12} (2020)
  104}, \href{http://arxiv.org/abs/2009.00929}{{\normalfont\ttfamily
  arXiv:2009.00929}}\relax
\mciteBstWouldAddEndPuncttrue
\mciteSetBstMidEndSepPunct{\mcitedefaultmidpunct}
{\mcitedefaultendpunct}{\mcitedefaultseppunct}\relax
\EndOfBibitem
\bibitem{Koppenburg:1027442}
P.~Koppenburg,  \ifthenelse{\boolean{articletitles}}{\emph{{Selection of \BllK
  at LHCb and Sensitvity to \RK}}}{},
  \href{https://cds.cern.ch/record/1027442}{LHCb-2007-034}, CERN, Geneva,
  2007\relax
\mciteBstWouldAddEndPuncttrue
\mciteSetBstMidEndSepPunct{\mcitedefaultmidpunct}
{\mcitedefaultendpunct}{\mcitedefaultseppunct}\relax
\EndOfBibitem
\bibitem{Dijkstra:691698}
H.~Dijkstra, H.~J. Hilke, T.~Nakada, and T.~Ypsilantis,
  \ifthenelse{\boolean{articletitles}}{\emph{{LHCb Letter of Intent, LHCb
  Collaboration}}, }{}
  \href{http://cdsweb.cern.ch/search?p=LHCb-95-001&f=reportnumber&action_search=Search&c=LHCb}
  {LHCb-95-001}, 1995\relax
\mciteBstWouldAddEndPuncttrue
\mciteSetBstMidEndSepPunct{\mcitedefaultmidpunct}
{\mcitedefaultendpunct}{\mcitedefaultseppunct}\relax
\EndOfBibitem
\bibitem{CERN-LHCC-98-004}
\ifthenelse{\boolean{articletitles}}{\emph{{LHCb : Technical Proposal}}, }{}
  \href{http://cdsweb.cern.ch/search?p=CERN-LHCC-98-004&f=reportnumber&action_search=Search&c=LHCb}
  {CERN-LHCC-98-004}, 1998\relax
\mciteBstWouldAddEndPuncttrue
\mciteSetBstMidEndSepPunct{\mcitedefaultmidpunct}
{\mcitedefaultendpunct}{\mcitedefaultseppunct}\relax
\EndOfBibitem
\bibitem{:2009ny}
LHCb collaboration, B.~Adeva {\em et~al.},
  \ifthenelse{\boolean{articletitles}}{\emph{{Roadmap for selected key
  measurements of LHCb}},
  }{}\href{http://arxiv.org/abs/0912.4179}{{\normalfont\ttfamily
  arXiv:0912.4179}}\relax
\mciteBstWouldAddEndPuncttrue
\mciteSetBstMidEndSepPunct{\mcitedefaultmidpunct}
{\mcitedefaultendpunct}{\mcitedefaultseppunct}\relax
\EndOfBibitem
\bibitem{LHCb-TDR-009}
LHCb collaboration, \ifthenelse{\boolean{articletitles}}{\emph{{LHCb
  reoptimized detector design and performance: Technical Design Report}}, }{}
  \href{http://cdsweb.cern.ch/search?p=CERN-LHCC-2003-030&f=reportnumber&action_search=Search&c=LHCb}
  {CERN-LHCC-2003-030}, 2003\relax
\mciteBstWouldAddEndPuncttrue
\mciteSetBstMidEndSepPunct{\mcitedefaultmidpunct}
{\mcitedefaultendpunct}{\mcitedefaultseppunct}\relax
\EndOfBibitem
\bibitem{Skidmore}
N.~A. Skidmore.
\newblock Unpublished\relax
\mciteBstWouldAddEndPuncttrue
\mciteSetBstMidEndSepPunct{\mcitedefaultmidpunct}
{\mcitedefaultendpunct}{\mcitedefaultseppunct}\relax
\EndOfBibitem
\bibitem{LHCb-PAPER-2014-024}
LHCb collaboration, R.~Aaij {\em et~al.},
  \ifthenelse{\boolean{articletitles}}{\emph{{Test of lepton universality using
  \mbox{\decay{\Bp}{\Kp\ellell}} decays}},
  }{}\href{https://doi.org/10.1103/PhysRevLett.113.151601}{Phys.\ Rev.\ Lett.\
  \textbf{113} (2014) 151601},
  \href{http://arxiv.org/abs/1406.6482}{{\normalfont\ttfamily
  arXiv:1406.6482}}\relax
\mciteBstWouldAddEndPuncttrue
\mciteSetBstMidEndSepPunct{\mcitedefaultmidpunct}
{\mcitedefaultendpunct}{\mcitedefaultseppunct}\relax
\EndOfBibitem
\bibitem{LHCb-PAPER-2019-040}
LHCb collaboration, R.~Aaij {\em et~al.},
  \ifthenelse{\boolean{articletitles}}{\emph{{Test of lepton universality using
  \mbox{\decay{\Lb}{p\Km\ellell}} decays}},
  }{}\href{https://doi.org/10.1007/JHEP05(2020)040}{JHEP \textbf{05} (2020)
  040}, \href{http://arxiv.org/abs/1912.08139}{{\normalfont\ttfamily
  arXiv:1912.08139}}\relax
\mciteBstWouldAddEndPuncttrue
\mciteSetBstMidEndSepPunct{\mcitedefaultmidpunct}
{\mcitedefaultendpunct}{\mcitedefaultseppunct}\relax
\EndOfBibitem
\bibitem{LHCb-PAPER-2019-009}
LHCb collaboration, R.~Aaij {\em et~al.},
  \ifthenelse{\boolean{articletitles}}{\emph{{Search for lepton-universality
  violation in \mbox{\decay{\Bp}{\Kp\ellell}} decays}},
  }{}\href{https://doi.org/10.1103/PhysRevLett.122.191801}{Phys.\ Rev.\ Lett.\
  \textbf{122} (2019) 191801},
  \href{http://arxiv.org/abs/1903.09252}{{\normalfont\ttfamily
  arXiv:1903.09252}}\relax
\mciteBstWouldAddEndPuncttrue
\mciteSetBstMidEndSepPunct{\mcitedefaultmidpunct}
{\mcitedefaultendpunct}{\mcitedefaultseppunct}\relax
\EndOfBibitem
\bibitem{LHCb-PAPER-2021-014}
LHCb collaboration, R.~Aaij {\em et~al.},
  \ifthenelse{\boolean{articletitles}}{\emph{{Branching fraction measurements
  of the rare $B^0_s \to \phi \mu^+\mu^-$ and $B^0_s \to f_2^\prime(1525)
  \mu^+\mu^-$ decays}},
  }{}\href{https://doi.org/10.1103/PhysRevLett.127.151801}{Phys.\ Rev.\ Lett.\
  \textbf{127} (2021) 151801},
  \href{http://arxiv.org/abs/2105.14007}{{\normalfont\ttfamily
  arXiv:2105.14007}}\relax
\mciteBstWouldAddEndPuncttrue
\mciteSetBstMidEndSepPunct{\mcitedefaultmidpunct}
{\mcitedefaultendpunct}{\mcitedefaultseppunct}\relax
\EndOfBibitem
\bibitem{Alguero:2023jeh}
M.~Alguer\'o {\em et~al.}, \ifthenelse{\boolean{articletitles}}{\emph{{To (b)e
  or not to (b)e: No electrons at LHCb}},
  }{}\href{http://arxiv.org/abs/2304.07330}{{\normalfont\ttfamily
  arXiv:2304.07330}}\relax
\mciteBstWouldAddEndPuncttrue
\mciteSetBstMidEndSepPunct{\mcitedefaultmidpunct}
{\mcitedefaultendpunct}{\mcitedefaultseppunct}\relax
\EndOfBibitem
\bibitem{Gubernari:2022hxn}
N.~Gubernari, M.~Reboud, D.~van Dyk, and J.~Virto,
  \ifthenelse{\boolean{articletitles}}{\emph{{Improved theory predictions and
  global analysis of exclusive $b \to s\mu^+\mu^-$ processes}},
  }{}\href{https://doi.org/10.1007/JHEP09(2022)133}{JHEP \textbf{09} (2022)
  133}, \href{http://arxiv.org/abs/2206.03797}{{\normalfont\ttfamily
  arXiv:2206.03797}}\relax
\mciteBstWouldAddEndPuncttrue
\mciteSetBstMidEndSepPunct{\mcitedefaultmidpunct}
{\mcitedefaultendpunct}{\mcitedefaultseppunct}\relax
\EndOfBibitem
\bibitem{DeBlas:2019ehy}
J.~De~Blas {\em et~al.},
  \ifthenelse{\boolean{articletitles}}{\emph{{$\texttt{HEPfit}$: a code for the
  combination of indirect and direct constraints on high energy physics
  models}}, }{}\href{https://doi.org/10.1140/epjc/s10052-020-7904-z}{Eur.\
  Phys.\ J.\  \textbf{C80} (2020) 456},
  \href{http://arxiv.org/abs/1910.14012}{{\normalfont\ttfamily
  arXiv:1910.14012}}\relax
\mciteBstWouldAddEndPuncttrue
\mciteSetBstMidEndSepPunct{\mcitedefaultmidpunct}
{\mcitedefaultendpunct}{\mcitedefaultseppunct}\relax
\EndOfBibitem
\bibitem{Isidori:2021vtc}
G.~Isidori, D.~Lancierini, P.~Owen, and N.~Serra,
  \ifthenelse{\boolean{articletitles}}{\emph{{On the significance of new
  physics in \blls decays}},
  }{}\href{https://doi.org/10.1016/j.physletb.2021.136644}{Phys.\ Lett.\
  \textbf{B822} (2021) 136644},
  \href{http://arxiv.org/abs/2104.05631}{{\normalfont\ttfamily
  arXiv:2104.05631}}\relax
\mciteBstWouldAddEndPuncttrue
\mciteSetBstMidEndSepPunct{\mcitedefaultmidpunct}
{\mcitedefaultendpunct}{\mcitedefaultseppunct}\relax
\EndOfBibitem
\bibitem{Hurth:2021nsi}
T.~Hurth, F.~Mahmoudi, D.~M. Santos, and S.~Neshatpour,
  \ifthenelse{\boolean{articletitles}}{\emph{{More indications for lepton
  bonuniversality in $b \to s \ell^+ \ell^-$}},
  }{}\href{https://doi.org/10.1016/j.physletb.2021.136838}{Phys.\ Lett.\
  \textbf{B824} (2022) 136838},
  \href{http://arxiv.org/abs/2104.10058}{{\normalfont\ttfamily
  arXiv:2104.10058}}\relax
\mciteBstWouldAddEndPuncttrue
\mciteSetBstMidEndSepPunct{\mcitedefaultmidpunct}
{\mcitedefaultendpunct}{\mcitedefaultseppunct}\relax
\EndOfBibitem
\bibitem{Altmannshofer:2021qrr}
W.~Altmannshofer and P.~Stangl, \ifthenelse{\boolean{articletitles}}{\emph{{New
  physics in rare B decays after Moriond 2021}},
  }{}\href{https://doi.org/10.1140/epjc/s10052-021-09725-1}{Eur.\ Phys.\ J.\ C
  \textbf{81} (2021) 952},
  \href{http://arxiv.org/abs/2103.13370}{{\normalfont\ttfamily
  arXiv:2103.13370}}\relax
\mciteBstWouldAddEndPuncttrue
\mciteSetBstMidEndSepPunct{\mcitedefaultmidpunct}
{\mcitedefaultendpunct}{\mcitedefaultseppunct}\relax
\EndOfBibitem
\bibitem{Geng:2021nhg}
L.-S. Geng {\em et~al.},
  \ifthenelse{\boolean{articletitles}}{\emph{{Implications of new evidence for
  lepton-universality violation in $b\to s\ell^+\ell^-$ decays}},
  }{}\href{http://arxiv.org/abs/2103.12738}{{\normalfont\ttfamily
  arXiv:2103.12738}}\relax
\mciteBstWouldAddEndPuncttrue
\mciteSetBstMidEndSepPunct{\mcitedefaultmidpunct}
{\mcitedefaultendpunct}{\mcitedefaultseppunct}\relax
\EndOfBibitem
\bibitem{Angelescu:2021lln}
A.~Angelescu {\em et~al.}, \ifthenelse{\boolean{articletitles}}{\emph{{Single
  leptoquark solutions to the B-physics anomalies}},
  }{}\href{https://doi.org/10.1103/PhysRevD.104.055017}{Phys.\ Rev.\
  \textbf{D104} (2021) 055017},
  \href{http://arxiv.org/abs/2103.12504}{{\normalfont\ttfamily
  arXiv:2103.12504}}\relax
\mciteBstWouldAddEndPuncttrue
\mciteSetBstMidEndSepPunct{\mcitedefaultmidpunct}
{\mcitedefaultendpunct}{\mcitedefaultseppunct}\relax
\EndOfBibitem
\bibitem{Ciuchini:2022wbq}
M.~Ciuchini {\em et~al.},
  \ifthenelse{\boolean{articletitles}}{\emph{{Constraints on lepton
  universality violation from rare B decays}},
  }{}\href{https://doi.org/10.1103/PhysRevD.107.055036}{Phys.\ Rev.\
  \textbf{D107} (2023) 055036},
  \href{http://arxiv.org/abs/2212.10516}{{\normalfont\ttfamily
  arXiv:2212.10516}}\relax
\mciteBstWouldAddEndPuncttrue
\mciteSetBstMidEndSepPunct{\mcitedefaultmidpunct}
{\mcitedefaultendpunct}{\mcitedefaultseppunct}\relax
\EndOfBibitem
\bibitem{Greljo:2022jac}
A.~Greljo, J.~Salko, A.~Smolkovi\v{c}, and P.~Stangl,
  \ifthenelse{\boolean{articletitles}}{\emph{{Rare b decays meet high-mass
  Drell-Yan}}, }{}\href{https://doi.org/10.1007/JHEP05(2023)087}{JHEP
  \textbf{05} (2023) 087},
  \href{http://arxiv.org/abs/2212.10497}{{\normalfont\ttfamily
  arXiv:2212.10497}}\relax
\mciteBstWouldAddEndPuncttrue
\mciteSetBstMidEndSepPunct{\mcitedefaultmidpunct}
{\mcitedefaultendpunct}{\mcitedefaultseppunct}\relax
\EndOfBibitem
\bibitem{Khodjamirian:2012rm}
A.~Khodjamirian, T.~Mannel, and Y.~M. Wang,
  \ifthenelse{\boolean{articletitles}}{\emph{{$B \to K \ell^{+}\ell^{-}$ decay
  at large hadronic recoil}},
  }{}\href{https://doi.org/10.1007/JHEP02(2013)010}{JHEP \textbf{02} (2013)
  010}, \href{http://arxiv.org/abs/1211.0234}{{\normalfont\ttfamily
  arXiv:1211.0234}}\relax
\mciteBstWouldAddEndPuncttrue
\mciteSetBstMidEndSepPunct{\mcitedefaultmidpunct}
{\mcitedefaultendpunct}{\mcitedefaultseppunct}\relax
\EndOfBibitem
\bibitem{B2Knunu}
Belle II collaboration, A.~Glazov,
  \ifthenelse{\boolean{articletitles}}{\emph{{Belle II highlights}}, }{}
  \href{https://indico.desy.de/event/34916/timetable/?view=standard\#833-news-from-belle-2}{Talk
  at EPS-HEP 2023}\relax
\mciteBstWouldAddEndPuncttrue
\mciteSetBstMidEndSepPunct{\mcitedefaultmidpunct}
{\mcitedefaultendpunct}{\mcitedefaultseppunct}\relax
\EndOfBibitem
\bibitem{BaBar:2012obs}
BaBar collaboration, J.~P. Lees {\em et~al.},
  \ifthenelse{\boolean{articletitles}}{\emph{{Evidence for an excess of $\Bb
  \to D^{(*)} \tau^-\bar{\nu}_\tau$ decays}},
  }{}\href{https://doi.org/10.1103/PhysRevLett.109.101802}{Phys.\ Rev.\ Lett.\
  \textbf{109} (2012) 101802},
  \href{http://arxiv.org/abs/1205.5442}{{\normalfont\ttfamily
  arXiv:1205.5442}}\relax
\mciteBstWouldAddEndPuncttrue
\mciteSetBstMidEndSepPunct{\mcitedefaultmidpunct}
{\mcitedefaultendpunct}{\mcitedefaultseppunct}\relax
\EndOfBibitem
\bibitem{BaBar:2013mob}
BaBar collaboration, J.~P. Lees {\em et~al.},
  \ifthenelse{\boolean{articletitles}}{\emph{{Measurement of an excess of $\Bb
  \to D^{(*)}\tau^- \bar{\nu}_\tau$ decays and implications for charged Higgs
  bosons}}, }{}\href{https://doi.org/10.1103/PhysRevD.88.072012}{Phys.\ Rev.\
  \textbf{D88} (2013) 072012},
  \href{http://arxiv.org/abs/1303.0571}{{\normalfont\ttfamily
  arXiv:1303.0571}}\relax
\mciteBstWouldAddEndPuncttrue
\mciteSetBstMidEndSepPunct{\mcitedefaultmidpunct}
{\mcitedefaultendpunct}{\mcitedefaultseppunct}\relax
\EndOfBibitem
\bibitem{LHCb-PAPER-2022-039}
LHCb collaboration, R.~Aaij {\em et~al.},
  \ifthenelse{\boolean{articletitles}}{\emph{{Measurement of the ratio of
  branching fractions $\mathcal{R}(\Dstar)$ and $\mathcal{R}(\Dz)$ }},
  }{}\href{https://doi.org/10.1103/PhysRevLett.131.111802}{Phys.\ Rev.\ Lett.\
  \textbf{131} (2023) 111802},
  \href{http://arxiv.org/abs/2302.02886}{{\normalfont\ttfamily
  arXiv:2302.02886}}\relax
\mciteBstWouldAddEndPuncttrue
\mciteSetBstMidEndSepPunct{\mcitedefaultmidpunct}
{\mcitedefaultendpunct}{\mcitedefaultseppunct}\relax
\EndOfBibitem
\bibitem{LHCb-PAPER-2022-052}
LHCb collaboration, R.~Aaij {\em et~al.},
  \ifthenelse{\boolean{articletitles}}{\emph{{Test of lepton flavour
  universality using \mbox{$\Bz\to \Dstarm\taup\neut$} decays, with hadronic
  $\tauon$ channels}},
  }{}\href{https://doi.org/10.1103/PhysRevD.108.012018}{Phys.\ Rev.\
  \textbf{D108} (2023) 012018},
  \href{http://arxiv.org/abs/2305.01463}{{\normalfont\ttfamily
  arXiv:2305.01463}}\relax
\mciteBstWouldAddEndPuncttrue
\mciteSetBstMidEndSepPunct{\mcitedefaultmidpunct}
{\mcitedefaultendpunct}{\mcitedefaultseppunct}\relax
\EndOfBibitem
\bibitem{PKanomalies}
P.~Koppenburg, \ifthenelse{\boolean{articletitles}}{\emph{Flavour anomalies},
  }{}
  \href{https://www.nikhef.nl/~pkoppenb/anomalies.html}{https://www.nikhef.nl/\~pkoppenb/anomalies.html},
  2023\relax
\mciteBstWouldAddEndPuncttrue
\mciteSetBstMidEndSepPunct{\mcitedefaultmidpunct}
{\mcitedefaultendpunct}{\mcitedefaultseppunct}\relax
\EndOfBibitem
\bibitem{LHCb-PAPER-2014-006}
LHCb collaboration, R.~Aaij {\em et~al.},
  \ifthenelse{\boolean{articletitles}}{\emph{{Differential branching fractions
  and isospin asymmetries of \mbox{\decay{\B}{K^{(*)}\mumu}} decays}},
  }{}\href{https://doi.org/10.1007/JHEP06(2014)133}{JHEP \textbf{06} (2014)
  133}, \href{http://arxiv.org/abs/1403.8044}{{\normalfont\ttfamily
  arXiv:1403.8044}}\relax
\mciteBstWouldAddEndPuncttrue
\mciteSetBstMidEndSepPunct{\mcitedefaultmidpunct}
{\mcitedefaultendpunct}{\mcitedefaultseppunct}\relax
\EndOfBibitem
\bibitem{Horgan:2015vla}
R.~R. Horgan, Z.~Liu, S.~Meinel, and M.~Wingate,
  \ifthenelse{\boolean{articletitles}}{\emph{{Rare $B$ decays using lattice QCD
  form factors}}, }{}\href{https://doi.org/10.22323/1.214.0372}{PoS
  \textbf{LATTICE2014} (2015) 372},
  \href{http://arxiv.org/abs/1501.00367}{{\normalfont\ttfamily
  arXiv:1501.00367}}\relax
\mciteBstWouldAddEndPuncttrue
\mciteSetBstMidEndSepPunct{\mcitedefaultmidpunct}
{\mcitedefaultendpunct}{\mcitedefaultseppunct}\relax
\EndOfBibitem
\bibitem{Muong-2:2021ojo}
Muon g-2 collaboration, B.~Abi {\em et~al.},
  \ifthenelse{\boolean{articletitles}}{\emph{{Measurement of the positive muon
  anomalous magnetic moment to 0.46 ppm}},
  }{}\href{https://doi.org/10.1103/PhysRevLett.126.141801}{Phys.\ Rev.\ Lett.\
  \textbf{126} (2021) 141801},
  \href{http://arxiv.org/abs/2104.03281}{{\normalfont\ttfamily
  arXiv:2104.03281}}\relax
\mciteBstWouldAddEndPuncttrue
\mciteSetBstMidEndSepPunct{\mcitedefaultmidpunct}
{\mcitedefaultendpunct}{\mcitedefaultseppunct}\relax
\EndOfBibitem
\bibitem{Aoyama:2020ynm}
T.~Aoyama {\em et~al.}, \ifthenelse{\boolean{articletitles}}{\emph{{The
  anomalous magnetic moment of the muon in the Standard Model}},
  }{}\href{https://doi.org/10.1016/j.physrep.2020.07.006}{Phys.\ Rept.\
  \textbf{887} (2020) 1},
  \href{http://arxiv.org/abs/2006.04822}{{\normalfont\ttfamily
  arXiv:2006.04822}}\relax
\mciteBstWouldAddEndPuncttrue
\mciteSetBstMidEndSepPunct{\mcitedefaultmidpunct}
{\mcitedefaultendpunct}{\mcitedefaultseppunct}\relax
\EndOfBibitem
\bibitem{LHCb-PAPER-2017-035}
LHCb collaboration, R.~Aaij {\em et~al.},
  \ifthenelse{\boolean{articletitles}}{\emph{{Measurement of the ratio of
  branching fractions
  \mbox{$\mathcal{B}(\decay{\Bc}{\jpsi\taup\nu_{\tau}})/\mathcal{B}(\decay{\Bc}{\jpsi\mup\nu_{\mu}})$}}},
  }{}\href{https://doi.org/10.1103/PhysRevLett.120.121801}{Phys.\ Rev.\ Lett.\
  \textbf{120} (2018) 121801},
  \href{http://arxiv.org/abs/1711.05623}{{\normalfont\ttfamily
  arXiv:1711.05623}}\relax
\mciteBstWouldAddEndPuncttrue
\mciteSetBstMidEndSepPunct{\mcitedefaultmidpunct}
{\mcitedefaultendpunct}{\mcitedefaultseppunct}\relax
\EndOfBibitem
\bibitem{Harrison:2020nrv}
HPQCD collaboration, J.~Harrison, C.~T.~H. Davies, and A.~Lytle,
  \ifthenelse{\boolean{articletitles}}{\emph{{$R(J/\psi)$ and $B_c^-
  \rightarrow J/\psi \ell^-\bar{\nu}_\ell$ lepton flavor universality violating
  observables from Lattice QCD}},
  }{}\href{https://doi.org/10.1103/PhysRevLett.125.222003}{Phys.\ Rev.\ Lett.\
  \textbf{125} (2020) 222003},
  \href{http://arxiv.org/abs/2007.06956}{{\normalfont\ttfamily
  arXiv:2007.06956}}\relax
\mciteBstWouldAddEndPuncttrue
\mciteSetBstMidEndSepPunct{\mcitedefaultmidpunct}
{\mcitedefaultendpunct}{\mcitedefaultseppunct}\relax
\EndOfBibitem
\bibitem{LHCb-PAPER-2021-044}
LHCb collaboration, R.~Aaij {\em et~al.},
  \ifthenelse{\boolean{articletitles}}{\emph{{Observation of the decay $\Lb \to
  \Lc \taum \neub_\tau$}},
  }{}\href{https://doi.org/10.1103/PhysRevLett.128.191803}{Phys.\ Rev.\ Lett.\
  \textbf{128} (2021) 191803},
  \href{http://arxiv.org/abs/2201.03497}{{\normalfont\ttfamily
  arXiv:2201.03497}}\relax
\mciteBstWouldAddEndPuncttrue
\mciteSetBstMidEndSepPunct{\mcitedefaultmidpunct}
{\mcitedefaultendpunct}{\mcitedefaultseppunct}\relax
\EndOfBibitem
\bibitem{Bernlochner:2018bfn}
F.~U. Bernlochner, Z.~Ligeti, D.~J. Robinson, and W.~L. Sutcliffe,
  \ifthenelse{\boolean{articletitles}}{\emph{{Precise predictions for $\Lb \to
  \Lc$ semileptonic decays}},
  }{}\href{https://doi.org/10.1103/PhysRevD.99.055008}{Phys.\ Rev.\
  \textbf{D99} (2019) 055008},
  \href{http://arxiv.org/abs/1812.07593}{{\normalfont\ttfamily
  arXiv:1812.07593}}\relax
\mciteBstWouldAddEndPuncttrue
\mciteSetBstMidEndSepPunct{\mcitedefaultmidpunct}
{\mcitedefaultendpunct}{\mcitedefaultseppunct}\relax
\EndOfBibitem
\bibitem{PDG2022}
Particle Data Group, R.~L. Workman {\em et~al.},
  \ifthenelse{\boolean{articletitles}}{\emph{{\href{http://pdg.lbl.gov/}{Review
  of particle physics}}}, }{}\href{https://doi.org/10.1093/ptep/ptac097}{Prog.\
  Theor.\ Exp.\ Phys.\  \textbf{2022} (2022) 083C01}\relax
\mciteBstWouldAddEndPuncttrue
\mciteSetBstMidEndSepPunct{\mcitedefaultmidpunct}
{\mcitedefaultendpunct}{\mcitedefaultseppunct}\relax
\EndOfBibitem
\bibitem{UTfit:2006vpt}
UTfit collaboration, M.~Bona {\em et~al.},
  \ifthenelse{\boolean{articletitles}}{\emph{{The Unitarity Triangle fit in the
  Standard Model and hadronic parameters from Lattice QCD: A reappraisal after
  the measurements of $\Delta m_s$ and $\BR(B\to\tau\nu_tau)$}},
  }{}\href{https://doi.org/10.1088/1126-6708/2006/10/081}{JHEP \textbf{10}
  (2006) 081}, \href{http://arxiv.org/abs/hep-ph/0606167}{{\normalfont\ttfamily
  arXiv:hep-ph/0606167}}, and updates at \url{http://www.utfit.org/}\relax
\mciteBstWouldAddEndPuncttrue
\mciteSetBstMidEndSepPunct{\mcitedefaultmidpunct}
{\mcitedefaultendpunct}{\mcitedefaultseppunct}\relax
\EndOfBibitem
\bibitem{DiLuzio:2019jyq}
L.~Di~Luzio, M.~Kirk, A.~Lenz, and T.~Rauh,
  \ifthenelse{\boolean{articletitles}}{\emph{{$\Delta M_s$ theory precision
  confronts flavour anomalies}},
  }{}\href{https://doi.org/10.1007/JHEP12(2019)009}{JHEP \textbf{12} (2019)
  009}, \href{http://arxiv.org/abs/1909.11087}{{\normalfont\ttfamily
  arXiv:1909.11087}}\relax
\mciteBstWouldAddEndPuncttrue
\mciteSetBstMidEndSepPunct{\mcitedefaultmidpunct}
{\mcitedefaultendpunct}{\mcitedefaultseppunct}\relax
\EndOfBibitem
\bibitem{ALEPH:2005ab}
SLD Electroweak Group, DELPHI, ALEPH, SLD, SLD Heavy Flavour Group, OPAL, LEP
  Electroweak Working Group, L3, S.~Schael {\em et~al.},
  \ifthenelse{\boolean{articletitles}}{\emph{{Precision electroweak
  measurements on the $Z$ resonance}},
  }{}\href{https://doi.org/10.1016/j.physrep.2005.12.006}{Phys.\ Rept.\
  \textbf{427} (2006) 257},
  \href{http://arxiv.org/abs/hep-ex/0509008}{{\normalfont\ttfamily
  arXiv:hep-ex/0509008}}\relax
\mciteBstWouldAddEndPuncttrue
\mciteSetBstMidEndSepPunct{\mcitedefaultmidpunct}
{\mcitedefaultendpunct}{\mcitedefaultseppunct}\relax
\EndOfBibitem
\bibitem{ConversationHadrons}
P.~Koppenburg and H.~Cliff, \ifthenelse{\boolean{articletitles}}{\emph{{CERN:
  scientists discover four new particles -- here's why they matter}},
  }{}\href{https://theconversation.\
  com/cern-scientists-discover-four-new-particles-heres-why-they-matter-155800}{The
  Conversation} (2021)\relax
\mciteBstWouldAddEndPuncttrue
\mciteSetBstMidEndSepPunct{\mcitedefaultmidpunct}
{\mcitedefaultendpunct}{\mcitedefaultseppunct}\relax
\EndOfBibitem
\bibitem{LHCb-FIGURE-2021-001}
LHCb collaboration, P.~Koppenburg,
  \ifthenelse{\boolean{articletitles}}{\emph{{List of hadrons observed at the
  LHC}}, }{}
  \href{http://cdsweb.cern.ch/search?p=LHCb-FIGURE-2021-001&f=reportnumber&action_search=Search&c=LHCb+Figures}
  {LHCb-FIGURE-2021-001}, 2021, and
  {\href{https://www.nikhef.nl/~pkoppenb/particles.html}{2023 updates}}\relax
\mciteBstWouldAddEndPuncttrue
\mciteSetBstMidEndSepPunct{\mcitedefaultmidpunct}
{\mcitedefaultendpunct}{\mcitedefaultseppunct}\relax
\EndOfBibitem
\bibitem{Choi:2003ue}
Belle collaboration, S.~K. Choi {\em et~al.},
  \ifthenelse{\boolean{articletitles}}{\emph{{Observation of a new narrow
  charmonium state in exclusive $\Bp\to\Kp{}\pip{}\pim{}\jpsi$ decays}},
  }{}Phys.\ Rev.\ Lett.\  \textbf{91} (2003) 262001,
  \href{http://arxiv.org/abs/hep-ex/0309032}{{\normalfont\ttfamily
  arXiv:hep-ex/0309032}}\relax
\mciteBstWouldAddEndPuncttrue
\mciteSetBstMidEndSepPunct{\mcitedefaultmidpunct}
{\mcitedefaultendpunct}{\mcitedefaultseppunct}\relax
\EndOfBibitem
\bibitem{Lebed:2016hpi}
R.~F. Lebed, R.~E. Mitchell, and E.~S. Swanson,
  \ifthenelse{\boolean{articletitles}}{\emph{{Heavy-Quark QCD Exotica}},
  }{}\href{https://doi.org/10.1016/j.ppnp.2016.11.003}{Prog.\ Part.\ Nucl.\
  Phys.\  \textbf{93} (2017) 143},
  \href{http://arxiv.org/abs/1610.04528}{{\normalfont\ttfamily
  arXiv:1610.04528}}\relax
\mciteBstWouldAddEndPuncttrue
\mciteSetBstMidEndSepPunct{\mcitedefaultmidpunct}
{\mcitedefaultendpunct}{\mcitedefaultseppunct}\relax
\EndOfBibitem
\bibitem{Belle:2006olv}
Belle collaboration, G.~Gokhroo {\em et~al.},
  \ifthenelse{\boolean{articletitles}}{\emph{{Observation of a near-threshold
  \Dz\Dzb\piz enhancement in \decay{\B}{\Dz\Dzb\piz K} decay}},
  }{}\href{https://doi.org/10.1103/PhysRevLett.97.162002}{Phys.\ Rev.\ Lett.\
  \textbf{97} (2006) 162002},
  \href{http://arxiv.org/abs/hep-ex/0606055}{{\normalfont\ttfamily
  arXiv:hep-ex/0606055}}\relax
\mciteBstWouldAddEndPuncttrue
\mciteSetBstMidEndSepPunct{\mcitedefaultmidpunct}
{\mcitedefaultendpunct}{\mcitedefaultseppunct}\relax
\EndOfBibitem
\bibitem{Belle:2011vlx}
Belle collaboration, S.-K. Choi {\em et~al.},
  \ifthenelse{\boolean{articletitles}}{\emph{{Bounds on the width, mass
  difference and other properties of $X(3872) \to \pi^+ \pi^- \jpsi$ decays}},
  }{}\href{https://doi.org/10.1103/PhysRevD.84.052004}{Phys.\ Rev.\
  \textbf{D84} (2011) 052004},
  \href{http://arxiv.org/abs/1107.0163}{{\normalfont\ttfamily
  arXiv:1107.0163}}\relax
\mciteBstWouldAddEndPuncttrue
\mciteSetBstMidEndSepPunct{\mcitedefaultmidpunct}
{\mcitedefaultendpunct}{\mcitedefaultseppunct}\relax
\EndOfBibitem
\bibitem{LHCb-PAPER-2021-045}
LHCb collaboration, R.~Aaij {\em et~al.},
  \ifthenelse{\boolean{articletitles}}{\emph{{Observation of sizeable $\omegaz$
  contribution to \mbox{$\chicone \to \pip\pim \jpsi$ decays}}},
  }{}\href{https://doi.org/10.1103/PhysRevD.108.L011103}{Phys.\ Rev.\
  \textbf{D108} (2023) L011103},
  \href{http://arxiv.org/abs/2204.12597}{{\normalfont\ttfamily
  arXiv:2204.12597}}\relax
\mciteBstWouldAddEndPuncttrue
\mciteSetBstMidEndSepPunct{\mcitedefaultmidpunct}
{\mcitedefaultendpunct}{\mcitedefaultseppunct}\relax
\EndOfBibitem
\bibitem{CMS:2013fpt}
CMS collaboration, S.~Chatrchyan {\em et~al.},
  \ifthenelse{\boolean{articletitles}}{\emph{{Measurement of the $X$(3872)
  production cross section via decays to $\jpsi \pi^+ \pi^-$ in $pp$ collisions
  at $\sqrt{s}$ = 7 TeV}},
  }{}\href{https://doi.org/10.1007/JHEP04(2013)154}{JHEP \textbf{04} (2013)
  154}, \href{http://arxiv.org/abs/1302.3968}{{\normalfont\ttfamily
  arXiv:1302.3968}}\relax
\mciteBstWouldAddEndPuncttrue
\mciteSetBstMidEndSepPunct{\mcitedefaultmidpunct}
{\mcitedefaultendpunct}{\mcitedefaultseppunct}\relax
\EndOfBibitem
\bibitem{ATLAS:2016kwu}
ATLAS collaboration, M.~Aaboud {\em et~al.},
  \ifthenelse{\boolean{articletitles}}{\emph{{Measurements of $\psi(2S)$ and
  $X(3872) \to \jpsi\pi^+\pi^-$ production in $pp$ collisions at $\sqrt{s} = 8$
  TeV with the ATLAS detector}},
  }{}\href{https://doi.org/10.1007/JHEP01(2017)117}{JHEP \textbf{01} (2017)
  117}, \href{http://arxiv.org/abs/1610.09303}{{\normalfont\ttfamily
  arXiv:1610.09303}}\relax
\mciteBstWouldAddEndPuncttrue
\mciteSetBstMidEndSepPunct{\mcitedefaultmidpunct}
{\mcitedefaultendpunct}{\mcitedefaultseppunct}\relax
\EndOfBibitem
\bibitem{LHCb-PAPER-2015-015}
LHCb collaboration, R.~Aaij {\em et~al.},
  \ifthenelse{\boolean{articletitles}}{\emph{{Quantum numbers of the $X(3872)$
  state and orbital angular momentum in its $\rhoz\jpsi$ decays}},
  }{}\href{https://doi.org/10.1103/PhysRevD.92.011102}{Phys.\ Rev.\
  \textbf{D92} (2015) 011102(R)},
  \href{http://arxiv.org/abs/1504.06339}{{\normalfont\ttfamily
  arXiv:1504.06339}}\relax
\mciteBstWouldAddEndPuncttrue
\mciteSetBstMidEndSepPunct{\mcitedefaultmidpunct}
{\mcitedefaultendpunct}{\mcitedefaultseppunct}\relax
\EndOfBibitem
\bibitem{LHCb-PAPER-2013-001}
LHCb collaboration, R.~Aaij {\em et~al.},
  \ifthenelse{\boolean{articletitles}}{\emph{{Determination of the $X(3872)$
  meson quantum numbers}},
  }{}\href{https://doi.org/10.1103/PhysRevLett.110.222001}{Phys.\ Rev.\ Lett.\
  \textbf{110} (2013) 222001},
  \href{http://arxiv.org/abs/1302.6269}{{\normalfont\ttfamily
  arXiv:1302.6269}}\relax
\mciteBstWouldAddEndPuncttrue
\mciteSetBstMidEndSepPunct{\mcitedefaultmidpunct}
{\mcitedefaultendpunct}{\mcitedefaultseppunct}\relax
\EndOfBibitem
\bibitem{LHCb-PAPER-2020-008}
LHCb collaboration, R.~Aaij {\em et~al.},
  \ifthenelse{\boolean{articletitles}}{\emph{{Study of the line shape of the
  $\chicone(3872)$ state}},
  }{}\href{https://doi.org/10.1103/PhysRevD.102.092005}{Phys.\ Rev.\
  \textbf{D102} (2020) 092005},
  \href{http://arxiv.org/abs/2005.13419}{{\normalfont\ttfamily
  arXiv:2005.13419}}\relax
\mciteBstWouldAddEndPuncttrue
\mciteSetBstMidEndSepPunct{\mcitedefaultmidpunct}
{\mcitedefaultendpunct}{\mcitedefaultseppunct}\relax
\EndOfBibitem
\bibitem{Belle:2023zxm}
Belle collaboration, H.~Hirata {\em et~al.},
  \ifthenelse{\boolean{articletitles}}{\emph{{Study of the lineshape of
  $X(3872)$ using \B decays to $\Dz\Dstarzb\kaon$}},
  }{}\href{https://doi.org/10.1103/PhysRevD.107.112011}{Phys.\ Rev.\
  \textbf{D107} (2023) 112011},
  \href{http://arxiv.org/abs/2302.02127}{{\normalfont\ttfamily
  arXiv:2302.02127}}\relax
\mciteBstWouldAddEndPuncttrue
\mciteSetBstMidEndSepPunct{\mcitedefaultmidpunct}
{\mcitedefaultendpunct}{\mcitedefaultseppunct}\relax
\EndOfBibitem
\bibitem{LHCb-PAPER-2021-026}
LHCb collaboration, R.~Aaij {\em et~al.},
  \ifthenelse{\boolean{articletitles}}{\emph{{Measurement of $\chi_{c1}(3872)$
  production in proton-proton collisions at $\sqrt{s} = 8$ and 13 TeV}},
  }{}\href{https://doi.org/10.1007/JHEP01(2022)131}{JHEP \textbf{01} (2022)
  131}, \href{http://arxiv.org/abs/2109.07360}{{\normalfont\ttfamily
  arXiv:2109.07360}}\relax
\mciteBstWouldAddEndPuncttrue
\mciteSetBstMidEndSepPunct{\mcitedefaultmidpunct}
{\mcitedefaultendpunct}{\mcitedefaultseppunct}\relax
\EndOfBibitem
\bibitem{Weinberg:1965zz}
S.~Weinberg, \ifthenelse{\boolean{articletitles}}{\emph{{Evidence that the
  deuteron is not an elementary particle}},
  }{}\href{https://doi.org/10.1103/PhysRev.137.B672}{Phys.\ Rev.\  \textbf{137}
  (1965) B672}\relax
\mciteBstWouldAddEndPuncttrue
\mciteSetBstMidEndSepPunct{\mcitedefaultmidpunct}
{\mcitedefaultendpunct}{\mcitedefaultseppunct}\relax
\EndOfBibitem
\bibitem{LHCb-PAPER-2011-033}
LHCb collaboration, R.~Aaij {\em et~al.},
  \ifthenelse{\boolean{articletitles}}{\emph{{Search for the $X(4140)$ state in
  \mbox{\decay{\Bp}{\jpsi\phiz\Kp}} decays}},
  }{}\href{https://doi.org/10.1103/PhysRevD.85.091103}{Phys.\ Rev.\
  \textbf{D85} (2012) 091103(R)},
  \href{http://arxiv.org/abs/1202.5087}{{\normalfont\ttfamily
  arXiv:1202.5087}}\relax
\mciteBstWouldAddEndPuncttrue
\mciteSetBstMidEndSepPunct{\mcitedefaultmidpunct}
{\mcitedefaultendpunct}{\mcitedefaultseppunct}\relax
\EndOfBibitem
\bibitem{LHCb-PAPER-2015-029}
LHCb collaboration, R.~Aaij {\em et~al.},
  \ifthenelse{\boolean{articletitles}}{\emph{{Observation of $\jpsi\proton$
  resonances consistent with pentaquark states in
  \mbox{\decay{\Lb}{\jpsi\proton\Km}} decays}},
  }{}\href{https://doi.org/10.1103/PhysRevLett.115.072001}{Phys.\ Rev.\ Lett.\
  \textbf{115} (2015) 072001},
  \href{http://arxiv.org/abs/1507.03414}{{\normalfont\ttfamily
  arXiv:1507.03414}}\relax
\mciteBstWouldAddEndPuncttrue
\mciteSetBstMidEndSepPunct{\mcitedefaultmidpunct}
{\mcitedefaultendpunct}{\mcitedefaultseppunct}\relax
\EndOfBibitem
\bibitem{Gershon:2022xnn}
LHCb collaboration, T.~Gershon,
  \ifthenelse{\boolean{articletitles}}{\emph{{Exotic hadron naming
  convention}}, }{}\href{http://arxiv.org/abs/2206.15233}{{\normalfont\ttfamily
  arXiv:2206.15233}}\relax
\mciteBstWouldAddEndPuncttrue
\mciteSetBstMidEndSepPunct{\mcitedefaultmidpunct}
{\mcitedefaultendpunct}{\mcitedefaultseppunct}\relax
\EndOfBibitem
\bibitem{LHCb-PAPER-2022-031}
LHCb collaboration, R.~Aaij {\em et~al.},
  \ifthenelse{\boolean{articletitles}}{\emph{{Observation of a
  $\jpsi\Lambdares$ resonance consistent with a strange pentaquark candidate in
  $\Bm\to \jpsi\Lambdares\antiproton$ decays}},
  }{}\href{https://doi.org/10.1103/PhysRevLett.131.031901}{Phys.\ Rev.\ Lett.\
  \textbf{131} (2023) 031901},
  \href{http://arxiv.org/abs/2210.10346}{{\normalfont\ttfamily
  arXiv:2210.10346}}\relax
\mciteBstWouldAddEndPuncttrue
\mciteSetBstMidEndSepPunct{\mcitedefaultmidpunct}
{\mcitedefaultendpunct}{\mcitedefaultseppunct}\relax
\EndOfBibitem
\bibitem{LHCb-PAPER-2020-011}
LHCb collaboration, R.~Aaij {\em et~al.},
  \ifthenelse{\boolean{articletitles}}{\emph{{Observation of structure in the
  \jpsi-pair mass spectrum}},
  }{}\href{https://doi.org/10.1016/j.scib.2020.08.032}{Science Bulletin
  \textbf{65} (2020) 1983},
  \href{http://arxiv.org/abs/2006.16957}{{\normalfont\ttfamily
  arXiv:2006.16957}}\relax
\mciteBstWouldAddEndPuncttrue
\mciteSetBstMidEndSepPunct{\mcitedefaultmidpunct}
{\mcitedefaultendpunct}{\mcitedefaultseppunct}\relax
\EndOfBibitem
\bibitem{CMS:2013jru}
CMS collaboration, S.~Chatrchyan {\em et~al.},
  \ifthenelse{\boolean{articletitles}}{\emph{{Observation of a peaking
  structure in the $\jpsi \phi$ mass spectrum from $B^{\pm} \to \jpsi \phi
  K^{\pm}$ decays}},
  }{}\href{https://doi.org/10.1016/j.physletb.2014.05.055}{Phys.\ Lett.\
  \textbf{B734} (2014) 261},
  \href{http://arxiv.org/abs/1309.6920}{{\normalfont\ttfamily
  arXiv:1309.6920}}\relax
\mciteBstWouldAddEndPuncttrue
\mciteSetBstMidEndSepPunct{\mcitedefaultmidpunct}
{\mcitedefaultendpunct}{\mcitedefaultseppunct}\relax
\EndOfBibitem
\bibitem{LHCb-PAPER-2016-019}
LHCb collaboration, R.~Aaij {\em et~al.},
  \ifthenelse{\boolean{articletitles}}{\emph{{Amplitude analysis of
  \mbox{\decay{\Bp}{\jpsi\phi\Kp}} decays}},
  }{}\href{https://doi.org/10.1103/PhysRevD.95.012002}{Phys.\ Rev.\
  \textbf{D95} (2017) 012002},
  \href{http://arxiv.org/abs/1606.07898}{{\normalfont\ttfamily
  arXiv:1606.07898}}\relax
\mciteBstWouldAddEndPuncttrue
\mciteSetBstMidEndSepPunct{\mcitedefaultmidpunct}
{\mcitedefaultendpunct}{\mcitedefaultseppunct}\relax
\EndOfBibitem
\bibitem{CMS:2023owd}
CMS collaboration, A.~Hayrapetyan {\em et~al.},
  \ifthenelse{\boolean{articletitles}}{\emph{{Observation of new structure in
  the J/$\psi$J/$\psi$ mass spectrum in proton-proton collisions at $\sqrt{s}$
  = 13 TeV}}, }{}\href{http://arxiv.org/abs/2306.07164}{{\normalfont\ttfamily
  arXiv:2306.07164}}\relax
\mciteBstWouldAddEndPuncttrue
\mciteSetBstMidEndSepPunct{\mcitedefaultmidpunct}
{\mcitedefaultendpunct}{\mcitedefaultseppunct}\relax
\EndOfBibitem
\bibitem{ATLAS:2023bft}
ATLAS collaboration, G.~Aad {\em et~al.},
  \ifthenelse{\boolean{articletitles}}{\emph{{Observation of an excess of
  di-charmonium events in the four-muon final state with the ATLAS detector}},
  }{}\href{http://arxiv.org/abs/2304.08962}{{\normalfont\ttfamily
  arXiv:2304.08962}}\relax
\mciteBstWouldAddEndPuncttrue
\mciteSetBstMidEndSepPunct{\mcitedefaultmidpunct}
{\mcitedefaultendpunct}{\mcitedefaultseppunct}\relax
\EndOfBibitem
\bibitem{LHCb-PAPER-2022-040}
LHCb collaboration, R.~Aaij {\em et~al.},
  \ifthenelse{\boolean{articletitles}}{\emph{{Evidence of a $\jpsi \KS $
  structure in $\Bz \to \jpsi \phiz \KS $ decays}},
  }{}\href{https://doi.org/10.1103/PhysRevLett.131.131901}{Phys.\ Rev.\ Lett.\
  \textbf{131} (2023) 131901},
  \href{http://arxiv.org/abs/2301.04899}{{\normalfont\ttfamily
  arXiv:2301.04899}}\relax
\mciteBstWouldAddEndPuncttrue
\mciteSetBstMidEndSepPunct{\mcitedefaultmidpunct}
{\mcitedefaultendpunct}{\mcitedefaultseppunct}\relax
\EndOfBibitem
\bibitem{LHCb-PAPER-2020-044}
LHCb collaboration, R.~Aaij {\em et~al.},
  \ifthenelse{\boolean{articletitles}}{\emph{{Observation of new resonances
  decaying to $ \jpsi K^+$ and $ \jpsi \phi$ }},
  }{}\href{https://doi.org/10.1103/PhysRevLett.127.082001}{Phys.\ Rev.\ Lett.\
  \textbf{127} (2021) 082001},
  \href{http://arxiv.org/abs/2103.01803}{{\normalfont\ttfamily
  arXiv:2103.01803}}\relax
\mciteBstWouldAddEndPuncttrue
\mciteSetBstMidEndSepPunct{\mcitedefaultmidpunct}
{\mcitedefaultendpunct}{\mcitedefaultseppunct}\relax
\EndOfBibitem
\bibitem{LHCb-PAPER-2022-026}
LHCb collaboration, R.~Aaij {\em et~al.},
  \ifthenelse{\boolean{articletitles}}{\emph{{First observation of a doubly
  charged tetraquark candidate and its neutral partner}},
  }{}\href{https://doi.org/10.1103/PhysRevLett.131.041902}{Phys.\ Rev.\ Lett.\
  \textbf{131} (2023) 041902},
  \href{http://arxiv.org/abs/2212.02716}{{\normalfont\ttfamily
  arXiv:2212.02716}}\relax
\mciteBstWouldAddEndPuncttrue
\mciteSetBstMidEndSepPunct{\mcitedefaultmidpunct}
{\mcitedefaultendpunct}{\mcitedefaultseppunct}\relax
\EndOfBibitem
\bibitem{LHCb-PAPER-2022-027}
LHCb collaboration, R.~Aaij {\em et~al.},
  \ifthenelse{\boolean{articletitles}}{\emph{{Amplitude analysis of $\Bz
  \rightarrow \Dzb \Dsp \pim$ and $\Bp \rightarrow \Dm \Dsp\pip$ decays}},
  }{}\href{https://doi.org/10.1103/PhysRevD.108.012017}{Phys.\ Rev.\
  \textbf{D108} (2023) 012017},
  \href{http://arxiv.org/abs/2212.02717}{{\normalfont\ttfamily
  arXiv:2212.02717}}\relax
\mciteBstWouldAddEndPuncttrue
\mciteSetBstMidEndSepPunct{\mcitedefaultmidpunct}
{\mcitedefaultendpunct}{\mcitedefaultseppunct}\relax
\EndOfBibitem
\bibitem{LHCb-PAPER-2020-025}
LHCb collaboration, R.~Aaij {\em et~al.},
  \ifthenelse{\boolean{articletitles}}{\emph{{Amplitude analysis of the $\Bp
  \to \Dp \Dm \Kp$ decay}},
  }{}\href{https://doi.org/10.1103/PhysRevD.102.112003}{Phys.\ Rev.\
  \textbf{D102} (2020) 112003},
  \href{http://arxiv.org/abs/2009.00026}{{\normalfont\ttfamily
  arXiv:2009.00026}}\relax
\mciteBstWouldAddEndPuncttrue
\mciteSetBstMidEndSepPunct{\mcitedefaultmidpunct}
{\mcitedefaultendpunct}{\mcitedefaultseppunct}\relax
\EndOfBibitem
\bibitem{LHCb-PAPER-2021-031}
LHCb collaboration, R.~Aaij {\em et~al.},
  \ifthenelse{\boolean{articletitles}}{\emph{{Observation of an exotic narrow
  doubly charmed tetraquark}},
  }{}\href{https://doi.org/10.1038/s41567-022-01614-y}{Nature Physics
  \textbf{18} (2022) 751},
  \href{http://arxiv.org/abs/2109.01038}{{\normalfont\ttfamily
  arXiv:2109.01038}}\relax
\mciteBstWouldAddEndPuncttrue
\mciteSetBstMidEndSepPunct{\mcitedefaultmidpunct}
{\mcitedefaultendpunct}{\mcitedefaultseppunct}\relax
\EndOfBibitem
\bibitem{LHCb-PAPER-2021-032}
LHCb collaboration, R.~Aaij {\em et~al.},
  \ifthenelse{\boolean{articletitles}}{\emph{{Study of the doubly charmed
  tetraquark $T^+_{cc}$}},
  }{}\href{https://doi.org/10.1038/s41467-022-30206-w}{Nature Communications
  \textbf{13} (2022) 3351},
  \href{http://arxiv.org/abs/2109.01056}{{\normalfont\ttfamily
  arXiv:2109.01056}}\relax
\mciteBstWouldAddEndPuncttrue
\mciteSetBstMidEndSepPunct{\mcitedefaultmidpunct}
{\mcitedefaultendpunct}{\mcitedefaultseppunct}\relax
\EndOfBibitem
\bibitem{Karliner}
M.~Karliner.
  \href{https://indico.cern.ch/event/1201720/\#13-theory-new-pentaquarks}{Talk
  at Exotic hadrons in LHCb workshop}\relax
\mciteBstWouldAddEndPuncttrue
\mciteSetBstMidEndSepPunct{\mcitedefaultmidpunct}
{\mcitedefaultendpunct}{\mcitedefaultseppunct}\relax
\EndOfBibitem
\bibitem{Brambilla:2022ura}
N.~Brambilla {\em et~al.},
  \ifthenelse{\boolean{articletitles}}{\emph{{Substructure of Multiquark
  Hadrons (Snowmass 2021 White Paper)}},
  }{}\href{http://arxiv.org/abs/2203.16583}{{\normalfont\ttfamily
  arXiv:2203.16583}}\relax
\mciteBstWouldAddEndPuncttrue
\mciteSetBstMidEndSepPunct{\mcitedefaultmidpunct}
{\mcitedefaultendpunct}{\mcitedefaultseppunct}\relax
\EndOfBibitem
\bibitem{LHCb-FIGURE-2020-016}
LHCbcollaboration, \ifthenelse{\boolean{articletitles}}{\emph{{RTA and DPA
  dataflow diagrams for Run 1, Run 2, and the upgraded LHCb detector }}, }{}
  \href{http://cdsweb.cern.ch/search?p=LHCb-FIGURE-2020-016&f=reportnumber&action_search=Search&c=LHCb+Figures}
  {LHCb-FIGURE-2020-016}, 2020\relax
\mciteBstWouldAddEndPuncttrue
\mciteSetBstMidEndSepPunct{\mcitedefaultmidpunct}
{\mcitedefaultendpunct}{\mcitedefaultseppunct}\relax
\EndOfBibitem
\bibitem{LHCb-PII-Physics}
LHCb collaboration, \ifthenelse{\boolean{articletitles}}{\emph{{Physics case
  for an LHCb Upgrade II --- Opportunities in flavour physics, and beyond, in
  the HL-LHC era}},
  }{}\href{http://arxiv.org/abs/1808.08865}{{\normalfont\ttfamily
  arXiv:1808.08865}}\relax
\mciteBstWouldAddEndPuncttrue
\mciteSetBstMidEndSepPunct{\mcitedefaultmidpunct}
{\mcitedefaultendpunct}{\mcitedefaultseppunct}\relax
\EndOfBibitem
\bibitem{LHCbCollaboration:2806113}
LHCb collaboration, \ifthenelse{\boolean{articletitles}}{\emph{{Future physics
  potential of LHCb}}, }{}
  \href{http://cdsweb.cern.ch/search?p=LHCb-PUB-2022-012&f=reportnumber&action_search=Search&c=LHCb+Notes}
  {LHCb-PUB-2022-012}, 2022\relax
\mciteBstWouldAddEndPuncttrue
\mciteSetBstMidEndSepPunct{\mcitedefaultmidpunct}
{\mcitedefaultendpunct}{\mcitedefaultseppunct}\relax
\EndOfBibitem
\bibitem{LHCb-PAPER-2014-059}
LHCb collaboration, R.~Aaij {\em et~al.},
  \ifthenelse{\boolean{articletitles}}{\emph{{Precision measurement of \CP
  violation in \mbox{\decay{\Bs}{\jpsi\Kp\Km}} decays}},
  }{}\href{https://doi.org/10.1103/PhysRevLett.114.041801}{Phys.\ Rev.\ Lett.\
  \textbf{114} (2015) 041801},
  \href{http://arxiv.org/abs/1411.3104}{{\normalfont\ttfamily
  arXiv:1411.3104}}\relax
\mciteBstWouldAddEndPuncttrue
\mciteSetBstMidEndSepPunct{\mcitedefaultmidpunct}
{\mcitedefaultendpunct}{\mcitedefaultseppunct}\relax
\EndOfBibitem
\bibitem{LHCb-PAPER-2015-013}
LHCb collaboration, R.~Aaij {\em et~al.},
  \ifthenelse{\boolean{articletitles}}{\emph{{Determination of the quark
  coupling strength $|\Vub|$ using baryonic decays}},
  }{}\href{https://doi.org/10.1038/nphys3415}{Nature Physics \textbf{11} (2015)
  743}, \href{http://arxiv.org/abs/1504.01568}{{\normalfont\ttfamily
  arXiv:1504.01568}}\relax
\mciteBstWouldAddEndPuncttrue
\mciteSetBstMidEndSepPunct{\mcitedefaultmidpunct}
{\mcitedefaultendpunct}{\mcitedefaultseppunct}\relax
\EndOfBibitem
\bibitem{LHCb-PAPER-2015-055}
LHCb collaboration, R.~Aaij {\em et~al.},
  \ifthenelse{\boolean{articletitles}}{\emph{{Measurement of the difference of
  time-integrated \CP asymmetries in \mbox{\decay{\Dz}{\Km\Kp}} and
  \mbox{\decay{\Dz}{\pim\pip}} decays}},
  }{}\href{https://doi.org/10.1103/PhysRevLett.116.191601}{Phys.\ Rev.\ Lett.\
  \textbf{116} (2016) 191601},
  \href{http://arxiv.org/abs/1602.03160}{{\normalfont\ttfamily
  arXiv:1602.03160}}\relax
\mciteBstWouldAddEndPuncttrue
\mciteSetBstMidEndSepPunct{\mcitedefaultmidpunct}
{\mcitedefaultendpunct}{\mcitedefaultseppunct}\relax
\EndOfBibitem
\bibitem{LHCb-TDR-023}
LHCb collaboration, \ifthenelse{\boolean{articletitles}}{\emph{{LHCb Framework
  TDR for the LHCb Upgrade II Opportunities in flavour physics, and beyond, in
  the HL-LHC era}}, }{}
  \href{http://cdsweb.cern.ch/search?p=CERN-LHCC-2021-012&f=reportnumber&action_search=Search&c=LHCb}
  {CERN-LHCC-2021-012}, 2022\relax
\mciteBstWouldAddEndPuncttrue
\mciteSetBstMidEndSepPunct{\mcitedefaultmidpunct}
{\mcitedefaultendpunct}{\mcitedefaultseppunct}\relax
\EndOfBibitem
\end{mcitethebibliography}

\end{document}